# Racah - Wigner quantum 6j Symbols, Ocneanu Cells for $A_N$ diagrams and quantum groupoids

R. Coquereaux *
*Centre de Physique Théorique*†
*Luminy, Marseille*

**Abstract**

We relate quantum 6J symbols of various types (quantum versions of Wigner and Racah symbols) to Ocneanu cells associated with $A_N$ Dynkin diagrams. We check explicitly the algebraic structure of the associated quantum groupoids and analyze several examples. Some features relative to cells associated with more general $ADE$ diagrams are also discussed.

# 1 Introduction

## 1.1 Contents of the paper

The first part of this paper provides a comparative discussion of different types of normalized and unnormalized quantum $6J$ symbols in the discrete case. Our discussion does not require any knowledge of Lie group theory, classical or quantum, since these symbols are here associated with graph data relative to $ADE$ diagrams (mostly $A_N$ diagrams in this paper). The symbols are then expressed in terms of another type of objects, called Ocneanu cells. The second part of this paper shows explicitly how such a system of cells gives rise to a finite dimensional bigebra properly described as a quantum groupoid. We illustrate those properties with diagrams $A_3$ and $A_4$. The last part contains complementary material that does not fit in previous sections but could be useful in view of several generalizations.

Our purpose in this paper is very modest. Indeed, all the objects that we shall manipulate have been already introduced and studied in the past, sometimes long ago: $6J$ symbols, quantum

* Email: Robert.Coquereaux@cpt.univ-mrs.fr
†Unité Mixte de Recherche (UMR 6207) du CNRS et des Universités Aix-Marseille I, Aix-Marseille II, et du Sud Toulon-Var; laboratoire affilié à la FRUMAM (FR 2291)

or classical, are considered to be standard material, cells and "double triangle algebras" have been invented in [29], [32] and analyzed for instance in [5], [38], [14] or [40], finally, quantum groupoids are studied in several other places like [7], [26] or [27]. However, it is so that many ideas and results presented in these quoted references are not easy to compare, not only at the level of conventions, but more importantly, at the level of concepts, despite of the existence of the same underlying mathematical "reality".

This article is not a review because it does not try to cover what has been done elsewhere and because it would require anyway many more pages. Moreover, it contains, we believe, a number of results not to be found in the literature. It should be considered, maybe, as the result of a (probably failed) attempt to present a pedagogical account on the subject, following the author's point of view. On purpose, the mathematical or physical background required for the reading of this article, which is a mixture of examples, remarks, general statements and comments, is very small. It is hoped that this article, which is almost self-contained and summarizes the necessary ingredients, can be of interest to readers belonging to different communities. The novice should certainly skip the general discussion that follows and proceed directly to the following sections.

## 1.2 General discussion

This article originated from the wish of obtaining an explicit description for the finite quantum groupoids associated with diagrams belonging to Coxeter - Dynkin systems[35]. Such a description involves five types of structure constants : one for the algebra, one for the cogebra, two others for the corresponding structures on the dual space, and a last type of structure constants (the so called Ocneanu cells) parametrizing the pairing between matrix units describing the two different products. These different types of structure constants appear like generalizations of quantum $6J$ symbols (classical $6J$ symbols describe, in usual quantum mechanics, the couplings of three angular momenta). In the simplest cases, the $A_N$ diagrams of the usual $ADE$ family of simply laced Dynkin diagrams, these different types of structure constants are simply related, and it is written in several places (for instance in [19]) that they can be expressed in terms of known quantum $6J$ symbols. One purpose of our paper is to clarify this problem.

Even before looking at the cell formalism, one has to face several issues about quantum $6J$ symbols themselves. The problems are of four types types: conceptual, terminological, notational, and computational. Conceptual: there are several nonequivalent definitions for the "non-normalized $6J$ symbols" (classical or quantum), the "normalized $6J$ symbols" being essentially unique. Terminological: the names associated with these non-normalized or normalized symbols fluctuate from one scientific community to the next, and even from one author to the other : "Racah $6J$ symbols" , "Wigner $6J$ symbols" or "recoupling coefficients" , for instance, seem to be used in a rather random way to denote one or another family of symbols. Notational: These objects refer to functions of 6 variables. Even taking into account invariance with respect to symmetries (that are not the same for "normalized" and "non-normalized" symbols), there are several ways to display the variables and several incompatible notations exist in the literature. Computational: in most references, quantum $6J$ symbols are given by recurrence relations; at best, they are expressed in terms of q-deformed hypergeometric functions. Only in [23] we found a close expression at roots of unity (derived from a formalism of spin networks), but the type of non normalized symbols used in this reference does not coincide with the more familiar type found for instance in [22]. A similar explicit formula which is difficult to use because of several misprints and undefined writing conventions can also be found in [32], which is nevertheless for us the most fundamental and inspiring article in this field.

Because of the above difficulties and although our purpose was not to study quantum $6J$ symbols but rather to use them in order to understand cells and quantum groupoids, we decided to devote a good part of the first section of this article to a quick survey of quantum $6J$ symbols, normalized or not. Since we had no wish to relate this study to the theory of representation of groups or quantum groups, we simply took the explicit formula established in [23] as a starting point providing an explicit definition of the normalized symbols (that we call Wigner $6J$ symbols). Since we work at roots of unity, the arguments of the symbols are therefore only points (vertices) on an $A_N$ diagram, and the admissible triangles that usually generalizes the law of composition of spins are only defined as non trivial entries in the multiplication table of the graph algebra of $A_N$. Among other results to be found in this first section, we establish relations between different types of non normalized $6J$ symbols (that we call Racah $6J$ symbols), for instance between those studied in reference [22] and those studied in reference [8]; none of these two references makes use of the close formula given in [23]. In our first section we also recall symmetry properties of normalized and non-normalized $6J$ symbols. The tetrahedral symmetries of the first are well known, the quadrilateral symmetries of the later do not seem to be well documented (but it would not be a surprise to learn that they have been described decennia ago, in books of quantum chemistry !) We therefore hope that this first section will be of interest for any reader interested in $6J$ symbols, classical or quantum.

Then, still in the first section, we turn to cells. After their introduction, long ago, by A. Ocneanu, in [29], together with the formalism of paragroups, in order to study systems of bimodules, cells have been used (mostly) in operator algebra and, to some extents, in statistical mechanics and conformal field theory (for instance [37]). One difficulty, here, is that there are also several types of cells, depending upon the kind of symmetries that one imposes in the axiomatics of the theory (for instance, the cells of reference [34] are not those of [29] or of [32]). In the last part of the first section, we make our own choice of symmetry factors and, in the case of $A_N$ diagrams, establish simple — and explicit — relations between cells and $6J$ symbols. Notice also that cells described in most older references are usually of horizontal and vertical length one; this is not so in the present article: our cells are not necessarily "basic" and the old "macrocells", for us, are just cells.

In references [31], [32] it was shown that, by studying spaces of paths on ADE Dynkin diagrams, and using the cell formalism introduced years before by the same author, one could define interesting spaces with two algebras structure (the "double triangle algebra"), whose theory of characters could be related to several interesting physical constructions (like the classification of modular invariant conformal field theories of affine $SU(2)$ type [9]). In [32] , the two multiplicative laws are defined on the same underlying vector space — rather than on a vector space and its dual, and this, in our opinion, complicates the description of the theory, for instance because the compatibility property between the two products — which is very easy to write when formalized in terms of product and coproduct —- is rather difficult to observe and to discuss. Reference [19] contains useful material about (basic) cells but does not mention the existence of associated bigebras. The so-called "double triangle algebras" however appear in reference [5] where they are discussed in the context of the theory of sectors and nets of subfactors (see also [6]); in this interesting paper, like in [32], the authors introduce the two multiplication laws on the same underlying vector space, with no study of the corresponding coproducts, and no mention of any algebraic structure described by a quantum groupoid.

Starting from another direction, a general definition for quantum groupoids (weak Hopf algebras) was proposed in [7], which was followed by a number of papers, in particular [26][28][27], describing and studying this structure. Several classes and lists of examples have been mentioned and studied in those articles, but the bigebras proposed by A. Ocneanu, although probably at the root of these developments (see [30]), do not appear in those lists, probably because of the lack of available written material. The particular quantum groupoids associated with Jones-Temperley-Lieb algebras and introduced in [26] should not be identified with the "Ocneanu quantum groupoids" that we discuss here (their structure and dimensions are very different), but it seems, according to a general property [4], that both types of examples can be generated from (different) Jones towers by a universal construction.

Reference [38], see also [41], starting from problems in conformal field theory, provides physical interpretations for several mathematical objects appearing in A. Ocneanu' s construction, and in particular relates the quantum symmetries associated with ADE diagrams to partition functions of boundary conformal field theory in presence of defects; in the same paper the authors strongly suggest the use of weak Hopf algebras to describe these structures and they indeed consider all the corresponding ingredients (coproduct, antipode, counit) in their framework. This reference also discusses many topics that we shall not mention at all here, but does not give explicit values for the cells. Because our approaches evolved independently, the formalism and notations differ. What we stress, in the present paper, is the conceptual interest of not defining the two products on the same vector space but rather, on a vector space and its dual, and keep in mind, at all times, the two adapted basis of elementary matrices, together with their dual basis. "Double triangle algebras" were also discussed and recognized as quantum groupoids in reference [14], where many proofs were given, starting from postulated properties of cells, taken from the axiomatics discussed in reference [32]. We can also mention [40] that analyses these algebras in terms of symmetries of faces models, and one section of [39], where an explicit calculation is carried out, relying on cell values obtained by explicit composition of basic cells.

We should finally mention a series of articles [12], [13], [39], [17], starting from [10], whose purpose is to provide explicit realizations for algebras of quantum symmetries and to show how to relate the modular invariant partition functions of conformal field theory (or more general toric matrices) without having to explicitly study the characters of associated bigebras, i.e., by exploiting some simple initial graph data. These results will not be summarized here.

In the second part, we explicitly describe the quantum groupoid structure of $A_N$ diagrams, i.e., we discuss the"double triangle algebras" relative to these examples. This is admitedly not general enough... but it has the advantage of being totally explicit. Contrarily to what was done in [14], we do not start from postulated properties of cells, but rather relate them — at least in the case of $A_N$ diagrams — to known properties of quantum Racah $6J$ symbols. Determination of the full structure of any such example amounts therefore to obtain values for all the cells, and this is done by using the results of the previous section. Using this totally explicit realization, we discuss[1] product, coproduct, antipode, counit, integrals, measures and adapted scalar products etc. both for the finite dimensional algebra $\mathcal{B}$ and for its dual $\widehat{\mathcal{B}}$. To illustrate the various objects and constructions given in the first and second part of this article, we systematically describe what happens in the case of the diagram $A_3$, and the fourth section describes the next non trivial example: the quantum groupoid associated with $A_4$, that we call the "golden quantum groupoid" because its norm is the golden number.

The final part of this article is a compendium of mostly unrelated comments and results about cells and associated structures, that do not properly fit in the previous sections. An explicit description of quantum groupoids associated with arbitrary members of Coxeter-Dynkin systems relies on the determination of all the corresponding cells (of various types). In a sense, the first three sections of the present article are devoted to the $A_N$ diagrams of the $SU(2)$ system and we hope that the amount of information given in the last part will help dedicated readers to find such explicit descriptions for quantum groupoids associated with other simply laced diagrams and also to go beyond the $SU(2)$ system. At the classical level, it is expected that any pair $(H \subset K)$ of Lie groups, or of finite subgroups of Lie groups, should give rise to a system of cells and to bigebra structures similar to those discussed here. At the quantum level, groups and subgroups are replaced by generalized Coxeter-Dynkin systems of diagrams, and we should obtain something similar. Strangely enough, in the classical situation, actually in the pure $SU(2)$ case, the existence of a space with two multiplication laws (called "Racah multiplication" and "Wigner multiplication") seems to be only mentioned in a rather ancient reference [2]. Values of $6J$ symbols for binary polyhedral systems $(H \subset K)$, where $H$ and $K$ are subgroups of $SU(2)$ (both can be equal and one may also take $K = SU(2)$) do not seem to be available for all cases, even in the classical literature of quantum chemistry, see for instance [3]; the corresponding analysis makes no use of the $SU(2)$ McKay correspondence [24], with no surprise since it had not been discovered at that time, and, for the same reason, does not use the algebraic structure provided by the theory of quantum groupoids.

Because of the lack of available references or because of the fact that relevant references are actually scattered among articles belonging to several independent communities, we tried to write a self-contained article that can be read by any reader. It certainly contains or refers to results originally obtained elsewhere and described in one or another specific language (opera-tor algebras, theory of sectors, conformal field theory, statistical mechanics, fusion categories, quantum groups, quantum groupoids, spin networks, quantum chemistry etc. ) but our presentation includes features or results that are of independent interest and discussed for the first time. Among the features of this article we could mention our comparative discussion of the different types of quantum $6J$ symbols, the explicit expression of $A_N$ Ocneanu cells in terms of specific Racah symbols, the fact of non using essential paths to define the double triangle algebras but rather using the module multiplication table associated with the chosen diagram, the description of these algebras in terms of quantum groupoids rather than in terms of a vector space endowed with several associative laws (to recover the later, one should choose some scalar product), the combinatorics giving the number of cells for the general case of an $ADE$ diagram, our discussion of quivers and quantum manifolds etc.

[1] This generalizes, for all $A_N$ the discussion given for the simplest case, $A_2$ in [15].

It is clear that quantum groupoids provide an appropriate framework for a description of the algebraic structures associated with double triangle algebras associated with $ADE$ diagrams and their systems of cells. It is expected that it is so, as well, for diagrams belonging to higher systems, like for the members of the Di Francesco – Zuber system of diagrams [18] (associated with $SU(3)$), but it seems that general proofs are not available in this context.

# 2 Quantum 6J symbols, quantum Racah symbols and Ocneanu cells for $A_N$ diagrams

In what follows, all the expressions are functions of integers. These integers can be thought as labels of irreducible representations of $SU(2)$ or $SU_q(2)$. Classically, the representation labelled by the integer $n$ has dimension $n+1$, so that $n$ refers to twice the usual spin. Warning: We are going to define several functions that depend on integers. For each such function $F$, one can define another function $f$ depending on half-integers, by setting $f(j_1, j_2, j_3, \ldots) = F(2j_1, 2j_2, 2j_3, \ldots)$. One should therefore be careful when comparing values of our functions (for instance the values of the quantum 6J symbols), with other definitions, tables or computer routines that can be found in the literature.

We choose an integer $N$ and consider the Dynkin diagram $A_N$. We call "level", the quantity $\ell = N-1$, set $\kappa = N+1$ (the Coxeter number of this diagram) and $q = exp(i\frac{\pi}{\kappa})$.

## 2.1 Triangular theta functions and tetrahedral normalizing factors

### 2.1.1 Quantum numbers and factorials

The quantum numbers and quantum factorial are defined by

$$[n] = \frac{q^n - q^{-n}}{q - q^{-1}} \qquad [n]! = [n][n-1][n-2]\ldots[1]$$

### 2.1.2 Admissible triplets or triangles and the fusion algebra of $A_N$

A triplet $(a, b, c)$ is called admissible whenever the irreducible representation labelled by $a$ belongs to the decomposition of the tensor product $b \otimes c$ of the representations $b$ and $c$ into irreps. This definition makes sense in $SU(2)$ and in the quantum group $SU_q(2)$ at roots of unity $q$ (warning: we set $q^{2(N+1)} = 1$). However one does not need to rely on group (or quantum group) theory knowledge since the boolean admissibility function $admissible[a, b, c]$ can be explicitly defined as follows [23] (this function is invariant with respect to any permutation of $a, b, c$):

$$(a+b+c) \bmod 2 = 0 \wedge a+b-c \geq 0 \wedge -a+b+c \geq 0 \wedge a-b+c \geq 0 \wedge a+b+c \leq 2\kappa - 4$$

A simple way to encode admissible triplets uses the adjacency matrix of the given diagram – here $A_N$. Indeed, the vector space $A_N$ spanned by the vertices of the $A_N$ diagram possesses an associative and commutative algebra structure: it is an algebra with unity, the vertex (0), and one generator, the vertex (1). Multiplication by the generator is encoded by the diagram $A_N$ itself : the product of a given vertex $\lambda$ by the vertex (1) is given by the sum of the neighbors of $\lambda$. Equivalently, this multiplication by (1) — also called the fundamental — is encoded by the adjacency matrix $N_1$ of the graph. Multiplication by linear generators $\lambda$ (arbitrary vertices of the diagram) is described by matrices $N_\lambda$, called fusion matrices. The identity is $N_0$, the identity matrix of dimension $N$. The other fusion matrices are obtained from $N_1$ by using the known recurrence relation for coupling and decomposition of irreducible $SU(2)$ representations (that we of course truncate at level $N-1$). The following can be taken as a definition of these matrices : the recurrence formula is $N_\lambda = N_1 N_{\lambda-1} - N_{\lambda-2}$. Matrices $N_\lambda$ are periodic in $\lambda$ and vanish when $\lambda = -1$ or $N$. They have positive integer entries called fusion coefficients.

Let us give for instance the three fusion matrices of the diagram $A_3$ and the multiplication table of its fusion algebra.

$$N_0 = \begin{pmatrix} 1 & 0 & 0 \\ 0 & 1 & 0 \\ 0 & 0 & 1 \end{pmatrix} N_1 = \begin{pmatrix} 0 & 1 & 0 \\ 1 & 0 & 1 \\ 0 & 1 & 0 \end{pmatrix} N_2 = \begin{pmatrix} 0 & 0 & 1 \\ 0 & 1 & 0 \\ 1 & 0 & 0 \end{pmatrix}$$

| · | 0 | 1 | 2 |
|---|---|---|---|
| 0 | 0 | 1 | 2 |
| 1 | 1 | 0 + 2 | 1 |
| 2 | 2 | 1 | 0 |

Any entry of the multiplication table – alternatively, any non-zero entry of a fusion matrix – determines an admissible triplet. In particular, the total number $d_n$ of admissible triplets, with one fixed edge (say $n$), is given by the sum of matrix elements of the fusion matrix $N_n$. With the above example, $d_0 = 3$, $d_1 = 4$, $d_2 = 3$. More generally, for $A_N$ diagrams, $d_n = (n+1)(N-n)$.

Actually we decide to introduce two copies of the set of admissible triplets: the first set is displayed in terms of triplets $\xi^n_{ab} = (a, n, b)$ with one horizontal edge (say $n$), the other set in terms of triplets $\alpha^x_{ac} = (a, x, c)$ with one vertical edge (say $x$). Using star - triangle duality , one can display admissible horizontal triangles as "vertical diffusion graphs" and admissible vertical triangles as "horizontal diffusion graphs".

Star-triangle duality :

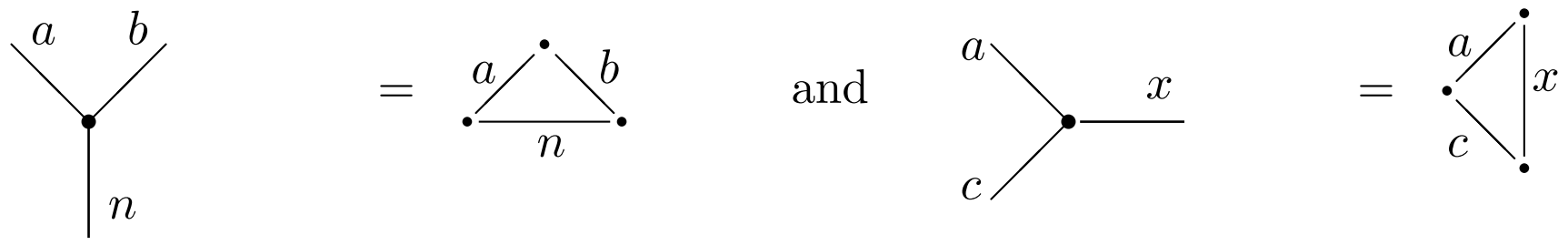

Diffusion graphs : the example of $A_3$.

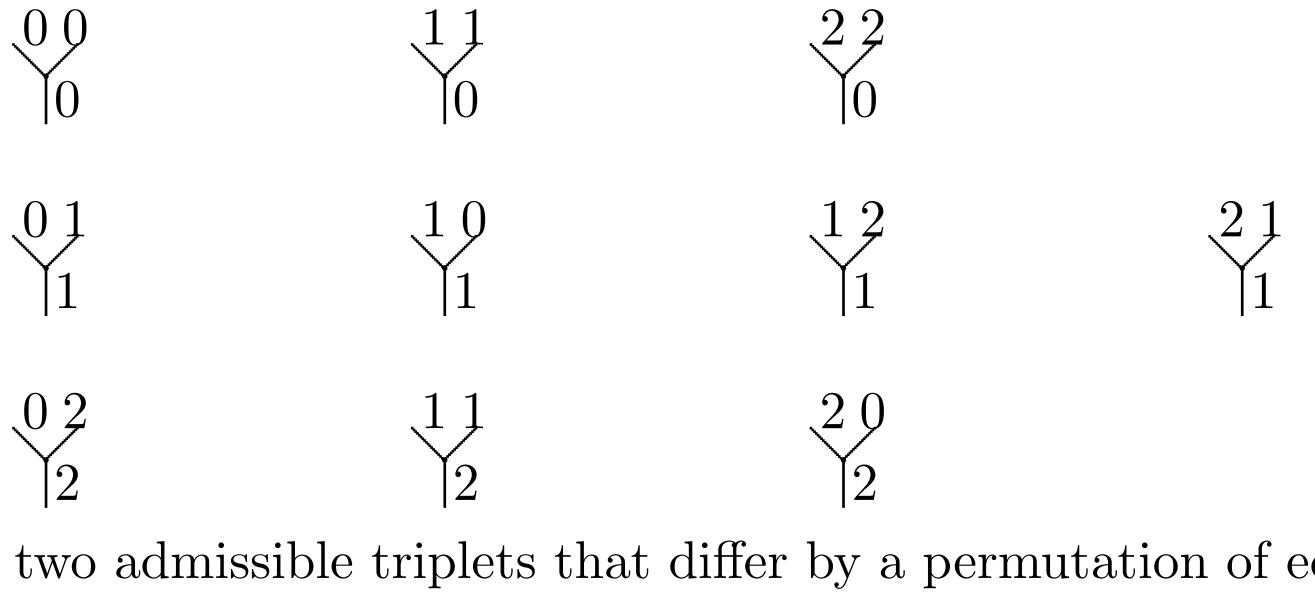

Warning: two admissible triplets that differ by a permutation of edges (for instance $(0,1,1)$ and $(1,0,1)$) are <u>not</u> identified. Permutations define equivalence classes of triplets, that we should call triangles rather than triplets, but we shall not stick to this terminology and will almost always use the word "triangle" when we mean "triplet". For reasons discussed later, horizontal triangles are also called essential paths on $A_N$ or horizontal paths, and vertical triangles are called "vertical paths". We shall later consider the graded vector spaces $\mathcal{H} = \sum_n \mathcal{H}_n$ spanned by horizontal triangles, and $\mathcal{V} = \sum_x \mathcal{V}_x$ spanned by vertical triangles. For $A_N$ diagrams, these two graded vector spaces are isomorphic, with common dimension $d_H = d_V$, but we do not identify them. In the case of $A_3$ for instance, the dimension is $\sum_n d_n = \sum_x d_x = 3+4+3 = 10$.

The quantum dimension $\mu_\lambda$ of the vertex $\lambda$ of $A_N$ is defined by the same recurrence formula as the fusion matrices, i.e., by $\mu_\lambda = \mu_1\mu_{\lambda-1} - \mu_{\lambda-2}$ with $\mu_0 = [1] = 1$ and $\mu_1 = [2] = 2\,cos(\pi/\kappa)$, so that $\mu_n = [n+1]$. One also defines the quantum mass of the diagram as $m = \sum_n \mu_n^2$.

### 2.1.3 Quantum triangular function associated with an admissible triangle

To every admissible triangle $\{k, l, m\}$ is associated a triangular function (also called theta function) denoted by the symbol $\theta$. In the theory of spin networks, it is equal to the value of the so - called theta net. This function is a symmetric function of its three variables.

$$\theta[k,\ell\ ,m] = (-1)^{j+p+r}\frac{[j+p+r+1]![j]![p]![r]!}{[j+p]![p+r]![r+j]!},$$

with

$$j = \frac{k+\ell\ -m}{2},\ p = \frac{\ell+m-k}{2},\ r = \frac{k+m-\ell}{2}$$

Example: we already determined the 10 admissible triangles of the $A_3$ case :

$$\{\{0,0,0\},\{0,1,1\},\{0,2,2\},\{1,0,1\},\{1,1,0\},\{1,1,2\},\{1,2,1\},\{2,0,2\},\{2,1,1\},\{2,2,0\}\}$$

Using permutations we have only 4 essentially distinct triangle classes : $\{0,0,0\}$, $\{0,1,1\}$, $\{0,2,2\}$, $\{1,1,2\}$. The corresponding values of the (quantum) triangular function are $1, -\sqrt{2}, 1, 1$. Results can be displayed as matrices:

$$\theta_0 = \begin{pmatrix} 1 & 0 & 0 \\ 0 & -\sqrt{2} & 0 \\ 0 & 0 & 1 \end{pmatrix} \theta_1 = \begin{pmatrix} 0 & -\sqrt{2} & 0 \\ -\sqrt{2} & 0 & 1 \\ 0 & 1 & 0 \end{pmatrix} \theta_2 = \begin{pmatrix} 0 & 0 & 1 \\ 0 & 1 & 0 \\ 1 & 0 & 0 \end{pmatrix}$$

### 2.1.4 The tetrahedral normalizing factor

Given a tetrahedra (four triangles), there is a quantity called the "tetrahedral normalizing factor" $NT[\{a,b,n\},\{d,c,x\}]$; it is equal to the square root of the product of the four theta functions associated with the faces of the tetrahedron.

$$NT[\begin{smallmatrix} a & b & n \\ d & c & x \end{smallmatrix}] = NT[\{a,b,n\},\{d,c,x\}] = \sqrt{|\theta[a,b,n]\,\theta[c,d,n]\,\theta[a,x,c]\,\theta[b,x,d]|}$$

The convention that we use for the argument [{a,b,n},{d,c,x}] is such that the upper line $(a,b,n)$ is an admissible triangle, that $d$ is opposite to $a$, that $c$ is opposite to $b$ and $x$ opposite to $n$. The lower line {d,c,x} is not an admissible triangle.

By construction, this function has tetrahedral symmetry (24 permutations). A tetrahedron is made of four triangles, so, we see that there are four possible upper lines (the choice of one triangular face) to denote a given tetrahedron, up to permutations of the three edges.

### 2.1.5 Comments

Warning : There are several possible choices (differing by a phase) of the tetrahedral normalizing factor. Indeed, triangular functions can be negative, and this introduces some freedom in the taking of the square root. One possibility is to define the tetrahedral normalizing factor as the square root of the absolute value of the product of the four triangles functions. We shall only use this "real" convention. However, the reader should be aware that another possibility (called the "complex convention") is to define the normalizing factor as the product of four square roots, but in that case one has to make a coherent choice for the complex determination of the square root, since each of the four arguments may be negative.

The admissible triangles used in the construction of the vector space $\mathcal{H}$ have been constructed by using the explicit multiplication table for the chosen $A_N$ diagram. There is another construction that uses a formalism of paths (actually essential paths) on graphs. One of the features of the present paper is precisely that one can go rather far, and in particular study the structure of the quantum groupoids associated with diagrams, without using the — rather involved — formalism of essential paths. It remains that the later has its own advantages, as well as a certain physical appeal. It will be briefly summarized later, in section 3.8.1.

### 2.1.6 Generalizations

There is a generalization of this construction where the $A_N$ diagram is replaced by an arbitrary simply laced Dynkin diagrams $G$. Such a diagram does not necessarily enjoy self-fusion but the vector space spanned by its vertices is always a module for the fusion algebra of $A_N$ where $N = \kappa - 1$, $\kappa$ being the Coxeter number of $G$. The associative algebra multiplication table $A_N \times A_N \mapsto A_N$ is replaced by a module multiplication table $A_N \times G \mapsto G$. Admissible triangles $(a, n, b)$ with $a, b \in G$ and $n \in A_N$ label the entries of this table. Also, when $G \neq A_N$ there is usually more than one triangle with given edges $a, n, b$ (i.e., there are multiplicities in the module multiplication table), so that one has sometimes to introduce an extra index (say $\xi$, like $\xi^n_{ab}$) to distinguish them.

The total number $d_n$ of admissible triangles, with one fixed edge of type $A_N$ (say $n$), is given by the sum of matrix elements of the annular[2] matrix $F_n$ : these matrices satisfy the same recurrence relation as the fusion matrices $N_n$ (it is a representation of the fusion algebra) and $F_0$ is the identity but the seed is now different : $F_1$ is the adjacency matrix of the diagram $G$.

Admissible triangles for the $A$ action on $G$ are drawn as horizontal triangles (or vertical diffusion graphs), and they span the graded horizontal vector space $\mathcal{H} = \sum_n \mathcal{H}_n$ of dimension $d_H = \sum_n d_n$. At a later stage, one also introduces vertical triangles but the construction is then more involved. One has to introduce another associative algebra (the algebra $Oc(G)$ of quantum symmetries of $G$ with a multiplication encoded by the so-called Ocneanu diagram of $G$) together with an action of $Oc(G)$ on $G$ described by dual annular matrices $S_x$. The vertical triangles correspond to the entries of the table describing the module action $Oc(G) \times G \mapsto G$. One introduces also a graded vertical vector space $\mathcal{V} = \sum_x \mathcal{V}_x$, of dimension $d_V = \sum_x d_x$. There exist both admissible triangles $\xi^n_{ab}$ of type $(a, n, b)$ and admissible triangles $\alpha^x_{ac}$ of type $(a, x, c)$ with $a, c \in G$ and $x \in Oc(G)$. The dimensions $d_x$ and $d_n$ usually differ. Since we have two associative algebras ($A_N$ and $Oc(G)$) and two distinct actions on $G$, we have four types of triangular functions With the same notation as before, their arguments (respectively of types $AGA$, $OGO$, $AAA$ and $OOO$ are $(a, n, b)$, $(a, x, b)$, $(m, n, p)$ and $(x, y, z)$. They can be displayed as triangles with colored vertices (black or white). In the particular case that we study in this article ($G = A$) all these concepts collapse. When $G$ enjoys self-fusion (this is not the case for $E_7$ and $D_{odd}$), one can furthermore introduce admissible triangles of type $(a, b, c)$, with $a, b, c \in G$.

Another type of generalization goes in a different direction : diagrams $A_N$ themselves are then replaced by the truncated Weyl chambers of the Lie group $SU(p)$ at some level $N - 1$. The theory that we are considering in this article deals only with the $A$ diagrams of the $SU(2)$ Coxeter-Dynkin system.

## 2.2 Tetrahedral function and quantum 6J symbols

### 2.2.1 The tetrahedral function $TET$

There is a function which, in theory of spin networks is given by the value of a tetrahedral net. We call it $TET$ and take the following formula (adapted from [23]) as a definition. Its arguments are the same as for the normalizing factor $NT$ : we consider $TET[\{a, b, n\}, \{d, c, x\}]$ , and the argument (specified by the two lists of three integers) is an admissible tetrahedron, ie the four triangles $\{a, b, n\}$, $\{d, c, n\}$, $\{d, b, x\}$, $\{a, c, x\}$ are admissible. In particular the top line $\{a, b, n\}$

[2]These matrices are also called "fused adjacency matrices" in other papers

$\left[\begin{smallmatrix} a & b & n \\ d & c & x \end{smallmatrix}\right] =$ 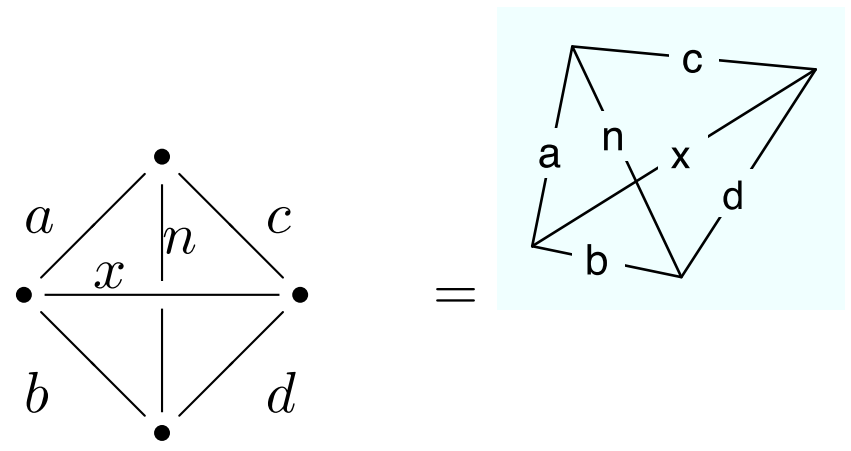

Figure 1: A quantum 6J function

is admissible[3]. With 4 triangles $A$ and 3 quadilaterals $B$, we define:

$$
\begin{aligned}
A[1] &= (a+c+x)/2;\ A[2] = (b+d+x)/2;\ A[3] = (a+b+n)/2; A[4] = (d+c+n)/2; \\
B[1] &= (b+c+x+n)/2;\ B[2] = (a+d+x+n)/2;\ B[3] = (a+b+d+c)/2; \\
m &= \text{Max}\ [A[1], A[2], A[3], A[4]]; \\
M &= \text{Min}\ [B[1], B[2], B[3]]; \\
I &= \prod_{i=1,j=1}^{i=4,j=3} [B[j] - A[i]]!, \quad \text{and} \quad E = [a]![b]![d]![c]![x]![n]!
\end{aligned}
$$

$$
\text{TET}\ \left[\begin{smallmatrix} a & b & n \\ d & c & x \end{smallmatrix}\right] = \text{TET}\ [\{a,b,n\},\{d,c,x\}] = \frac{I}{E} \sum_{s=m}^{M} \frac{(-1)^s [s+1]!}{\prod_{i=1}^{4}[s - A[i]]! \prod_{j=1}^{3}[B[j] - s]!}
$$

The function $TET$, like $NT$ has tetrahedral symmetry.

### 2.2.2 The quantum 6J symbols

They are defined as the quotient of the function $TET$ by the tetrahedral normalizing factor $NT$.

$$
\text{qSIXJ}\ [\{a,b,n\},\{d,c,x\}] = \begin{bmatrix} a & b & n \\ d & c & x \end{bmatrix} = \frac{\text{TET}\ [\{a,b,n\},\{d,c,x\}]}{\text{NT}\ [\{a,b,n\},\{d,c,x\}]}
$$

By construction 6J symbols have tetrahedral symmetry. They are denoted with square braces, as above. Because of this symmetry, one can denote this function of six variables ($q$ is fixed) by a tetrahedron, since the notation itself encodes the symmetry properties.

### 2.2.3 Symmetries of quantum 6J symbols and notations

If we denote the tetrahedral 6j function by the rectangular array $\left[\begin{smallmatrix} a & b & n \\ d & c & x \end{smallmatrix}\right]$ (the 6J symbol itself), we have to remember its symmetries : the upper line is always an admissible triangle and its ordering does not matter but it determines the entries of the lower line since $a, b, n$ edges are respectively opposite to $d, c, x$. There are four possible admissible triangles ($(a,b,n)$, $(b,d,x)$, $(a,c,x)$, $(c,d,n)$) and six permutations ($= 3!$) for the vertices of each of these triangles. We therefore recover the 24 symmetries of the tetrahedron. It is useful to remember the symmetries in terms of the admissible triangles that appear connected as follows in a given 6J symbol :

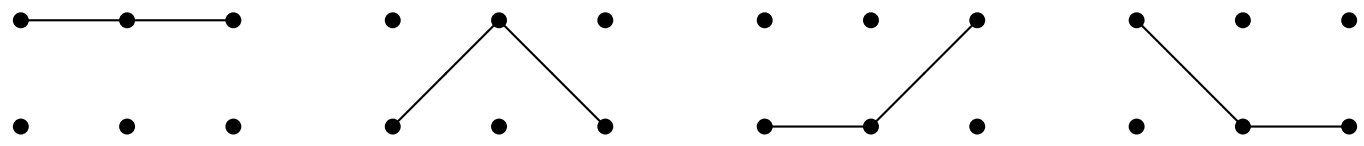

[3]This does not mean that a tetrahedron with edges equal to the prescribed integers can be constructed as a geometrical tetrahedron of the euclidean space $\mathbb{R}^3$.

A tetrahedron is admissible if all its faces are admissible triangles. If it is not admissible, we do not use the previously given explicit formulae and the value of the corresponding quantum 6J symbol is defined to be zero. One can think of the "6J function" as defined on the set of admissible tetrahedra (remember that $q$ is fixed), or as a function defined on the set of 6J symbols (the 2-dimensional arrays), rather than as a function of 6 variables. The 6J symbols themselves refer to the rectangular arrays of size $2 \times 3$ displaying the chosen arguments. It is sometimes necessary to distinguish between the 6J symbols themselves, the admissible tetrahedra, the 6J function (defined on the set of 6J symbols) and the value they take, but the context should in general be clear to decide which is which.

As we shall see later, the number of 6J symbols (here we mean admissible arrays), for a given graph $A_N$ i.e., a given choice of $q$, is $\sum_x \sum_n c(x,n)$ with $c(x,n) = Tr(N_n \cdot N_x \cdot N_n \cdot N_x)$ where $N$ are the fusion matrices. We also set $c(x) = \sum_n c(x,n)$. Using symmetries, one observes that the number of distinct tetrahedra classes is much smaller.

#### 2.2.4 Example : the $A_3$ case

With $n = 0,1,2$ we have $c(0,n) = \{3,4,3\}$, $c(1,n) = \{4,8,4\}$, $c(2,n) = \{3,4,3\}$, a total of $10 + 16 + 10 = 36$ admissible 6J symbols (arrays). Using symmetries, there are only 9 distinct classes of tetrahedra, namely those characterized for instance by the 6J symbols

$$\begin{bmatrix} 0 & 0 & 0 \\ 0 & 0 & 0 \end{bmatrix}, \begin{bmatrix} 0 & 1 & 1 \\ 1 & 0 & 0, \end{bmatrix} \begin{bmatrix} 0 & 2 & 2 \\ 2 & 0 & 0 \end{bmatrix}, \begin{bmatrix} 1 & 1 & 0 \\ 1 & 1 & 0 \end{bmatrix}, \begin{bmatrix} 1 & 1 & 2 \\ 1 & 1 & 0 \end{bmatrix}, \begin{bmatrix} 1 & 2 & 1 \\ 2 & 1 & 0 \end{bmatrix}, \begin{bmatrix} 2 & 2 & 0 \\ 2 & 2 & 0 \end{bmatrix}, \begin{bmatrix} 0 & 2 & 2 \\ 1 & 1 & 1 \end{bmatrix}, \begin{bmatrix} 1 & 2 & 1 \\ 1 & 2 & 1 \end{bmatrix}$$

Their respective values are :

$$\left\{1, -\frac{1}{\sqrt[4]{2}}, 1, -\frac{1}{\sqrt{2}}, \frac{1}{\sqrt{2}}, \frac{1}{\sqrt[4]{2}}, 1, \frac{1}{\sqrt[4]{2}}, \frac{1}{\sqrt{2}}\right\}$$

#### 2.2.5 Comments

Warning: It is certainly interesting to discuss how to extend the definition of quantities like $NT$, $TET$ or $qSIXJ$ (and the later functions to come) beyond the prescribed family of admissible tetrahedra. Such possible extended definitions do not necessarily formally coincide with the given explicit expression given in this paper. If the argument of $TET$ is not an admissible tetrahedron, we set its value to zero. The same convention applies to $qSIXJ$ and all the functions to be defined later in this article.

Since we had a phase freedom in the definition of $NT$, we have the same phase freedom in the definition of the function $qSIXJ$. As before, we use the real convention (because its classical limit coincides with the usual 6J symbols, that are real), but sometimes we shall make comments about what happens if one chooses the complex convention.

The quantum 6J symbols (with square brackets) that we just defined could be called "Wigner quantum 6J symbols" or "normalized 6J symbols". They coincide[4] with the normalized symbols (square brackets) used in the book [8], however the authors use spin variables rather than twice the spin and do not give any explicit expression for the function $TET$ (or for the function $qSIXJ$). Our own explicit expression for the tetrahedral function $TET$ can be found in the book [23], however we had to change the order of its entries to make it compatible with the conventions of [8]; another explicit expression for $TET$ (with other conventions and several misprints) can also be found in [32]. When $q$ is generic, the classical limit of our function $qSIXJ$ coincides with the pre-defined function SixJSymbol of Mathematica, provided one uses spin variables rather than twice the spin variables. Explicitly : SixJSymbol $_{\text{Mathematica}}$ [{a,b,f},{e,d,c}] = $\lim_{q\to 1}$ qSIXJ[{2a,2b,2f},{2e,2d,2c}].

[4]Actually, one should be cautious since these authors use the real convention for the normalizing factor in the classical case, and the complex convention in the quantum case, see the remark 3.11.5 in this reference, because the authors wanted to make contact with several results from the book [23] that uses anyway very different conventions.

### 2.2.6 Generalizations

When the graph $G$ is not of type $A_N$ we have four types of triangles with black or white vertices, and therefore five types of tetrahedra (they have $0, 1, 2, 3$ or $4$ black vertices).

To our knowledge, explicit expressions for the different types of 6J symbols associated with diagrams of the $D_N$ series, or with the three exceptional diagrams $E_6$, $E_7$ and $E_8$ are not known — they would be quantum hypergeometric functions with special properties.

When one moves from the $SU(2)$ system of diagrams (the usual $ADE$), to higher systems, for instance to the Di Francesco - Zuber system of diagrams, associated with $SU(3)$, the situation is of course even more involved and we are not aware of any explicit formula giving the $6J$ symbols for diagrams belonging to the $\mathcal{A}$ family.

## 2.3 The (standard) quantum Racah functions

### 2.3.1 Quantum standard Racah function associated with a quadrilateral : definition

Given an admissible tetrahedron described by some 6J symbol, for instance $\left[\begin{smallmatrix} a & b & n \\ d & c & x \end{smallmatrix}\right]$ we choose a (skew) quadrilateral — there are three of them. It is useful to project it on a plane in such a way that its shape forms a convex quadrilateral (the diagonals, that we represent as dotted lines, are inside). The sides of this quadrilateral correspond to two pairs of opposite sides of the given tetrahedron. To these three quadrilaterals one associates[5] "standard Racah symbols", $qRACAH[\{a, b, n\}, \{d, c, x\}] \equiv C[a, b, d, c\ ; n, x]$, as follows:

$$C[a, b, d, c\,; n, x] = \left\{\begin{smallmatrix} a & b & n \\ d & c & x \end{smallmatrix}\right\} =$$

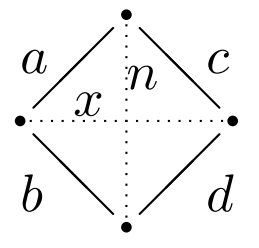

$$= (-1)^{\frac{a+b+c+d}{2}} \sqrt{[n+1][x+1]}\, \left[\begin{smallmatrix} a & b & n \\ d & c & x \end{smallmatrix}\right]$$

$$C[a, n, d, x\,; b, c] = \left\{\begin{smallmatrix} a & n & b \\ d & x & c \end{smallmatrix}\right\} =$$

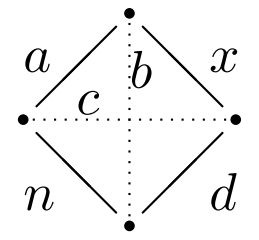

$$= (-1)^{\frac{a+n+d+x}{2}} \sqrt{[b+1][c+1]}\, \left[\begin{smallmatrix} a & b & n \\ d & c & x \end{smallmatrix}\right]$$

$$C[n, b, x, c\,; a, d] = \left\{\begin{smallmatrix} n & b & a \\ x & c & d \end{smallmatrix}\right\} =$$

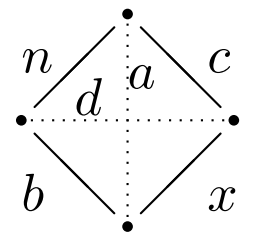

$$= (-1)^{\frac{n+b+x+c}{2}} \sqrt{[a+1][d+1]}\, \left[\begin{smallmatrix} a & b & n \\ d & c & x \end{smallmatrix}\right]$$

Notice the phase factor, associated with the perimeter of the quadrilateral, and the square root, associated with the pair of opposite diagonals. With the $C$ notation (see above first case) the first diagonal $n$ "follows" the last edge $c$ of the quadrilateral $a, b, d, c$. In the symbol $\left\{\begin{smallmatrix} \cdot & \cdot & \cdot \\ \cdot & \cdot & \cdot \end{smallmatrix}\right\}$, the upper line is always an admissible triangle and the last column refers to the two diagonals.

### 2.3.2 Symmetries of quantum Racah functions and notations

By construction, the Racah symbols are not invariant under tetrahedral symmetries but are invariant under the transformations that respect the chosen quadrilateral. For instance one can write $C[a, n, d, x\,; b, c]$ in eight possible ways :

$$\left\{\begin{smallmatrix} a & n & b \\ d & x & c \end{smallmatrix}\right\} = \left\{\begin{smallmatrix} n & d & c \\ x & a & b \end{smallmatrix}\right\} = \left\{\begin{smallmatrix} d & x & b \\ a & n & c \end{smallmatrix}\right\} = \left\{\begin{smallmatrix} x & a & c \\ n & d & b \end{smallmatrix}\right\} =$$
$$\left\{\begin{smallmatrix} n & a & b \\ x & d & c \end{smallmatrix}\right\} = \left\{\begin{smallmatrix} a & x & c \\ d & n & b \end{smallmatrix}\right\} = \left\{\begin{smallmatrix} x & d & b \\ n & a & c \end{smallmatrix}\right\} = \left\{\begin{smallmatrix} d & n & c \\ a & x & b \end{smallmatrix}\right\}$$

The above two lines correspond to the two distinct orientations of the same quadrilateral $a, n, d, x$. Each line corresponds to the cyclic permutations of its edges. The values of these eight symbols are equal. All together, we of course recover the 24 ($= 3 \times 8$) symmetries of the underlying tetrahedron. Notice that one does not discuss Regge symmetries in the discrete case.

[5]Warning: several authors use the same $2 \times 3$ brace notation but permute the last two columns

### 2.3.3 Relations between the three Racah functions associated with a tetrahedron

Since the three essentially distinct Racah symbols are associated with the same (tetrahedral) 6J symbol, we have a way to compare them. We find immediately :

$$\left\{\begin{smallmatrix} a & n & b \\ d & x & c \end{smallmatrix}\right\} = \frac{(-1)^{\frac{n+x}{2}}\sqrt{[b+1][c+1]}}{(-1)^{\frac{b+c}{2}}\sqrt{[n+1][x+1]}}\left\{\begin{smallmatrix} a & b & n \\ d & c & x \end{smallmatrix}\right\}$$

and

$$\left\{\begin{smallmatrix} n & b & a \\ x & c & d \end{smallmatrix}\right\} = \frac{(-1)^{\frac{n+x}{2}}\sqrt{[a+1][d+1]}}{(-1)^{\frac{a+d}{2}}\sqrt{[n+1][x+1]}}\left\{\begin{smallmatrix} a & b & n \\ d & c & x \end{smallmatrix}\right\}$$

### 2.3.4 Example : the $A_3$ case

We give here the list of Racah symbols (arrays). Their values will be given at the end of the next section, using cell notations. We display these symbols $\left\{\begin{smallmatrix} a & n & b \\ d & x & c \end{smallmatrix}\right\}$, with the column $n, x$ in middle position, into three matrix blocs labelled $x = 0, 1, 2$; each line is itself labelled by a pair $(a, c)$ and each column by a pair $(b, d)$. When there is more than one Racah symbol with given $a, b, c, d, x$ differing by the value of $n$, they appear in the same entry of the $x$-blocks. Warning: in the cell formalism (next section) lower indices $d$ and $c$ will be flipped.

$$\left\{\begin{matrix} \left\{\begin{smallmatrix} 0&0&0\\0&0&0 \end{smallmatrix}\right\} & & \left\{\begin{smallmatrix} 0&1&1\\1&0&0 \end{smallmatrix}\right\} & & \left\{\begin{smallmatrix} 0&2&2\\2&0&0 \end{smallmatrix}\right\} \\ \left\{\begin{smallmatrix} 1&1&0\\0&0&1 \end{smallmatrix}\right\} & , & \left( \left\{\begin{smallmatrix} 1&0&1\\1&0&1 \end{smallmatrix}\right\}, \left\{\begin{smallmatrix} 1&2&1\\1&0&1 \end{smallmatrix}\right\} \right) & , & \left\{\begin{smallmatrix} 1&1&2\\2&0&1 \end{smallmatrix}\right\} \\ \left\{\begin{smallmatrix} 2&2&0\\0&0&2 \end{smallmatrix}\right\} & & \left\{\begin{smallmatrix} 2&1&1\\1&0&2 \end{smallmatrix}\right\} & & \left\{\begin{smallmatrix} 2&0&2\\2&0&2 \end{smallmatrix}\right\} \end{matrix}\right\}$$

$$\left\{\begin{matrix} \left\{\begin{smallmatrix} 0&0&0\\1&1&1 \end{smallmatrix}\right\} & & \left\{\begin{smallmatrix} 0&1&1\\0&1&1 \end{smallmatrix}\right\} & & \left\{\begin{smallmatrix} 0&1&1\\2&1&1 \end{smallmatrix}\right\} & & \left\{\begin{smallmatrix} 0&2&2\\1&1&1 \end{smallmatrix}\right\} \\ \left\{\begin{smallmatrix} 1&1&0\\1&1&0 \end{smallmatrix}\right\} & & \left\{\begin{smallmatrix} 1&0&1\\0&1&0 \end{smallmatrix}\right\} & & \left\{\begin{smallmatrix} 1&2&1\\2&1&0 \end{smallmatrix}\right\} & & \left\{\begin{smallmatrix} 1&1&2\\1&1&0 \end{smallmatrix}\right\} \\ \left\{\begin{smallmatrix} 1&1&0\\1&1&2 \end{smallmatrix}\right\} & , & \left\{\begin{smallmatrix} 1&2&1\\0&1&2 \end{smallmatrix}\right\} & , & \left\{\begin{smallmatrix} 1&0&1\\2&1&2 \end{smallmatrix}\right\} & , & \left\{\begin{smallmatrix} 1&1&2\\1&1&2 \end{smallmatrix}\right\} \\ \left\{\begin{smallmatrix} 2&2&0\\1&1&1 \end{smallmatrix}\right\} & & \left\{\begin{smallmatrix} 2&1&1\\0&1&1 \end{smallmatrix}\right\} & & \left\{\begin{smallmatrix} 2&1&1\\2&1&1 \end{smallmatrix}\right\} & & \left\{\begin{smallmatrix} 2&0&2\\1&1&1 \end{smallmatrix}\right\} \end{matrix}\right\}$$

$$\left\{\begin{matrix} \left\{\begin{smallmatrix} 0&0&0\\2&2&2 \end{smallmatrix}\right\} & & \left\{\begin{smallmatrix} 0&1&1\\1&2&2 \end{smallmatrix}\right\} & & \left\{\begin{smallmatrix} 0&2&2\\0&2&2 \end{smallmatrix}\right\} \\ \left\{\begin{smallmatrix} 1&1&0\\2&2&1 \end{smallmatrix}\right\} & , & \left( \left\{\begin{smallmatrix} 1&0&1\\1&2&1 \end{smallmatrix}\right\}, \left\{\begin{smallmatrix} 1&2&1\\1&2&1 \end{smallmatrix}\right\} \right) & , & \left\{\begin{smallmatrix} 1&1&2\\0&2&1 \end{smallmatrix}\right\} \\ \left\{\begin{smallmatrix} 2&2&0\\2&2&0 \end{smallmatrix}\right\} & & \left\{\begin{smallmatrix} 2&1&1\\1&2&0 \end{smallmatrix}\right\} & & \left\{\begin{smallmatrix} 2&0&2\\0&2&0 \end{smallmatrix}\right\} \end{matrix}\right\}$$

### 2.3.5 Comments and generalizations

In a more general situation where $A_N$ is replaced by another $ADE$ diagram $G$, the formula giving the number $c(x, n)$ of 6J symbols (or of Racah symbols, or of cells) of type $(x, n)$ is replaced by $c(x, n) = Tr(F_n \cdot S_x \cdot F_n \cdot S_x)$ where $F$ and $S$ are respectively annular matrices and dual annular matrices respectively encoding the actions of $A_N$ and $Oc(G)$ on $G$.

## 2.4 Cells

Ocneanu cells, at least those that we use in this article devoted to the $A_N$ case, are simply related to our quantum standard Racah symbols at roots of unity. When the entries $n$ and $x$ are kept fixed, we simply denote a cell by a square with four corners $(a, b, c, d)$. Another useful mental picture is to think of $n$ as a length between $a$ and $b$ (or between $c$ and $d$) and to think of $x$ as a length between $a$ and $c$ (or between $b$ and $d$). Cells with fixed $(n, x)$ depend on four arguments and are drawn as squares, they have a simple identification in terms of standard Racah symbols, provided we flip the bottom labels $c, d$ and provided we display the horizontal and vertical lengths $(n, x)$ of the cell in the middle column position. This identification between particular Racah symbols and cells is justified by the study of their symmetries, as we shall see.

### 2.4.1 Cells with given horizontal length $n$ and vertical length $x$ : definition

We draw and define cells as follows:

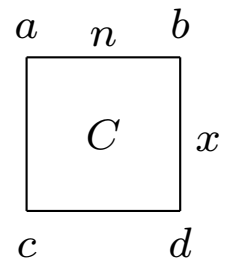

$$= \{\begin{smallmatrix} a & n & b \\ d & x & c \end{smallmatrix}\} =$$

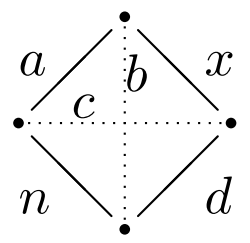

$$= (-1)^{\frac{a+n+d+x}{2}} \sqrt{[b+1][c+1]}\, [\begin{smallmatrix} a & n & b \\ d & x & c \end{smallmatrix}] = (-1)^{\frac{a+n+d+x}{2}} \sqrt{[b+1][c+1]}$$

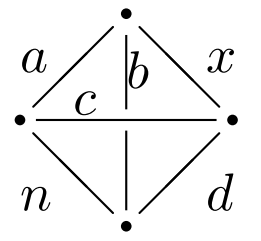

At this moment, this is only a fancy re - writing (or re - drawing) of the Racah symbols. Cells can of course be displayed in terms of tetrahedra but the later denote normalized quantum 6J, so that writing cells in terms of tetrahedra involve numerical prefactors (see above). They can also be expressed in terms of the "geometrical Racah symbols" (see a later section) by using the relation between the so-called standard and geometrical types of Racah symbols.

### 2.4.2 Symmetries of cells

From the quadrilateral symmetries of Racah symbols, we immediately obtain the symmetry properties of cells (warning: in the last two, $x$ and $n$ are permuted):

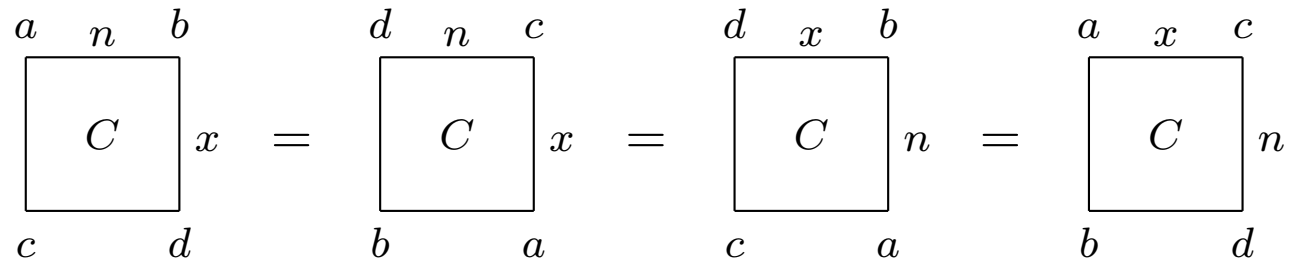

Keeping $n$ and $x$ fixed but permuting the two vertical or the two horizontal sides of the cell gives :

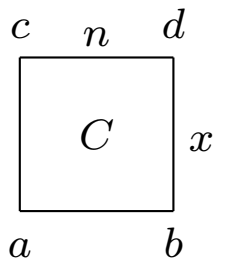

$$= \{\begin{smallmatrix} c & n & d \\ b & x & a \end{smallmatrix}\} =$$

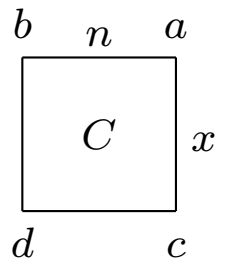

$$= \{\begin{smallmatrix} b & n & a \\ c & x & d \end{smallmatrix}\} = \frac{(-1)^{\frac{c+b}{2}} \sqrt{[a+1][d+1]}}{(-1)^{\frac{a+d}{2}} \sqrt{[c+1][b+1]}}$$

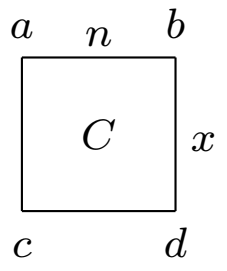

These symmetry properties of cells under horizontal and vertical reflections (which are here obtained as a re-writing of the known symmetry properties of Racah symbols), lead us to precisely identify[6] the Ocneanu cells mentioned in various papers – at least those appearing in the study of $A_N$ graphs – with the above objects. We shall see that they also obey the expected bi-unitarity properties. One could be tempted to redefine cells in order to get rid of the phase factor that appears in the reflection properties, but this leads to the appearance of unnecessary complex entries in cells values. Another possibility, even more drastic, would be to redefine the notion of cells in such a way that the symmetry properties are as simple as possible (in which case the unitarity properties and gluing properties, that we shall discuss next, would look more complicated) but one of our purposes, in this paper, was to establish relations between several objects of the literature... so we refrained from creating new ones.

### 2.4.3 Example : the $A_3$ case

We display the results on figures 2 and 3 using the cell notation (compare with 2.3.4):

[6]The convention used in [14] or [39] is the same as [29] but differs from the one used here by a sign

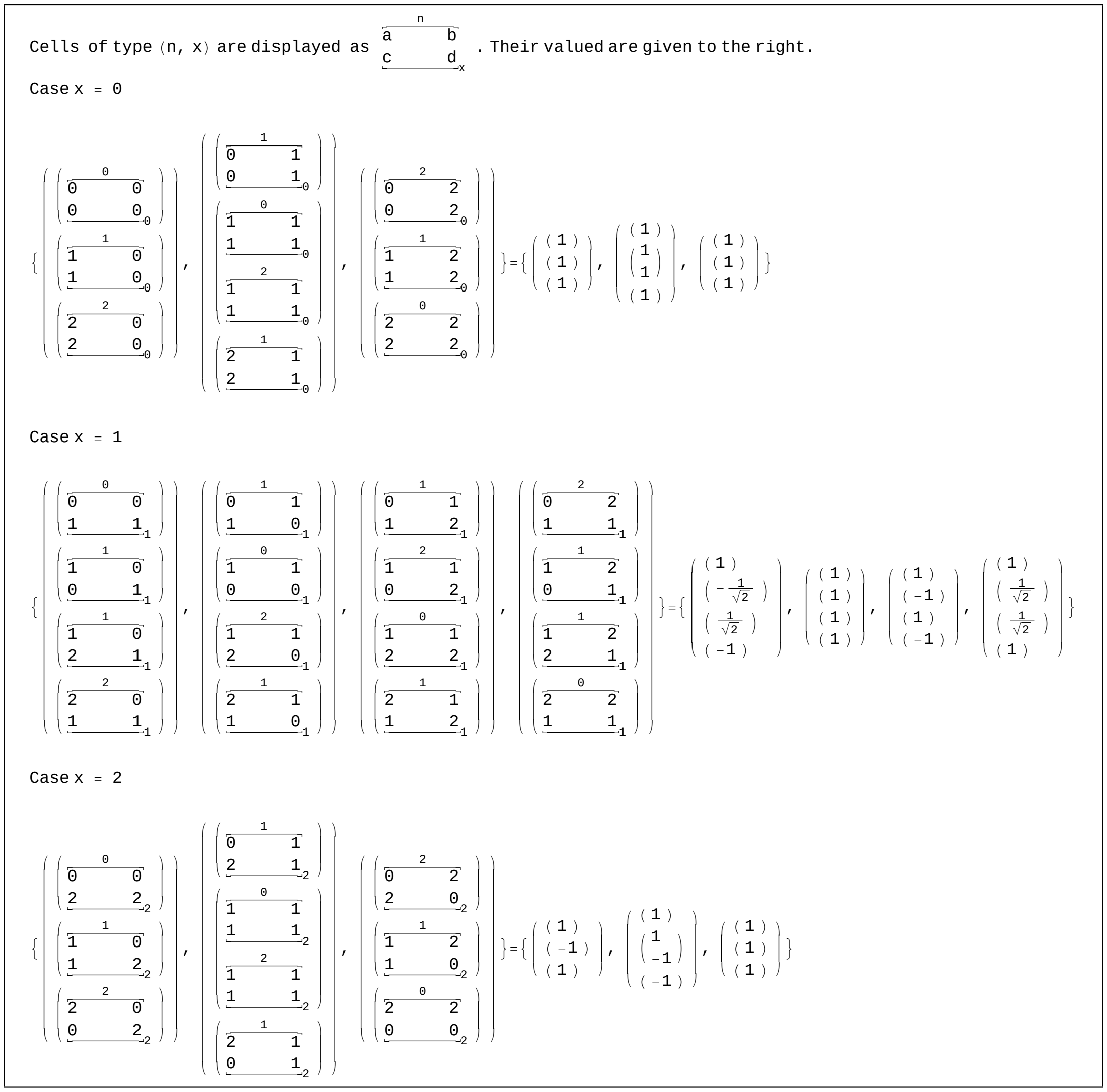

Figure 2: Cells for $A_3$

## 2.5 The geometrical quantum Racah symbols

We decided to define the quantum 6J symbols as explicit functions of six integers and of $q$. In the literature, people often prefer to start from Lie group theory or quantum groups, choose a base, actually a family of bases in a family of representative vector spaces (for the classical Lie group, this could be done by choosing a scalar product in its Lie algebra ), construct the so-called Racah symbols first and the so-called normalized Racah symbols (or Wigner 6J) next. The later turn out to be the same, anyway, independently of the choice of the conventions used to defined the Racah symbols themselves. We preferred to reverse the machine and start from explicitly given (normalized) 6J symbols. We should remember that the 6J symbols – with square brackets – should not be confused with Racah symbols – the later use braces. In this paper, we have made a choice, the one explicitly defined in the previous section, a choice that is in agreement with the usual normalization used in quantum mechanical textbooks, where one imposes unitarity of generators, and leading to what we called the "standard Racah symbols, or their quantum deformations. However, many authors working in knot theory or in spin networks, (see for instance [8]) use another normalization and obtain symbols that we shall call

$$\left\{\begin{pmatrix}(1)\\(1)\\(1)\end{pmatrix},\begin{pmatrix}(1)\\(\frac{1}{2})\\(\frac{1}{2})\\(1)\end{pmatrix},\begin{pmatrix}(1)\\(1)\\(1)\end{pmatrix}\right\}\left\{\begin{pmatrix}(1)\\(-\sqrt{2})\\(\sqrt{2})\\(-1)\end{pmatrix},\begin{pmatrix}(1)\\(1)\\(1)\\(1)\end{pmatrix},\begin{pmatrix}(1)\\(-1)\\(1)\\(-1)\end{pmatrix},\begin{pmatrix}(1)\\(\sqrt{2})\\(\sqrt{2})\\(1)\end{pmatrix}\right\}\left\{\begin{pmatrix}(1)\\(-1)\\(1)\end{pmatrix},\begin{pmatrix}(1)\\(\frac{1}{2})\\(-\frac{1}{2})\\(-1)\end{pmatrix},\begin{pmatrix}(1)\\(1)\\(1)\end{pmatrix}\right\}$$

Figure 3: Inverse cells for $A_3$. Arguments are the same as for cells.

geometrical[7] Racah symbols. They do not coincide with the previous "standard Racah symbols". These geometrical Racah symbols are used, for instance, in the book [8], where they are written with usual braces (warning: this reference moreover uses half-integer variables and calls "6J symbols" what we call "geometrical Racah symbols", and "normalized symbols" what we just call "6J symbols"). These geometrical Racah symbols – that we shall denote with double braces – are the objects that are used, mostly, by people doing geometry of 3-manifolds, whereas the standard Racah symbols – actually their classical limit – appear in most Physics textbooks (or in reference [22]). We shall only give the relation existing between these two types of Racah symbols since we shall only use the "standard" ones in the following, but we could use the "geometrical" ones, as well, with essentially the same properties.

$$\begin{aligned}\left\{\!\!\left\{\begin{array}{ccc} a & b & c \\ d & e & f\end{array}\right\}\!\!\right\} &= \frac{[c+1]}{(-1)^{\frac{a+b+d+e}{2}}}\sqrt{\text{Abs}\left[\frac{\theta[a,f,e]\theta[b,f,d]}{\theta[a,b,c]\theta[e,d,c]}\right]}\left[\begin{array}{ccc} a & b & c \\ d & e & f\end{array}\right] \\ &= \sqrt{\frac{[c+1]}{[f+1]}}\sqrt{\text{Abs}\left[\frac{\theta[a,f,e]\theta[b,f,d]}{\theta[a,b,c]\theta[e,d,c]}\right]}\left\{\begin{array}{ccc} a & b & c \\ d & e & f\end{array}\right\}\end{aligned}$$

## 2.6 Identities

### 2.6.1 Orthogonality

**For quantum 6J symbols.** We start from an admissible tetrahedron, represented for instance by the 6J symbol $\left[\begin{smallmatrix} a & b & n \\ d & c & x\end{smallmatrix}\right]$. For every choice of a quadrilateral (three possibilities), we can write an orthogonality relation involving a summation over one of the diagonals.

Orthogonality associated with the quadrilateral $(a,b,d,c)$.

$$\sum_x (-1)^{a+b+c+d}\;[n+1][n'+1]$$

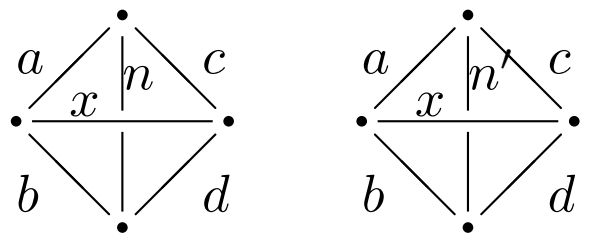

$$= \delta_{n,n'}$$

Orthogonality associated with the quadrilateral $(a,n,d,x)$.

$$\sum_b (-1)^{a+n+d+x}\;[c+1][c'+1]$$

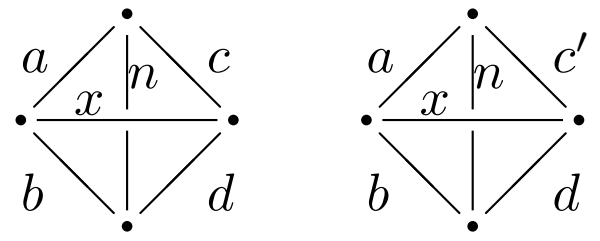

$$= \delta_{c,c'}$$

Orthogonality associated with the quadrilateral $(c,n,b,x)$.

$$\sum_d (-1)^{c+n+b+x}\;[a+1][a'+1]$$

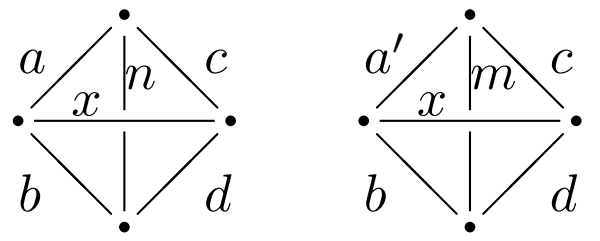

$$= \delta_{a,a'}$$

[7]Of course, these geometrical Racah symbols are no more geometrical than the standard ones... but we needed to invent a name, for the purpose of this article.

**For quantum Racah symbols** The orthogonality relations are deduced from the previous ones. What happens is that the pre-factor disappears (both for standard Racah symbols and geometrical Racah symbols). In order to avoid possible mistakes, one should re-write the previous tetrahedra by choosing a projection such that the selected quadrilateral is convex. The orthogonality relations are then immediately written in terms of Racah symbols (braces), but such a writing involves a lot of freedom since one can use symmetry properties.

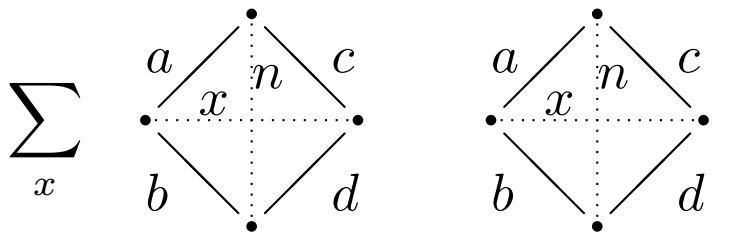

$$= \delta_{n,n'} = \sum_x \left\{ \begin{smallmatrix} a & b & n \\ d & c & x \end{smallmatrix} \right\} \left\{ \begin{smallmatrix} a & b & n' \\ d & c & x \end{smallmatrix} \right\}$$

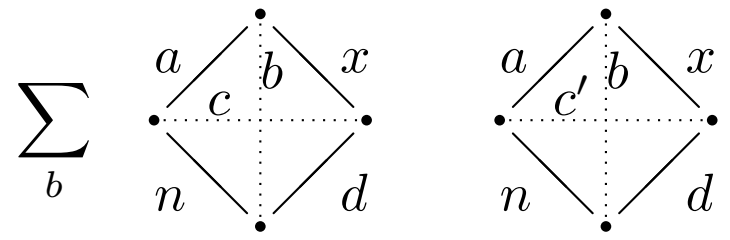

$$= \delta_{c,c'} = \sum_b \left\{ \begin{smallmatrix} a & n & b \\ d & x & c \end{smallmatrix} \right\} \left\{ \begin{smallmatrix} a & n & b \\ d & x & c' \end{smallmatrix} \right\}$$

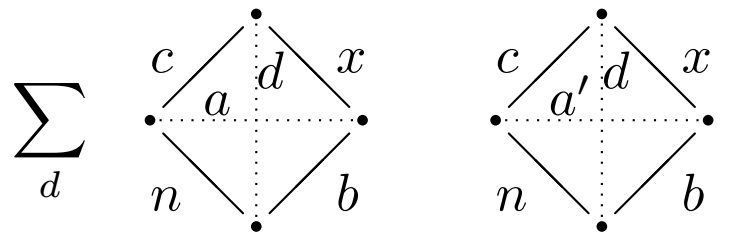

$$= \delta_{a,a'} = \sum_d \left\{ \begin{smallmatrix} c & n & d \\ b & x & a \end{smallmatrix} \right\} \left\{ \begin{smallmatrix} c & n & d \\ b & x & a' \end{smallmatrix} \right\}$$

**For cells** Again, this is a re-writing of the previous identities. However, because of the asymmetrical nature of variables $n, x$, and $a, b, c, d$, in the definition of cells, the first orthogonal identity looks very different from the other two. For cells written with corners $a, b, c, d$, horizontal edge $n$ and vertical edge $x$, the above relations for Racah symbols, together with the freedom associated with symmetry properties, lead to two identities:

In the first, the summation is over a corner and the first diagonal of the cell is fixed :

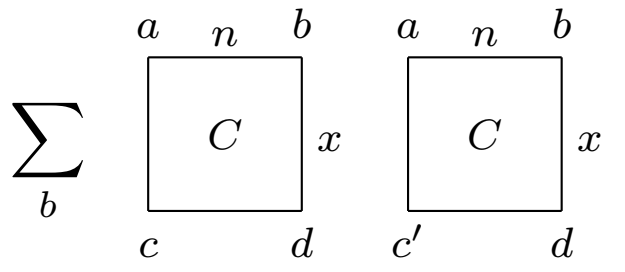

$$= \delta_{c,c'}$$

In the next, the summation is over a corner but the second diagonal of the cell is fixed :

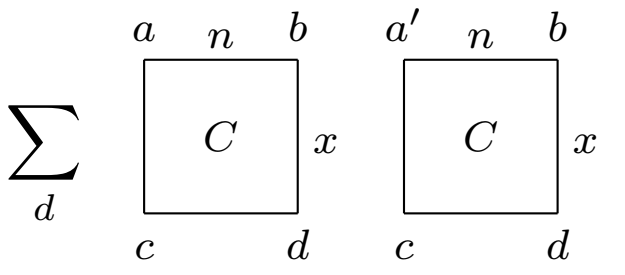

$$= \delta_{a,a'}$$

### 2.6.2 Pentagonal identity : quantum generalization of the Biedenharn - Elliott identity

**For quantum 6J symbols.** For every choice of nine elements $a, b, c, d, e, f, g, h, k$ belonging to the set $\{0, 1, 2, \ldots, N-1\}$, the following equation holds :

$$\begin{bmatrix} c & d & h \\ g & e & f \end{bmatrix} \begin{bmatrix} b & h & k \\ g & a & e \end{bmatrix} = s \sum_{x=0}^{N-1} (-1)^{\frac{x}{2}} [x+1] \begin{bmatrix} b & c & x \\ f & a & e \end{bmatrix} \begin{bmatrix} x & d & k \\ g & a & f \end{bmatrix} \begin{bmatrix} c & d & h \\ k & b & x \end{bmatrix}$$

where $s$, in front is the sign $s = (-1)^{\frac{a+b+c+d+e+f+g+h+k}{2}}$.

Up to the presence of prefactors (phases and $q$ - numbers), this identity is geometrically interpreted as follows : gluing two tetrahedra along a common face, in order to build a bipyramid, is equivalent to the gluing of three tetrahedra, sharing a new edge $x$ drawn between the opposite apexes of the bipyramid (see figure 4). Using tetrahedral symmetries, one can of course write this equation in many different ways if it is written in terms of 6J symbols.

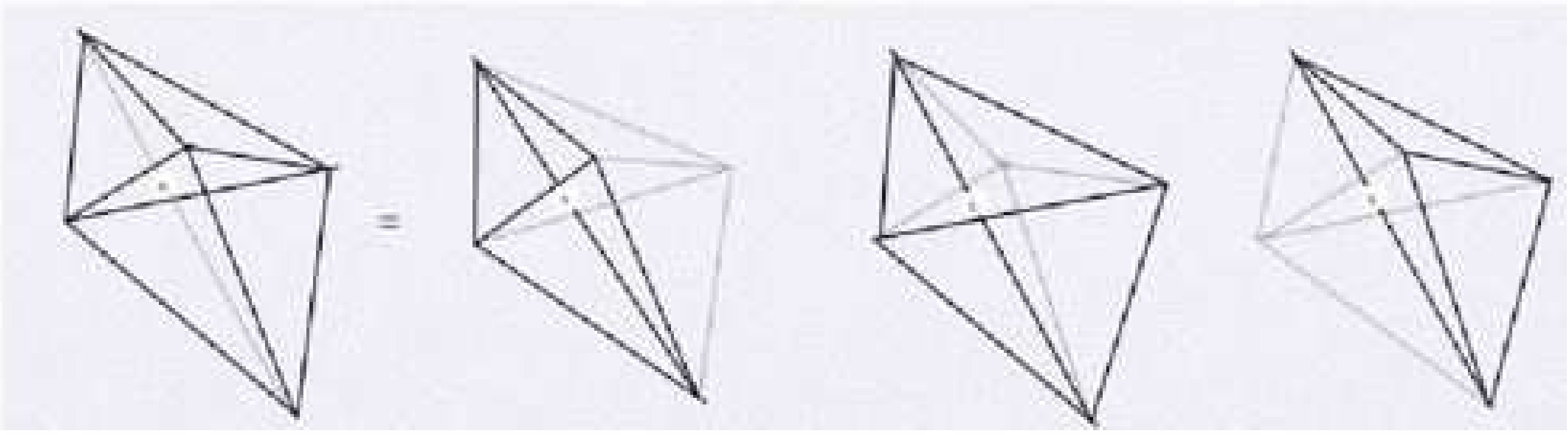

Figure 4: Pentagonal equation for 6J

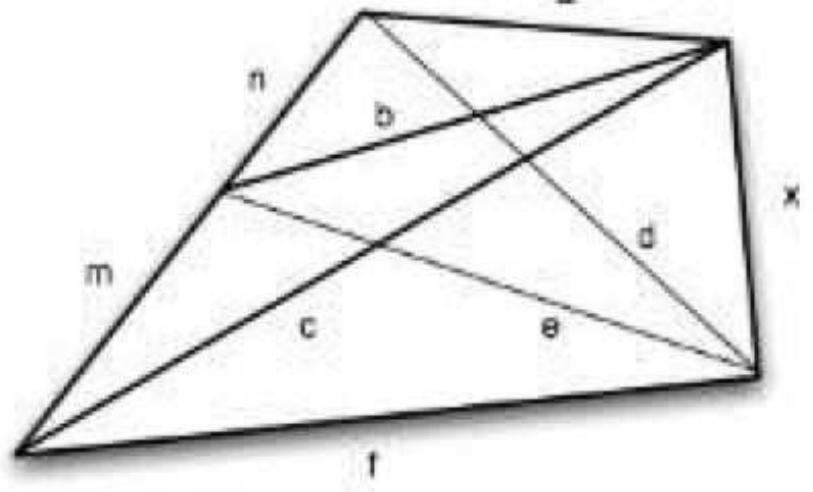

Figure 5: Racah identity for Racah symbols

**For quantum Racah symbols** the appearance of prefactors in the pentagonal equations — see below disappears. This also happens with orthogonality relations. A given 6J symbols gives rise to three –usually distinct – Racah symbols, so that, using relations between them and symmetries, we have several variants of this equation. Here is one of them:

$$\left\{\begin{matrix} c & d & h \\ g & e & f \end{matrix}\right\}\left\{\begin{matrix} b & h & k \\ g & a & e \end{matrix}\right\} = \sum_{x=0}^{N-1} \left\{\begin{matrix} b & c & x \\ f & a & e \end{matrix}\right\} \left\{\begin{matrix} x & d & k \\ g & a & f \end{matrix}\right\}\left\{\begin{matrix} c & d & h \\ k & b & x \end{matrix}\right\}$$

### 2.6.3 The Racah identity

In the classical situation, there are many identities attributed to Racah. One such set of equations allows to calculate the values of symbols for which the arguments are "big" (high spin) in terms of "smaller" symbols, i.e., symbols for which the arguments only involve representations of small dimension. In the quantum case, the situation is analoguous. For quantum Racah symbols and for cells this equation reads:

$$\left\{\begin{smallmatrix} a & p & c \\ f & x & d \end{smallmatrix}\right\} = \sum_{n,m,e,b} (\text{prefactors}) \left\{\begin{smallmatrix} a & n & b \\ e & x & d \end{smallmatrix}\right\}\left\{\begin{smallmatrix} b & m & c \\ f & x & e \end{smallmatrix}\right\}$$

The previous identity allows one to build large tetrahedra in terms of smaller ones (take the union of the two tetrahedra sharing a common triangular face $\{b, e, x\}$ in cases where the resulting structure is a bigger tetrahedron rather than a bi-pyramid : see figure 5. Using cells, the same relation can be read as a composition rule:

$$\begin{matrix} a \;\; p \;\; c \\ \boxed{\;C\;}\, x \\ d \qquad f \end{matrix} = \sum_{n,m,e,b} (\text{prefactors}) \begin{matrix} a \;\; n \;\; b \\ \boxed{\;C\;}\, x \\ d \qquad e \end{matrix} \begin{matrix} b \;\; m \;\; c \\ \boxed{\;C\;}\, x \\ e \qquad f \end{matrix}$$

Trivial cells, i.e., those for which $n = 0$ or $x = 0$, are equal to 1. A set of basic cells – those for which both $n = 1$ and $x = 1$ – can be obtained by imposing symmetry and orthogonality constraints. Repeated use of the Racah identity allows one to obtain any cell as sum of products of basic cells. The difficulty is to determine the correct prefactors. We shall not describe them since, in the present paper, values of Racah symbols for diagrams of type $A_N$ are obtained from explicit expressions for the 6J's. Let us nevertheless mention that one possibility is to use, simultaneously, a set of constraints coming from orthogonal and pentagonal identities (this would be a quantum generalization of the method described by [3]), that another possibility is to use the wire model of [23] (values of the triangular function can also be obtained in this way) and that a last possibility, that we briefly summarise now, is to use the concept of essential paths described in section 3.8.1: in the path model, admissible triangles are in one to one correspondance with normalized essential paths.

Consider for instance the cell $C$ displayed above and suppose $x = 1$ and $p > 1$. Since $x = 1$, vertical sides $ad$ and $cf$ correspond to normalized essential paths of length 1, i.e., also to (normalized) elementary paths of length 1. It usually happens that the essential path $\xi^p_{ac}$ corresponding to the cell top edge $p$ from $a$ to $c$ is a non trivial linear combination of (possibly backtracking) elementary paths. Same remark for the cell bottom edge $\eta^p_{df}$. In those cases, the prefactors appearing in the composition rule for cells are not equal to 1. Let us suppose, to further simplify, that $\eta^p_{df}$ is elementary (it is a succession of vertices) but that $\xi^p_{ac}$ is not, for instance take $\xi^p_{ac} = \sum_i \alpha_i\, [a, a_1, a_2, \ldots, c]_i$ where $[a, a_1, a_2, \ldots, c]_i$ are elementary paths (we shall give an example below). When calculating the value of the cell $C$, each term of the previous sum gives rise to the multiplication of $p$ basic cells, and the prefactor associated with it is equal to $|\alpha_i|$. In the same way, if $\eta^p_{df}$ is not elementary, we decompose it along elementary paths and keep the corresponding coefficients in the expression of $C$. Cases $n = 1$ and $x > 1$, or, more generally $n > 1$ and $x > 1$ can be treated similarly, at least for diagrams of type $A_N$, which are those that we consider in this paper. Examples:

$A_3$. The admissible triangle $\xi^2_{11}$ is identified with the essential path $(121 - 101)/\sqrt{2}$, and $\eta^2_{02}$ with 012 (which is elementary), cf. also Fig 2.

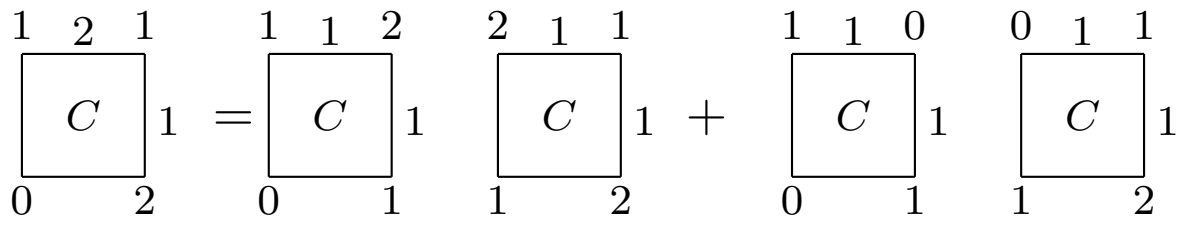

$$= \frac{1}{\sqrt{2}}\left(\frac{1}{\sqrt{2}}(-1) + \left(-\frac{1}{\sqrt{2}}(+1)\right)\right) = -1$$

$A_4$. The admissible triangle $\xi^3_{21}$ is identified with the essential path $\frac{1}{\varphi}2101 - \frac{1}{\varphi^{3/2}}2121 + \frac{1}{\varphi}2321$ and $\eta^3_{03}$ with 0123 (which is elementary, cf. section 4 and Fig. 17).

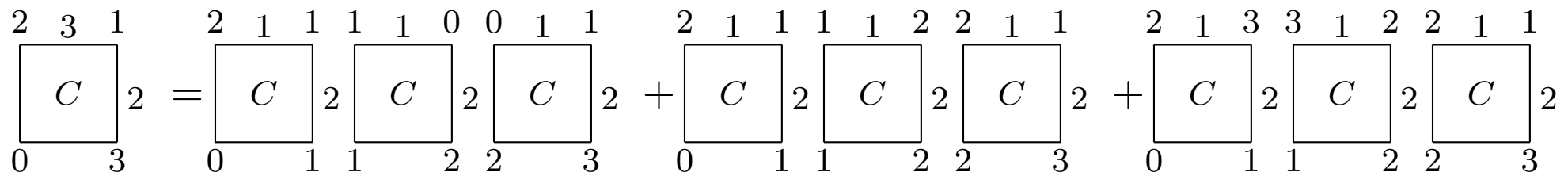

$$= \frac{1}{\varphi}\left(\frac{-1}{\sqrt{\varphi}}\right)\left(\frac{-1}{\sqrt{\varphi}}\right)(1) + \frac{1}{\varphi^{3/2}}\left(\frac{-1}{\sqrt{\varphi}}\right)\left(\frac{1}{\varphi}\right)(-1) + \frac{1}{\varphi}\left(\frac{1}{\varphi}\right)(-1)(-1) = \frac{1}{\varphi^2} + \frac{1}{\varphi^3} + \frac{1}{\varphi^2} = 1$$

### 2.6.4 Comments and generalizations

When the graph $G$ is not of $A$ type, and as discussed in section 2.1.6, there exist usually more than a single triangle with prescribed sides $(a, n, b)$ (or $(a, x, c)$), so that one has to introduce extra indices to distinguish between them. These new indices, in turn, show up in the writing of double triangles, cells and identities between them. Moreover, and as it was also discussed in section 2.1.6, there are usually four types of triangles, one can introduce two colors for vertices of those triangles in order to distinguish between them, or several types of lines if we use the dual notation. We have therefore also five types of tetrahedra (with $1, 2, 3, 4$ or 0 black vertices) and therefore five type of cells. Actually, only those with two black and two white vertices should be called "Ocneanu cells". By drawing all possible pyramids and using two kind of colored vertices one can see that the previous quantum pentagonal identity is replaced by a set of five coupled pentagonal identities nicknamed “the Big Pentagon” equation. This was commented in [30] and in reference [7].
In the $A_N$ cases, it was possible to make real the 6J or Racah symbols (hence the cells), this is not so in general, and the orthogonality relation becomes a unitarity relation – actually a bi-unitarity relation since one can always keep fixed one of the two diagonals.
The concept of cells, and the corresponding terminology, has changed along the years since their introduction in [29] . We call basic cells, those cells that are such that both $n$ and $x$ are equal to 1. Years ago, the non-basic cells were sometimes called “macrocells” or even “partition functions” because the rule of multiplication allowing one to compute macrocells from basic ones is reminiscent of statistical sums over Boltzman weights in two - dimensional statistical mechanics. In the present paper, “cells” can be of any horizontal length $n$ or of any vertical length $x$, and the macrocell terminology is unnecessary. See more comments about the general concept of cells in the last section.

### 2.7 Inverse quantum (standard) Racah functions and inverse cells

The justification for the introduction of “inverse Racah symbols” (and inverse cells) will only come at a later stage. At the moment it is enough to say that we need to introduce a function such that the following relation holds :

$$\sum_n \left\{\begin{smallmatrix} a & n & b \\ d & x & c \end{smallmatrix}\right\} \left\{\begin{smallmatrix} a & n & b \\ d & y & c \end{smallmatrix}\right\}_{-1} = \delta_{xy}$$

This is reminiscent of the orthogonality equation for Racah symbols, but it is nevertheless quite different since we are now performing the summation over one edge of the quadrilateral $\{a, n, d, x\}$ and not over a diagonal; this also shows that inverse cells and direct cells cannot be identified (even if we permute the arguments in all possible ways). From the known tetrahedral symmetry properties of the 6J symbols and the definition of the (direct) Racah symbols, we see that we can take one of the following equivalent definitions:

$$\begin{aligned}
\left\{\begin{smallmatrix} a & n & b \\ d & x & c \end{smallmatrix}\right\}_{-1} &= \frac{(-1)^{b+c}}{(-1)^{n+x}} \frac{[n+1][x+1]}{[b+1][c+1]} \left\{\begin{smallmatrix} a & n & b \\ d & x & c \end{smallmatrix}\right\} \\
&= \frac{(-1)^{\frac{b+c}{2}}}{(-1)^{\frac{n+x}{2}}} \frac{\sqrt{[n+1][x+1]}}{\sqrt{[b+1][c+1]}} \left\{\begin{smallmatrix} a & b & n \\ d & c & x \end{smallmatrix}\right\} \\
&= \frac{(-1)^{\frac{a+b+c+d}{2}}}{(-1)^{n+x}} \frac{[n+1][x+1]}{\sqrt{[a+1][b+1][c+1][d+1]}} \left\{\begin{smallmatrix} c & n & a \\ b & x & d \end{smallmatrix}\right\} \\
&= \left\{\begin{smallmatrix} a & b & n \\ d & c & x \end{smallmatrix}\right\}^2 / \left\{\begin{smallmatrix} a & n & b \\ d & x & c \end{smallmatrix}\right\}
\end{aligned}$$

Inverse (standard) Racah symbols are denoted as the Racah symbols themselves, but we add a lower $-1$ subscript to the pair of braces; we can also use a one-dimensional notation, replacing $C$ by $\Lambda$, and since we have inverse Racah symbols, we have also inverse cells. Notations are summarized as follows.

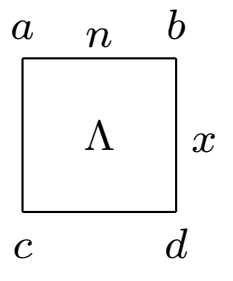

$$= \left\{\begin{smallmatrix} a & n & b \\ d & x & c \end{smallmatrix}\right\}_{-1} = \Lambda[a, n, d, x\ ; b, c]$$

# 3 Quantum groupoid structure

To every diagram $G$, member of a generalized Coxeter-Dynkin system, one associates a (particular type of) quantum groupoid $\mathcal{B}(G)$. In the present paper, we are interested in the $SU(2)$ system, i.e., the usual ADE Dynkin diagrams, and more particularly, in the $A$ diagrams. Our purpose is not to discuss the general theory but to show explicitly how the structure constants of this quantum groupoid are related to known quantum $6J$ symbols or to the Racah symbols defined in the first section, and to discuss examples. What we do now is to briefly recall how the bialgebra $\mathcal{B}(G)$ is constructed, with a particular emphasis on what happens when $G$ is a diagram of type $A_N$. We often write $\mathcal{B} = \mathcal{B}(G)$ since $G$ is usually given.

## 3.1 The algebra $\mathcal{B}$ of horizontal double triangles (or vertical diffusion graphs) and the algebra $\widehat{\mathcal{B}}$ of vertical double triangles (or horizontal diffusion graphs)

### 3.1.1 The vector space $\mathcal{B}$

The graded vector space $\mathcal{B}$ is defined as the algebra spanned by (admissible) double triangles (we recall that "triangle", here, actually means "triplet" since equivalent triplets are not identified). We already discussed the graded horizontal space $\mathcal{H} = \sum_n \mathcal{H}_n$, spanned by these triangles. A double triangle of type $n$ (the grade) is a pair of two triangles sharing the same edge $n$. Since these double triangles share a common horizontal edge, we shall call them horizontal double triangles (we shall introduce vertical ones later). These objects can be displayed as double-triangles or, using star-triangle duality, as diffusion graphs (the drawing below should be explicit); in given cases, one of these representation may be more intuitive than the other. The picture below is valid for any type of graph $G$ (where $n$-lines are not necessarily of the same type as $a$ and $b$ - lines), and in terms of triangles, one should also distinguish between two kinds of vertices (black or white dots). However, for $A_N$ diagrams, this distinction can be forgotten.

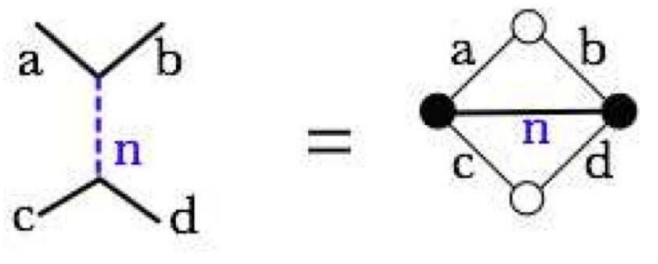

Figure 6: Horizontal double triangle for a graph $G$

The vector space $\mathcal{B}$ can be identified with the graded endomorphism algebra of $\mathcal{H}$. We shall come back to this algebra structure in a moment, but for now, we are only interested in the vector space structure. Admissible triangles for $A_3$ were drawn in the previous section in such a way that the bottom edge was always the same for each line of the display (i.e., for each component of the graded space $\mathcal{H}$). Double triangles can be readily constructed by pairing these triangles on all possible ways on their bottom edge. Elementary combinatorics leads again to the correct dimension $d_n^2$ for the grade $n$ component of $\mathcal{B}$. We display below four such basis elements chosen among the $4^2 = 16$ double triangles of type $n = 1$.

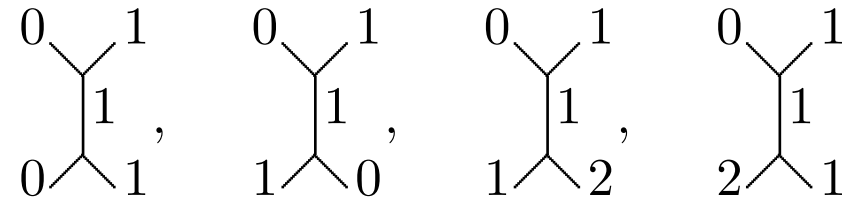

### 3.1.2 The algebra $(\mathcal{B}, \circ)$

The algebra structure on $\mathcal{B}$ is obtained by choosing the set of horizontal double triangles of type $n$ as a basis $\{e_I\}$ of elementary matrices[8] $e_I = e_{\xi\eta}$ for an associative product that we call $\circ$. Multi-indices $\xi$ are like $\xi = (a, b, n)$. Since $e_{\xi\eta} \circ e_{\kappa\lambda} = \delta_{\eta\kappa}\, e_{\xi\lambda}$, this multiplication can be

[8]Warning: "matrix units" of a matrix algebra often denote what we call "elementary matrices", but some people call "matrix units" the corresponding elements in the dual; to avoid ambiguities we shall not use this terminology.

interpreted graphically in terms of vertical concatenation (for instance using diffusion graphs) as follows[9] :

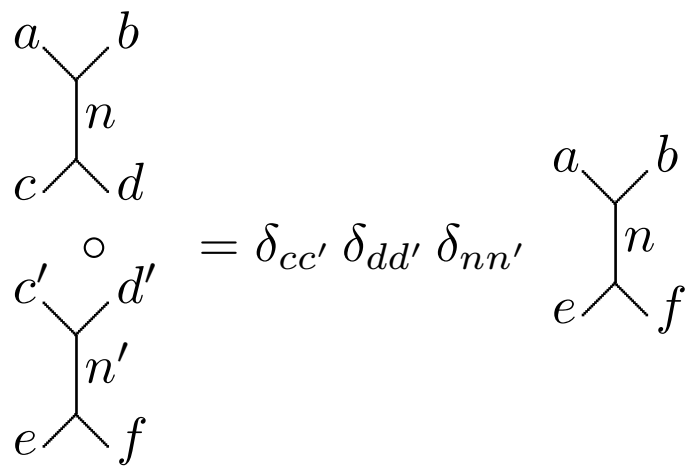

### 3.1.3 The vector space $\widehat{\mathcal{B}}$

Horizontal double triangles constitute a basis $e_I$ of the vector space $\mathcal{B}$. The corresponding dual basis (a basis of the dual vector space $\widehat{\mathcal{B}}$) is therefore well defined. Its elements $e^I$ are displayed as double triangles with a wide hat (see below), $e^I = \widehat{e}_I$, and we have, by definition, the pairing $< e^I, e_J >= \delta^I_J$.

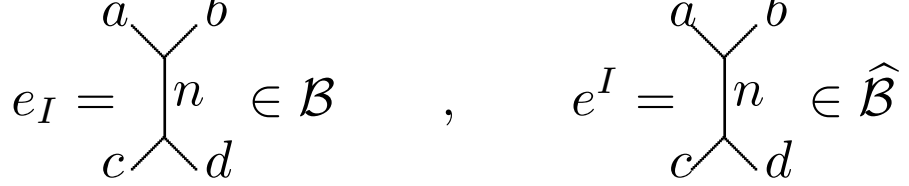

We are going to consider (and construct) another very specific base $f^A$ of the dual space $\widehat{\mathcal{B}}$. Its own dual basis $f_A = \widehat{f}^A$ is a basis of the bidual of $\mathcal{B}$ but it is identified with $\mathcal{B}$ since we are in finite dimension.

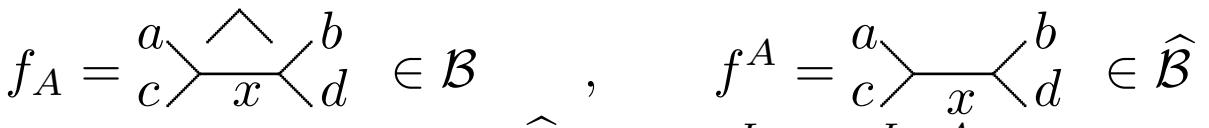

The change of basis in the vector space $\widehat{\mathcal{B}}$ reads $e^I = C^I_A f^A$, and the coefficients $C^I_A$ called “cells” are taken to be the cells defined in the previous section in terms of standard quantum Racah symbols. This is a very particular change of basis since potentially non zero coefficients occur only whenever external labels $a, b, c, d$ are the same for $e^I$ and for $f^A$. Elementary linear algebra gives $C^I_A =< e^I, f_A >$ and one can also write $f_A = C^I_A e_I$. Geometrical interpretation of this pairing : any tetrahedron (four triangular faces), see the one displayed on figure 1 can be built from the gluing of two double triangles, for instance the double triangle $(abn), (cnd)$ (articulated around the common edge $n$) and the the double triangle $(abn), (cnd)$ (articulated around the common edge $n$) and the double triangle $(acx), (bdx)$ (articulated around the common edge $x$).

The change of basis in the vector space $\mathcal{B}$ reads $e_I = \Lambda^A_I f_A$, and the coefficients $\Lambda^A_I$ are the “inverse cells”. Elementary linear algebra gives $\Lambda^A_I =< f^A, e_I >$. At the level of matrices we have $[\Lambda] = [C]^{-1}$. One can also write $f^A = \Lambda^A_I e^I$.

The basis $f^A$ is displayed in terms of diffusion graphs with an horizontal edge called $x$, or, using star-triangle duality, in terms of double triangles sharing a vertical edge (called $x$). Since these triangles share a common vertical edge, they will be called vertical double triangles.

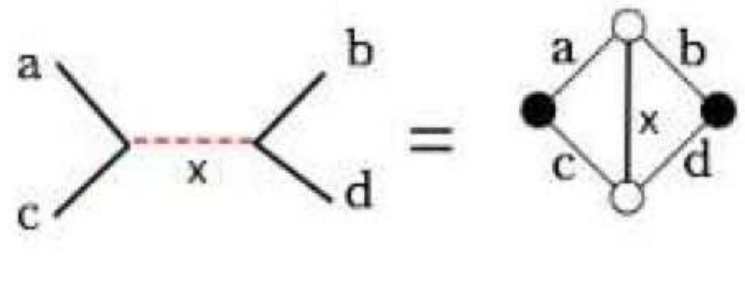

Figure 7: Vertical double triangle for a graph $G$

These changes of basis $e^I = C^I_A f^A$ , $f^A = \Lambda^A_I e^I$, $e_I = \Lambda^A_I f_A$, $f_A = C^I_A e_I$ are also written[10]:

[9]Several authors introduce double triangles differing from ours by a multiplicative factor.

[10]Notice the left-right flip of indices $c$ and $d$ and the middle vertical position of the pair $(n, x)$ in the Racah symbols

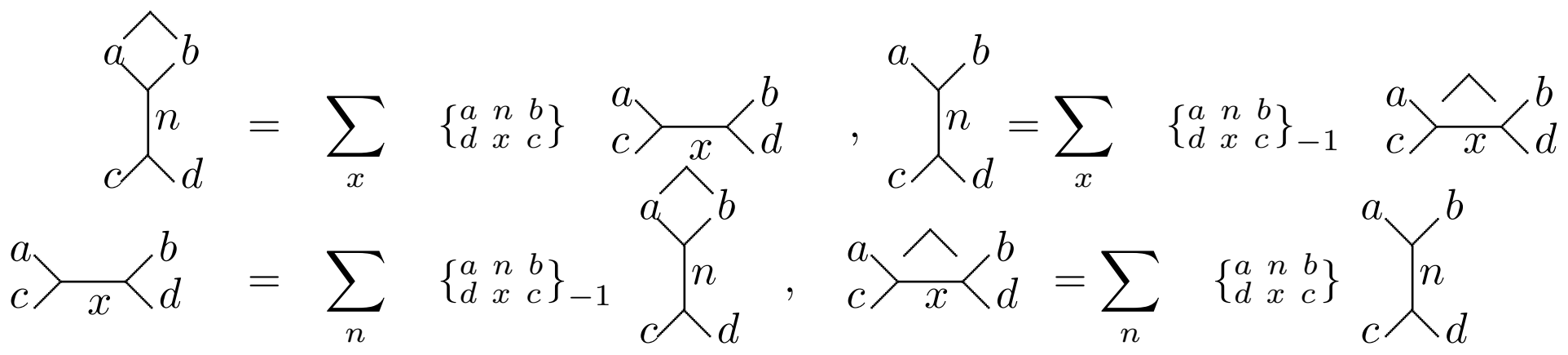

Remember that

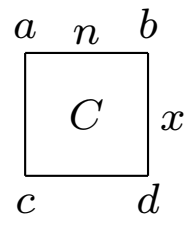

$$= \left\{\begin{smallmatrix} a & n & b \\ d & x & c \end{smallmatrix}\right\} = (-1)^{\frac{a+n+d+x}{2}} \sqrt{[b+1][c+1]} \left[\begin{smallmatrix} a & n & b \\ d & x & c \end{smallmatrix}\right]$$

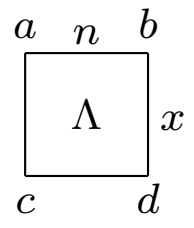

$$= \left\{\begin{smallmatrix} a & n & b \\ d & x & c \end{smallmatrix}\right\}_{-1} = \frac{(-1)^{b+c}(-1)^{\frac{a+d}{2}}}{(-1)^{\frac{n+x}{2}}} \frac{[n+1][x+1]}{\sqrt{[b+1][c+1]}} \left[\begin{smallmatrix} a & n & b \\ d & x & c \end{smallmatrix}\right]$$

In the classical $SU(2)$ recoupling theory, these relations — for instance the first — can be interpreted, in terms of the coupling of three angular momenta $j_1, j_2, j_3$, as a change of basis in the vector space $j_1 \otimes j_2 \otimes j_3$ and read as follows : you either couple $j_1, j_2$ and the result $j_{12}$ with $j_3$ or first couple $j_2, j_3$ and the result $j_{23}$ with $j_1$ (in this interpretation, we orient edges and suppress the "hat" on the left hand side by using some identification between the representation space and its dual, this amounts to choose a particular scalar product).

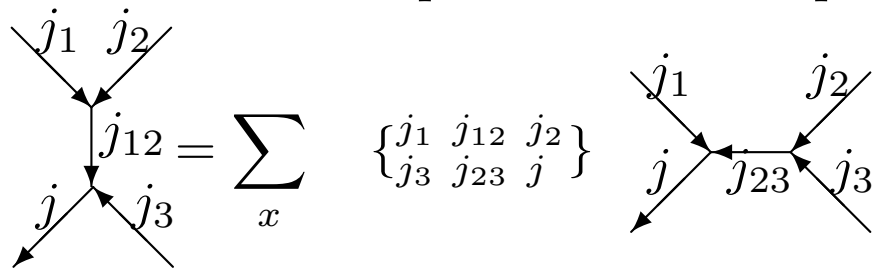

### 3.1.4 The algebra $(\widehat{\mathcal{B}}, \widehat{\circ})$

This algebra structure on $\widehat{\mathcal{B}}$ is obtained by choosing the set of vertical double triangles as a basis of elementary matrices $f^A = f^{\alpha\beta}$ for an associative product that we call $\widehat{\circ}$. Therefore $f^{\alpha\beta} \widehat{\circ} f^{\gamma\delta} = \delta^{\beta\gamma} f^{\alpha\delta}$. Multi-indices are like $\alpha = (a, c, x)$.This product looks therefore very simple in terms of the $f^A$ basis, but it is of course quite complicated when written in terms of the dual of the $e_I$ basis. This new multiplication can be interpreted graphically in terms of horizontal concatenation , as

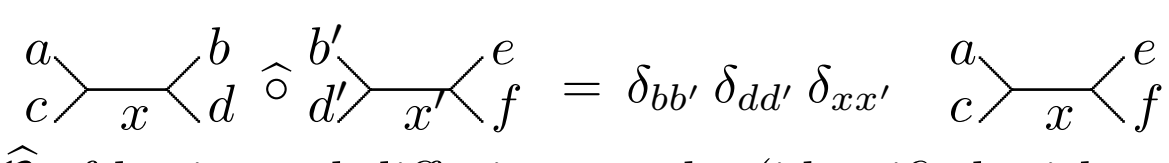

The vector space $\widehat{\mathcal{B}}$ of horizontal diffusion graphs (identified with vertical double triangles) can be considered as the graded endomorphism algebra of a vector space $\mathcal{V}$ whose elements are "vertical triangles" or "vertical paths".

### 3.1.5 Example of results : the bigebra structure of $A_3$

Figure 8 summarizes the algebra structure of $(\mathcal{B}, \circ)$, which is a direct sum of three simple components (blocks $n = 0, 1, 2$). Each entry denotes an elementary matrix $e_I$, i.e., one identifies any single entry of the following table with the particular elementary matrix possessing a unique non-zero matrix element equal to 1 at this location (all other entries are equal to 0). We also introduce in this table a shorthand notation for horizontal double triangles relative to the diagram $A_3$. This notation (inherited from the theory of paths and essential paths, see section 3.8.1) comes from denoting the admissible triangles themselves as follows : the three triangles of length $n = 0$ are $v_a = \xi^0_{aa}$, the three triangles of length $n = 1$ are $r_0 = \xi^1_{01}$ ("right" from 0), $r_1 = \xi^1_{12}$ ("right" from 1), $l_1 = \xi^1_{10}$ ("left" from 1), $l_2 = \xi^1_{21}$ ("left from 2), and the three triangles of length $n = 2$ are $g = \xi^2_{20}$ path "gauche", $d = \xi^2_{02}$ path "droit" , and $\gamma = \xi^2_{11}$.

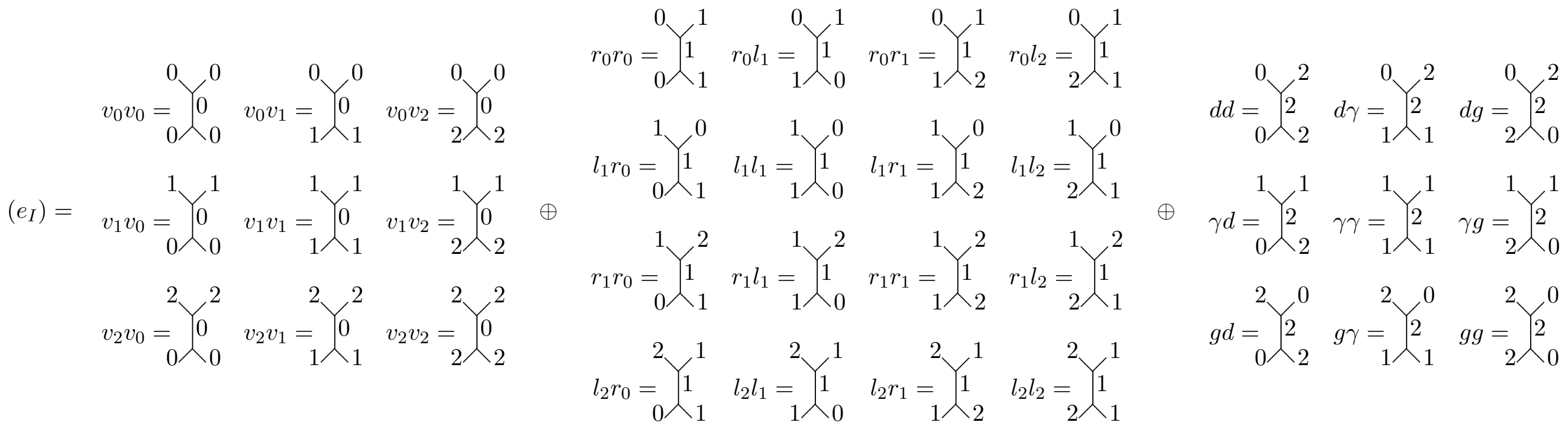

Figure 8: Elementary matrices $e_I$ of $(\mathcal{B}, \circ)$ for the $A_3$ diagram

Figure 9 summarizes the algebra structure of $(\widehat{\mathcal{B}}, \widehat{\circ})$, which is also a direct sum of three simple components (blocks $n = 0, 1, 2$). Each entry denotes an elementary matrix $f^A$, i.e., one identifies single entries of the following table with the particular elementary matrices possessing a unique non-zero matrix element equal to 1 at this location (all other entries are equal to 0). Table 10 expresses the same elementary matrices $f^A$ written in terms of the dual basis $e^I$ (coefficients are inverse cells). For instance, the elementary matrix $f^A$ which has a matrix element equal to 1 in position $(2, 2)$ of the third block (all others being zeros), is

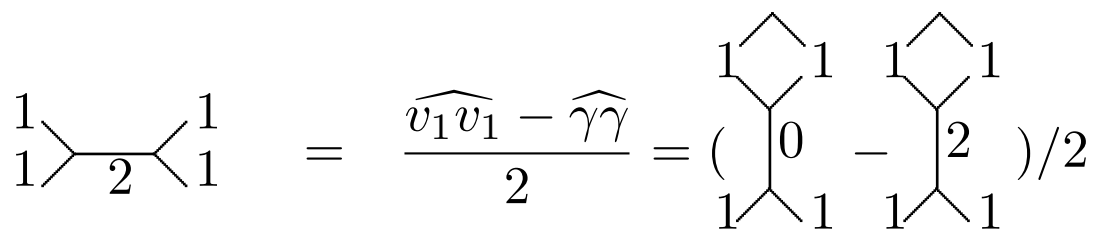

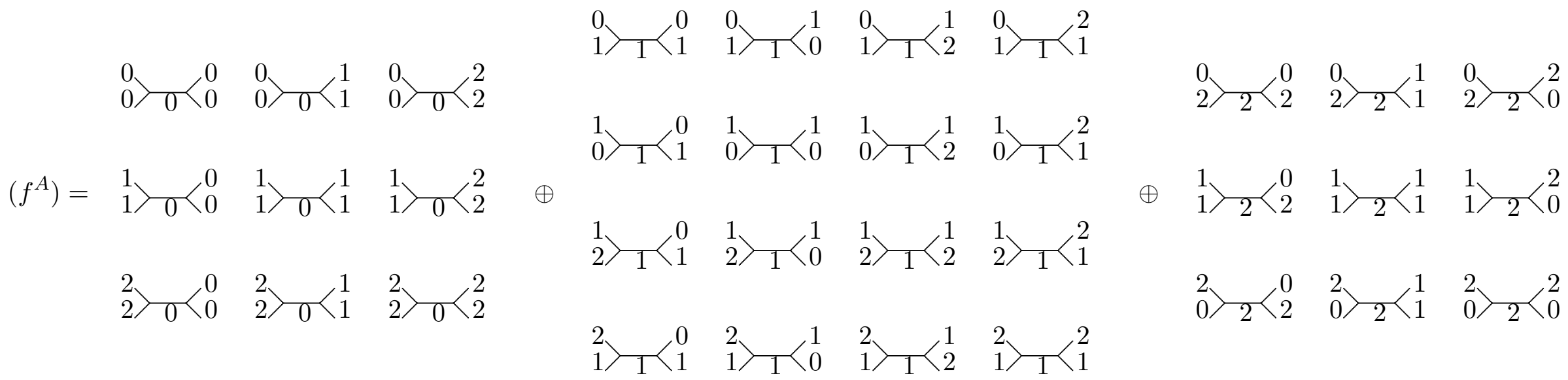

Figure 9: Elementary matrices $f^A$ of $(\widehat{\mathcal{B}}, \widehat{\circ})$ for the $A_3$ diagram (table 1 of 2)

We now give two other tables displaying the dual basis $e^I$ in $\widehat{\mathcal{B}}$ and the dual basis $f_A$ in $\mathcal{B}$. These two tables will turn out to be very useful when we determine the characters of the two algebras. Warnings (1) : the $e^I$ and $f_A$ are <u>not</u> elementary matrices, (2) : the hat that we put over basis vectors is only a notation used for distinguishing vectors belonging to dually paired basis ! For instance,

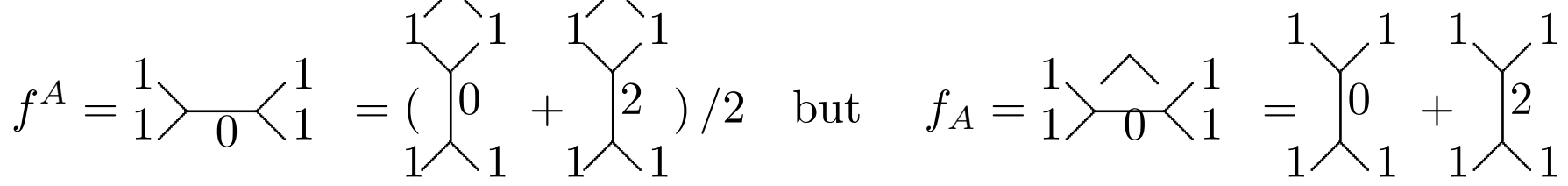

Tables describing the $A_3$ bigebra structure already appear in [39] and in [14] with slightly different sign conventions describing properties of cells under reflexion, they give rise to isomor-

$$
(f^A) = \begin{pmatrix} \widehat{v_0v_0} & \widehat{r_0r_0} & \widehat{dd} \\ \widehat{l_1l_1} & \frac{\widehat{v_1v_1}+\widehat{\gamma\gamma}}{2} & \widehat{r_1r_1} \\ \widehat{gg} & \widehat{l_2l_2} & \widehat{v_2v_2} \end{pmatrix} \oplus \begin{pmatrix} \widehat{v_0v_1} & \widehat{r_0l_1} & \widehat{r_0r_1} & \widehat{d\gamma} \\ -\sqrt{2}\,\widehat{l_1r_0} & \widehat{v_1v_0} & -\widehat{\gamma d} & \sqrt{2}\,\widehat{r_1r_0} \\ \sqrt{2}\,\widehat{l_1l_2} & \widehat{\gamma g} & \widehat{v_1v_2} & \sqrt{2}\,\widehat{r_1l_2} \\ -\widehat{g\gamma} & \widehat{l_2l_1} & -\widehat{l_2r_1} & \widehat{v_2v_1} \end{pmatrix} \oplus \begin{pmatrix} \widehat{v_0v_2} & \widehat{r_0l_2} & \widehat{dg} \\ -\widehat{l_1r_1} & \frac{\widehat{v_1v_1}-\widehat{\gamma\gamma}}{2} & \widehat{r_1l_1} \\ \widehat{gd} & -\widehat{l_2r_0} & \widehat{v_2v_0} \end{pmatrix}
$$

Figure 10: Elementary matrices $f^A$ of $(\widehat{\mathcal{B}}, \widehat{\circ})$ for the $A_3$ diagram (table 2 of 2)

$$
(f_A) = \begin{pmatrix} v_0v_0 & r_0r_0 & dd \\ l_1l_1 & v_1v_1+\gamma\gamma & r_1r_1 \\ gg & l_2l_2 & v_2v_2 \end{pmatrix} \oplus \begin{pmatrix} v_0v_1 & r_0l_1 & r_0r_1 & d\gamma \\ -\frac{l_1r_0}{\sqrt{2}} & v_1v_0 & -\gamma d & \frac{r_1r_0}{\sqrt{2}} \\ \frac{l_1l_2}{\sqrt{2}} & \gamma g & v_1v_2 & \frac{r_1l_2}{\sqrt{2}} \\ -g\gamma & l_2l_1 & -l_2r_1 & v_2v_1 \end{pmatrix} \oplus \begin{pmatrix} v_0v_2 & r_0l_2 & dg \\ -l_1r_1 & v_1v_1-\gamma\gamma & r_1l_1 \\ gd & -l_2r_0 & v_2v_0 \end{pmatrix}
$$

Figure 11: The dual basis $f_A \in \mathcal{B}$ in terms of elementary matrices $e_I$

phic bigebras; in particular the homomorphism property for the coproduct – that we describe now– can be explicitly checked there, as well.

### 3.2 Cogebra structures for $\mathcal{B}$, $\widehat{\mathcal{B}}$ and compatibility equations

At the moment, we have an algebra $\mathcal{B}$ with a product $\circ$ and another algebra structure $\widehat{\circ}$ on the dual $\widehat{\mathcal{B}}$ of $\mathcal{B}$. These two associative algebras are finite dimensional since, for a given $N$, the number of admissible triangles (or the number of admissible tetrahedra) is finite. By construction, we have therefore two finite dimensional semi - simple associative algebras. Simple blocks for the first law are labelled by $n$, and simple blocks for the second law are labelled by $x$. If $d_n$ and $d_x$ respectively denote the dimensions of these blocks, we have $\sum_n d_n^2 = \sum_x d_x^2$. Actually, for the case of $A_N$ diagrams, that we consider mostly in this paper, one can establish a correspondence between $n$ and $x$ values such that $d_n = d_x$ (see the explicit $A_3$ and $A_4$ examples for illustration).

However, what makes this construction particularly interesting is that there exists a compatibility relation between the two structures that make $\mathcal{B}$ not only an algebra and a cogebra, but a bigebra (same thing for $\widehat{B}$). We shall now describe the cogebra structures and the compatibility relation.

#### 3.2.1 The cogebra $\mathcal{B}$ and its coproduct $\Delta$

Since we have a product $\widehat{\circ}$ on $\widehat{\mathcal{B}}$, we have a coproduct $\Delta$ on $\mathcal{B}$ defined as follows (we use the particular basis $e_I$ introduced before). The coproduct $\Delta e_I$ of $e_I$ is the element of $\mathcal{B} \otimes \mathcal{B}$ such that $< e^K \otimes e^L, \Delta e_I > = < e^K \,\widehat{\circ}\, e^L, e_I >$.

$\Delta$, determined by the algebra law $\widehat{\circ}$ on the dual of $\mathcal{B}$ obeys the following compatibility relation: $\Delta(e_I \circ e_J) = \Delta e_I \circ \Delta e_J$. This compatibility relation between $\Delta$ and $\circ$ makes $\mathcal{B}$

$$
(e^I) = \begin{array}{ccc} \widehat{v_0v_0} & \widehat{v_0v_1} & \widehat{v_0v_2} \\ \widehat{v_1v_0} & \widehat{v_1v_1} & \widehat{v_1v_2} \\ \widehat{v_2v_0} & \widehat{v_2v_1} & \widehat{v_2v_2} \end{array} \oplus \begin{array}{cccc} \widehat{r_0r_0} & \widehat{r_0l_1} & \widehat{r_0r_1} & \widehat{r_0l_2} \\ \widehat{l_1r_0} & \widehat{l_1l_1} & \widehat{l_1r_1} & \widehat{l_1l_2} \\ \widehat{r_1r_0} & \widehat{r_1l_1} & \widehat{r_1r_1} & \widehat{r_1l_2} \\ \widehat{l_2r_0} & \widehat{l_2l_1} & \widehat{l_2r_1} & \widehat{l_2l_2} \end{array} \oplus \begin{array}{ccc} \widehat{dd} & \widehat{d\gamma} & \widehat{dg} \\ \widehat{\gamma d} & \widehat{\gamma\gamma} & \widehat{\gamma g} \\ \widehat{gd} & \widehat{g\gamma} & \widehat{gg} \end{array}
$$

Figure 12: The dual basis $e^I \in \widehat{\mathcal{B}}$

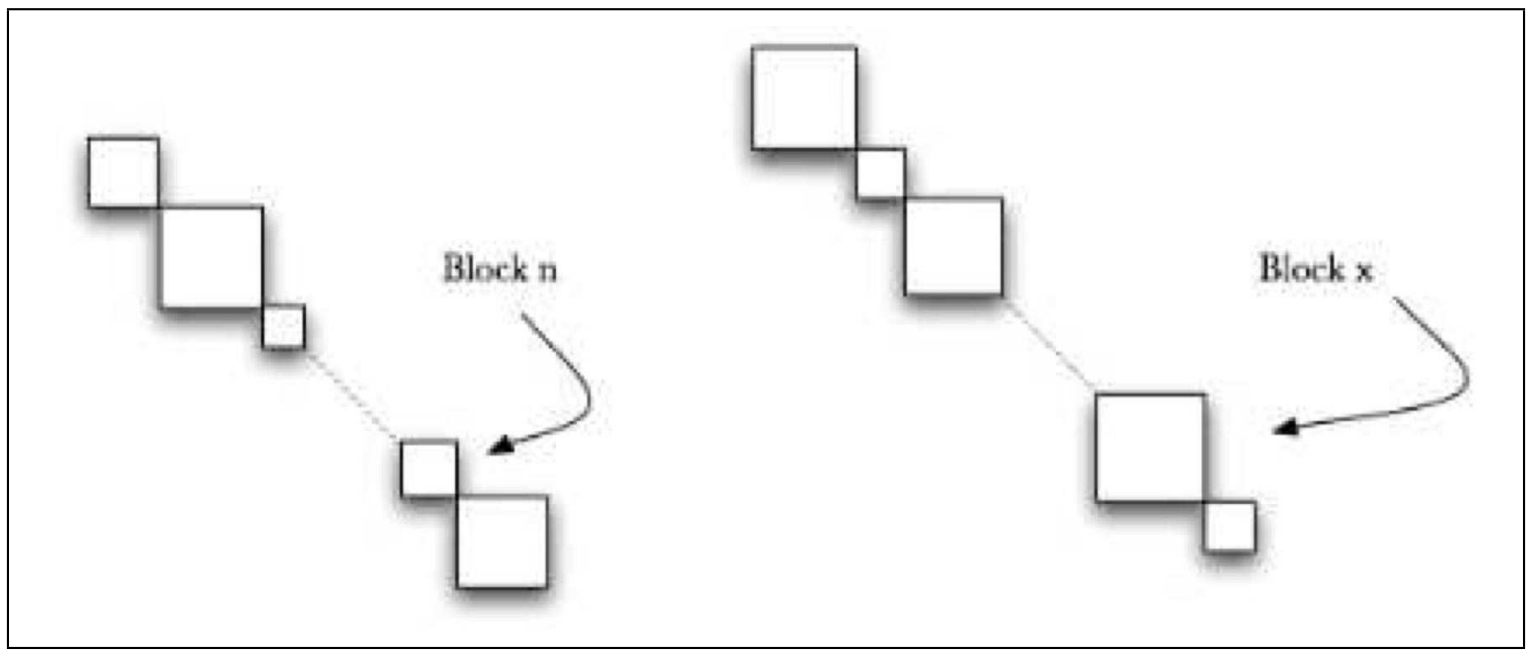

Figure 13: General block structures for $\mathcal{B}$ and $\widehat{\mathcal{B}}$

a bigebra. This bigebra actually satisfies other nice properties (recalled below), so that $\mathcal{B}$ is actually a weak Hopf algebra (a quantum groupoid).

The concept of bigebra is self-dual, so that instead of defining a coproduct $\Delta$ on $\mathcal{B}$, using the associative law on the dual of $\mathcal{B}$, we could have constructed a coproduct $\widehat{\Delta}$ on the dual of $\mathcal{B}$ by using the associative law defined on $\mathcal{B}$. Of course this coproduct $\widehat{\Delta}$ is compatible with $\widehat{\circ}$.

To find an explicit expression of $\Delta$ on the basis $e_I$, we compute the right hand side of the pairing $< e^K \otimes e^L, \Delta e_I > = < e^K \,\widehat{\circ}\, e^L, e_I >$ by expressing the dual of the vertical basis in terms of the horizontal basis. The right hand side is

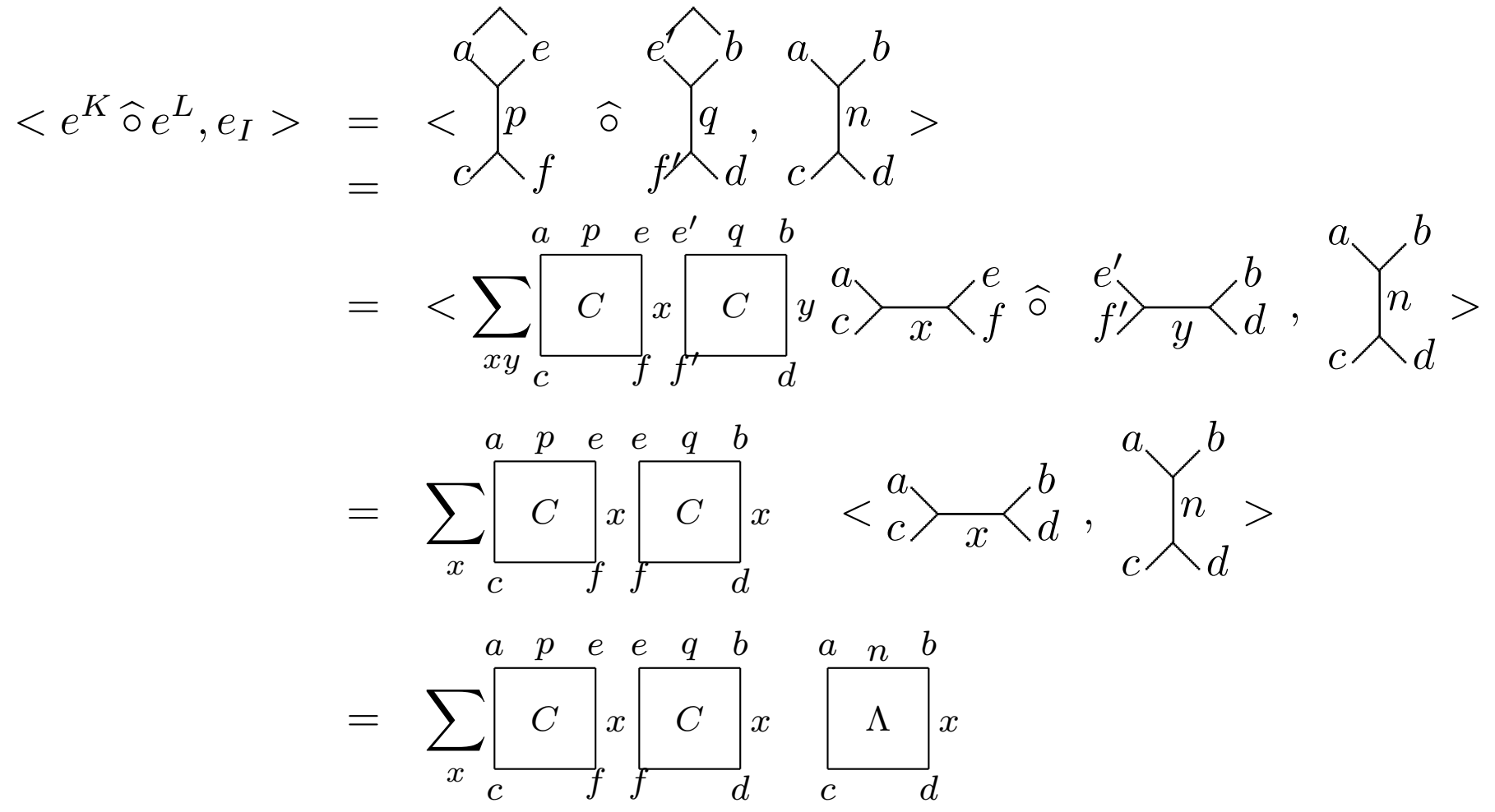

Therefore, the coproduct reads

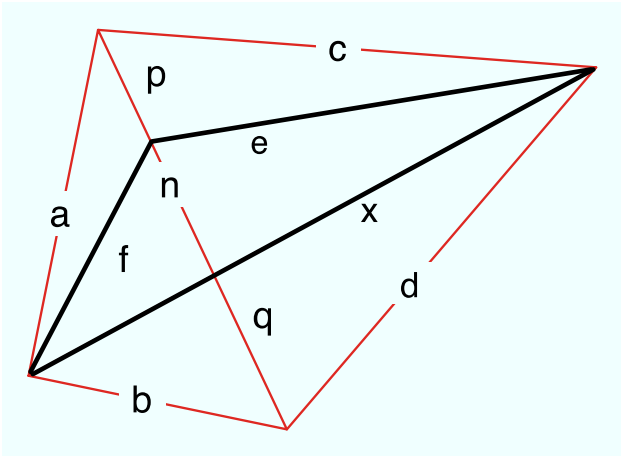

Figure 14: Graphical interpretation of the coproduct

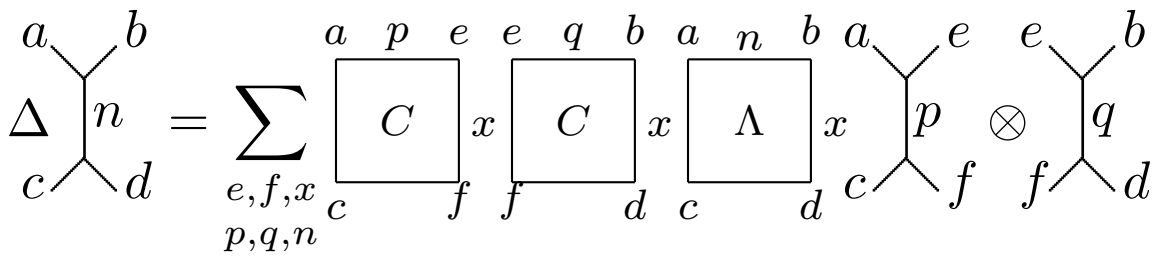

Notice that we cannot use the Racah identity to concatenate the first two (direct) cells appearing on the r.h.s. since there is a summation over tensor products of double triangles labelled by $p$ and $q$. The previous equation actually looks quite simple and natural when written in terms of tensors with multi-indices. Indeed, if $\nu$ denotes the set of structure constants for the product $\widehat{\circ}$ of $\widehat{\mathcal{B}}$, for instance $f^A \widehat{\circ} f^B = \nu^{AB}_C\, f^C$, then, using $e_M = \Lambda^A_M\, f_A$ and $e^M = C^M_A\, f^A$, we find that the coproduct in $\mathcal{B}$ reads $\Delta\, e_M = \nu^{KL}_M\, e_K \otimes e_L$, where $\nu^{KL}_M = C^K_A\, C^L_B\, \Lambda^C_M\, \nu^{AB}_C$.

Geometrical interpretation of the coproduct: consider the tetrahedron displayed on figure 1. Now, slice it into two pieces like on figure 14. We start from the double triangle $(abn), (cdn)$ on which we apply $\Delta$ (the left hand side of the above equation); to do that, we choose an arbitrary point on the common edge $n$ and slice the tetrahedron as indicated. This decomposes $n$ into a sum $n = p + q$ and determines three new edges $e, f, x$. This also creates two new double triangles $(afp), (cep)$, articulated along $p$, and $(bfq), (deq)$, articulated along $q$. The summation on the right hand side corresponds to these freedom of slicing, the two direct cells that appear correspond to the two resulting new tetrahedra, and the inverse cell corresponds to the full tetrahedron.

### 3.2.2 The cogebra $\widehat{\mathcal{B}}$ and its coproduct $\widehat{\Delta}$

In the same way, one defines $< \widehat{\Delta} f^A, f_B \otimes f_C > = < f^A, f_B \circ f_C >$, and proves the formula :

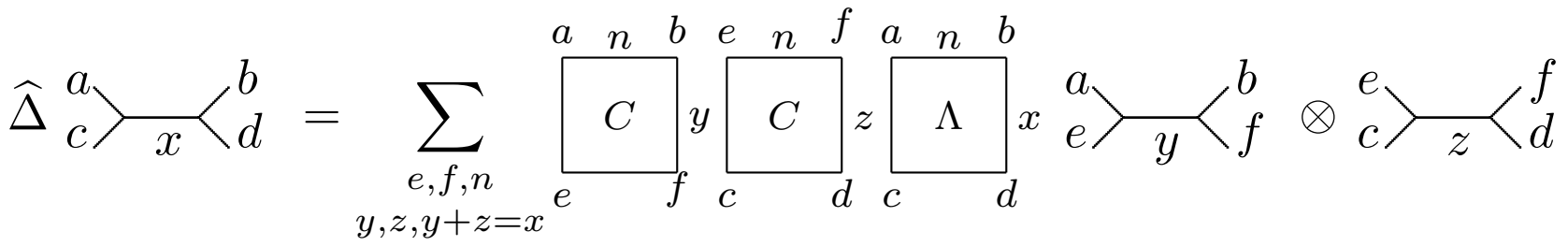

### 3.2.3 Compatibility equations

The compatibility equations tell us that $\Delta$ is a homomorphism from the algebra $\mathcal{B}$ to its tensor square $\mathcal{B} \otimes \mathcal{B}$ and that $\widehat{\Delta}$ is a homomorphism from the algebra $\widehat{\mathcal{B}}$ to its tensor square $\widehat{\mathcal{B}} \otimes \widehat{\mathcal{B}}$, i.e., for any $u, v \in \mathcal{B}$, we have $\Delta(u \circ v) = \Delta u \circ \Delta v$, and for any $\sigma, \tau$ in $\widehat{B}$, we have $\widehat{\Delta}(\sigma \,\widehat{\circ}\, \tau) = \widehat{\Delta}\sigma \,\widehat{\circ}\, \widehat{\Delta}\tau$. Here the product in the tensor square of $\mathcal{B}$ is of course $(u \otimes v) \circ (u' \otimes v') = (u \circ u') \otimes (v \circ v')$ and similarly for the product used in the tensor square of $\widehat{\mathcal{B}}$. If $\mu$ denotes the structure constants of the product $\circ$ in $\mathcal{B}$ , and $\nu$, those of the product $\widehat{\circ}$ in $\widehat{\mathcal{B}}$, the compatibility equation reads $\mu^K_{IJ}\, \nu^{LM}_K = \nu^{PP'}_I\, \nu^{QQ'}_J\, \mu^L_{PQ}\, \mu^M_{P'Q'}$. Let us illustrate that property in the case of $A_3$.

### 3.2.4 An example in $A_3$

We have, for instance 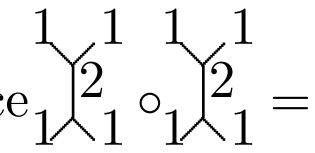 $= 0$, and using the following expressions for the coproducts, one can check that in the tensor square algebra of $(\mathcal{B}(A_3), \circ)$ we have also 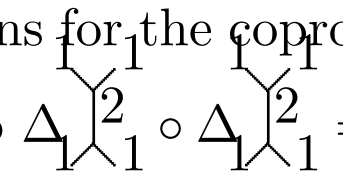 $= 0$. There are $34^2$ such identities to check in $A_3$.

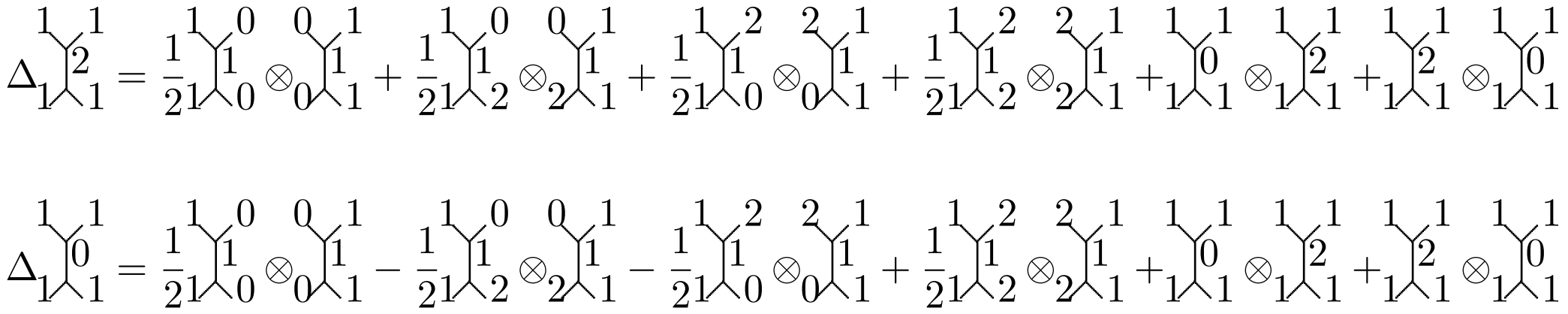

## 3.3 Units and counits

### 3.3.1 Minimal central projectors

It will be useful to consider the following quantities:

In $\mathcal{B}$, $\pi_n$ denotes the projector on the block $n$ of dimension $d_n \times d_n$ i.e., its restriction to this block is the identity and its restriction to all other blocks is zero.

In $\widehat{\mathcal{B}}$, $\omega^x$ denotes the projector on the block $x$ of dimension $d_x \times d_x$ i.e., its restriction to this block is the identity and its restriction to all other blocks is zero. Since the $e_I$ are elementary matrices for $\circ$ and $f^A$ elementary matrices for $\widehat{\circ}$ we have explicitly

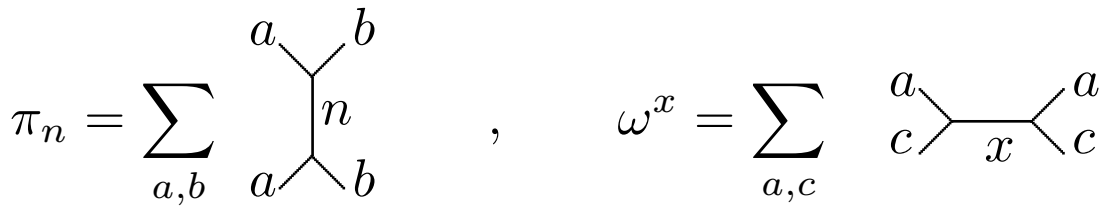

where the sum , in the case of $\pi$, is over all $(a,b)$ such that the triangle $(a,n,b)$ is admissible and, in the case of $\omega$ over all $(a,c)$ such that the triangle $(a,x,c)$ is admissible. In other words, each term of those sums is a double triangle (horizontal for $\pi$, vertical for $\omega$) made of two identical triangles with a common edge (respectively $n$ or $x$). Of course one could write the projectors $\pi_n$ in terms of the basis $f_A$ and projectors $\omega^x$ in terms of the basis $e^I$, but these expressions are less simple since they involve cells (direct of inverse).

The $\pi_n$ are minimal central projectors for $\circ$ so that $\pi_m \circ \pi_n = \delta_{mn}$ and the $\omega^x$ are minimal central projectors for $\widehat{\circ}$ so that $\omega^x \,\widehat{\circ}\, \omega^y = \delta^{xy}$.

To illustrate this, we take the case of $A_3$ and use, for the base $e_I$, the path notation introduced in 3.1.5. We have :

$$
\begin{aligned}
\pi_0 &= v_0 v_0 + v_1 v_1 + v_2 v_2 \\
\pi_1 &= l_1 l_1 + l_2 l_2 + r_0 r_0 + r_1 r_1 \\
\pi_2 &= dd + gg + \gamma\gamma
\end{aligned}
$$

Rather than writing the projectors $\omega$ in the base $f^A$ (elementary matrices for $\widehat{\circ}$), we write them in terms of the dual basis $e^I$ of $e_I$.

$$
\begin{aligned}
\omega^0 &= \widehat{v_0 v_0} + \widehat{v_1 v_1}/2 + \widehat{v_2 v_2} + \widehat{\gamma\gamma}/2 \\
\omega^1 &= \widehat{v_0 v_1} + \widehat{v_1 v_0} + \widehat{v_1 v_2} + \widehat{v_2 v_1} \\
\omega^2 &= \widehat{v_0 v_2} + \widehat{v_1 v_1}/2 + \widehat{v_2 v_0} - \widehat{\gamma\gamma}/2
\end{aligned}
$$

### 3.3.2 The unit of $\mathcal{B}$

The unit $\mathbb{1}$ of $B$ is immediately obtained from the fact that $\mathcal{B}$ is explicitly isomorphic with a direct sum of $d_n \times d_n$ matrices. The unit is $\mathbb{1} = \sum_n \pi_n$. For example, in the case of $A_3$, $\mathbb{1} = \pi_1 \oplus \pi_2 \oplus \pi_3$.

From the above definition, we calculate the expression of $\Delta\mathbb{1}$. We find

$$\Delta\mathbb{1} = \sum_v t_v \otimes s_v$$

where the summation runs over all vertices $v$ of $A_N$ and where $t_v = \sum_{n,a}$ 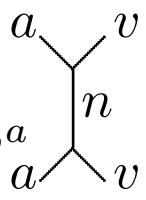

, $s_v = \sum_{n,b}$ 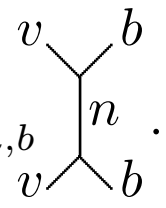

. In the case of $A_3$, for instance, we have

$$\begin{array}{rcl} t_0 = gg + l_1 l_1 + v_0 v_0 & , & s_0 = dd + r_0 r_0 + v_0 v_0 \\ t_1 = l_2 l_2 + r_0 r_0 + v_1 v_1 + \gamma\gamma & , & s_1 = l_1 l_1 + r_1 r_1 + v_1 v_1 + \gamma\gamma \\ t_2 = dd + r_1 r_1 + v_2 v_2 & , & s_2 = gg + l_2 l_2 + v_2 v_2 \end{array}$$

Since $\Delta\mathbb{1}$ is not equal to $\mathbb{1}\otimes\mathbb{1}$, $\mathcal{B}$ cannot be a Hopf algebra; it is a quantum groupoid (a weak Hopf algebra). For any $u \in \mathcal{B}$, one usually writes $\Delta u = u_{(1)} \otimes u_{(2)}$, with a hidden summation over (1) and (2). In particular, $\Delta\mathbb{1} = \mathbb{1}_{(1)} \otimes \mathbb{1}_{(2)}$, but in the present situation, the right hand side is explicitly obtained in terms of the target and source vectors $t_v$ and $s_v$ previously given.

The linear spans $\mathcal{B}_t$ of vectors $s_v$ and $\mathcal{B}_s$ of vectors $t_v$ are actually subalgebras called source and target subalgebras. Their existence results from the general theory of weak Hopf algebras but their explicit realization, found here, depends on the specific class of examples that we consider in this paper. These subalgebras are also characterized by the following :

$$\begin{array}{rcl} \mathcal{B}_t &=& \{< \sigma \otimes id, \Delta\mathbb{1} >, \sigma \in \widehat{B}\} \\ \mathcal{B}_s &=& \{< id \otimes \sigma, \Delta\mathbb{1} >, \sigma \in \widehat{B}\} \end{array}$$

These subalgebras are not commutative but they commute with each other in $\mathcal{B}$. Their common dimension is equal to $N$, the number of vertices of $A_N$. In quantum groupoids, it is useful to define two maps, called target and source counital maps (see [26]) that, in the present case, are as follows:

$$\begin{array}{rcl} \epsilon_t(u) &=& \sum_v \epsilon(t_v \circ u)\, s_v \\ \epsilon_s(u) &=& \sum_v t_v\, \epsilon(u \circ s_v) \end{array}$$

Their images are the two particular subalgebras introduced above and one can prove, as in [26][11], a useful property that in the present case reads:

$$\begin{array}{rcl} \epsilon_t(\mathcal{B}) &=& \mathcal{B}_t = \{u \in \mathcal{B} / \Delta u = \sum_v t_v \circ u \otimes s_v\} \\ \epsilon_s(\mathcal{B}) &=& \mathcal{B}_s = \{u \in \mathcal{B} / \Delta u = \sum_s t_v \otimes u \circ s_v\} \end{array}$$

[11]The finite quantum groupoids that we consider here do not appear in the class of examples studied in that reference.

### 3.3.3 The unit of $\widehat{B}$

The unit $\widehat{\mathbb{1}}$ of $\widehat{\mathcal{B}}$ is immediately obtained from the fact that it $\widehat{B}$ is explicitly isomorphic with a direct sum of $d_x \times d_x$ matrices. The unit is $\widehat{\mathbb{1}} = \sum_x \omega^x$. For example, in the case of $A_3$, $\widehat{\mathbb{1}} = \omega^1 \oplus \omega^2 \oplus \omega^3$ and it is instructive to notice that the term $\widehat{\gamma\gamma}$ disappears from the sum, so that in terms of the dual basis $e^I$, the unit of $\widehat{B}$ is precisely equal to the sum $\sum_{I \in 0} e^I$, the notation meaning that we sum over *all* elements of the block $n = 0$ (the corresponding $e_I$'s are the elementary matrices of the first block of $\mathcal{B}$). This is a generic feature. Properties of the coproduct $\widehat{\Delta}\widehat{\mathbb{1}}$ could be discussed in a way similar to those of $\Delta\mathbb{1}$.

### 3.3.4 The counit of $\mathcal{B}$

From the study of the unit in $\widehat{\mathcal{B}}$ we define the counit $\epsilon$ of $\mathcal{B}$ as a linear map from $\mathcal{B}$ to the complex field equal to 1 on all the horizontal double triangles of type $n = 0$ and vanishing on all the others. In the case of weak Hopf algebras, the counit is not an homomorphism and the usual compatibility axiom is modified as [7]:

$$\epsilon(u \circ v) \quad = \quad \epsilon(u\mathbb{1}_{(1)})\epsilon(\mathbb{1}_{(2)}v)$$

These equations can be checked here in the form $\epsilon(e_I \circ e_J) = \sum_n \epsilon(e_I \circ t_n)\epsilon(s_n \circ e_I)$ for all $e_I$. Therefore $(\mathcal{B}, \circ, \Delta, \mathbb{1}, \epsilon)$ is a unital and counital bialgebra.

### 3.3.5 The counit of $\widehat{\mathcal{B}}$

In an analogous way, one can define and characterize the counit $\widehat{\epsilon}$ of $\widehat{\mathcal{B}}$. It is a linear map equal to 1 on all the vertical double triangles of type $x = 0$ and it vanishes on all others. It obeys the required compatibility condition and $(\widehat{\mathcal{B}}, \widehat{\circ}, \widehat{\Delta}, \widehat{\mathbb{1}}, \widehat{\epsilon})$ is a unital and counital bialgebra.

## 3.4 Antipodes

A priori one may introduce several kinds of natural conjugations in the basis $e_I$, the basis of horizontal triangles, i.e., of *vertical* diffusion graphs. One may flip globally the top and bottom labels, this operation corresponds to the adjoint operation in the matrix realization of $(\mathcal{B}, \circ)$. One may flip globally the left and right labels, this operation corresponds to the adjoint operation in the matrix realization of its dual. These two operations can be extended by linearity (for $A_N$ diagrams, the basis elements $e_I$ are real, but on the field of complex numbers, we should take complex conjugates). It is therefore tempting to define the antipode $S$ of $\mathcal{B}$ as a map from $\mathcal{B}$ to $\mathcal{B}$ that simultaneously flips top and bottom labels, as well as left and right labels. Such a map would be an algebra anti-homomorphism, as it should for an antipode. However, in a quantum groupoid, the following equality should hold ([26]) for every element $u$ of $\mathcal{B}$

$$\begin{aligned} m_\circ(id \otimes S)\, \Delta(u) \quad &= \quad (\epsilon \otimes id)((\Delta\mathbb{1}) \circ (u \otimes \mathbb{1})) \\ m_\circ(S \otimes id)\, \Delta(u) \quad &= \quad (id \otimes \epsilon)((u \otimes \mathbb{1}) \circ (\Delta\mathbb{1})) \end{aligned}$$

Here $m_\circ$ denotes the multiplication map $u \otimes v \in \mathcal{B} \otimes \mathcal{B} \mapsto u \circ v \in \mathcal{B}$. In the present case, if we set $\Delta u = u_{(1)} \otimes u_{(2)}$, we should therefore have, for all $u \in \mathcal{B}$,

$$\begin{aligned} u_{(1)} \circ S(u_{(2)}) \quad &= \quad \epsilon_t(u) \\ S(u_{(1)}) \circ u_{(2)} \quad &= \quad \epsilon_s(u) \end{aligned}$$

The map defined on basis elements $e_I$ by simultaneously transposing up and down labels together with left and right labels does not obey this condition. However, one can find a scalar function $\mu_{abncd}$ such that the map defined as follows obeys the required properties:

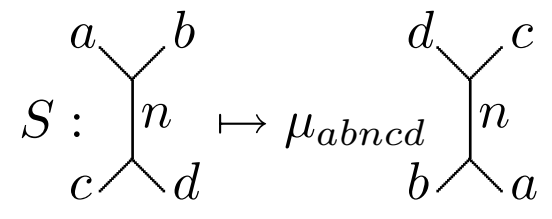

This scalar function $\mu$ is determined by imposing that the previous weak Hopf algebra constraints holds on basis elements (this is a set of linear equations). In general $\mu$ is not constant on the chosen basis (and a fortiori it is not equal to the number 1), so no power of $S$ is equal to $S$ and the antipode is of infinite order.

For instance, in the case or $A3$, solving these equations gives

$$\mu_{0,0,0,0,0}=1,\mu_{0,0,0,1,1}=1,\mu_{1,0,1,1,0}=1,\mu_{2,0,2,2,0}=1,\mu_{2,0,2,1,1}=-1,\mu_{1,0,1,1,2}=-1$$
$$\mu_{0,0,0,2,2}=1,\mu_{2,0,2,0,2}=1,\mu_{1,1,2,2,0}=-1,\mu_{2,1,1,1,0}=\sqrt{2},\mu_{0,1,1,1,0}=-\sqrt{2},\mu_{1,1,0,0,0}=1$$
$$\mu_{1,1,0,1,1}=1,\mu_{1,1,2,1,1}=1,\mu_{0,1,1,1,2}=\sqrt{2},\mu_{1,1,0,2,2}=1,\mu_{1,1,2,0,2}=-1,\mu_{2,1,1,1,2}=-\sqrt{2},$$
$$\mu_{0,2,2,2,0}=1,\mu_{1,2,1,1,0}=-1,\mu_{2,2,0,0,0}=1,\mu_{0,2,2,1,1}=-1,\mu_{2,2,0,1,1}=1,\mu_{0,2,2,0,2}=1$$
$$\mu_{1,2,1,1,2}=1,\mu_{2,2,0,2,2}=1,\mu_{1,2,1,0,1}=\frac{\sqrt{2}}{2},\mu_{1,2,1,2,1}=-\frac{\sqrt{2}}{2},\mu_{1,0,1,2,1}=\frac{\sqrt{2}}{2},\mu_{1,0,1,0,1}=-\frac{\sqrt{2}}{2}$$
$$\mu_{0,1,1,0,1}=1,\mu_{0,1,1,2,1}=-1,\mu_{2,1,1,2,1}=1,\mu_{2,1,1,0,1}=-1$$

Actually, for $A_N$ diagrams, the following holds[12] :

$$\mu_{abncd}=\frac{(-1)^{\frac{b+c}{2}}}{(-1)^{\frac{a+d}{2}}}\frac{\sqrt{[a+1][d+1]}}{\sqrt{[b+1][c+1]}}$$

For more general diagrams, $q$-numbers are replaced by $q$ - dimensions of vertices (like in [38]).

We can also find the antipode on the dual $\widehat{\mathcal{B}}$. As expected, when acting on basis $f^A$ which is adapted for the $\widehat{\circ}$ product, the antipode also permutes top and bottom, as well as left and right labels, however, it also multiplies the obtained element by appropriate constants $\hat{\mu}_{acxbd}$ in order to obey the required axioms for an antipode. $\widehat{S}$ is again of infinite order.

$\widehat{S}:$ 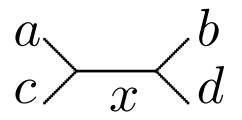 $\mapsto \widehat{\mu}_{acsbd}$ 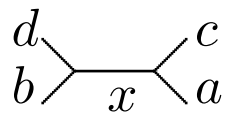

## 3.5 Integrals and measures

### 3.5.1 Integrals in $\mathcal{B}$

Left integrals (resp. right integrals) are elements $\ell \in \mathcal{B}$ (resp. $r \in \mathcal{B}$) such that for all $u \in \mathcal{B}$, we have $u \circ \ell = \epsilon_t(u) \circ \ell$ (resp. $r \circ u = r \circ \epsilon_s(u)$). For $A_N$ diagrams, we find that left integrals span the $N$-dimensional subspace $\int^\ell = \{\sum_{v \in A_N}$ 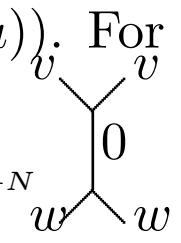 $, w \in A_N\}$ and right integrals the subspace $\int^r = \{\sum_{w \in A_N}$ 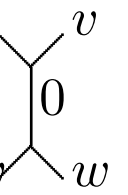 $, v \in A_N\}$. Double-sided integrals $\int$ span therefore a one-dimensional subspace proportional to $\sum_{v,w \in A_N}$ 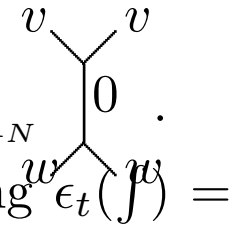 .

We can normalize by imposing $\epsilon_t(\int) = \mathbb{1}$; this is compatible with the condition $\epsilon_s(\int) = \mathbb{1}$ and fixes the proportionality factor: we find a unique double-sided normalized integral equal to $1/N$ times the previous sum (the sum of all elementary matrices for the first block ($n = 0$)). In $A_3$, and using the path notation, it reads

$$\int = (v_0v_0 + v_0v_1 + v_0v_2 + v_1v_0 + v_1v_1 + v_1v_2 + v_2v_0 + v_2v_1 + v_2v_2)/3$$

### 3.5.2 Measures on $\mathcal{B}$

Left invariant measures (resp. right invariant measures) on $\mathcal{B}$ are elements $\mu$ of $\widehat{\mathcal{B}}$ such that, for all $u \in \mathcal{B}$, $u_{(1)} < \mu, u_{(2)} >= \epsilon_t(\mu_{(1)}) < \mu, u_{(2)} >$ (resp. $< \mu, u_{(1)} > u_{(2)} =< \mu, u_{(1)} > \epsilon_s(\mu_{(2)})$. Bi-invariant measures are both left and right invariant. Integrals in $\mathcal{B}$ and measures on $\mathcal{B}$ are dual notions, and one can determine measures on the algebra $(\mathcal{B}, \circ)$ by looking at the integrals in the algebra $(\widehat{\mathcal{B}}, \widehat{\circ})$. With no surprise, we find a unique normalized bi-invariant measures on $\mathcal{B}$ equal to to $\mu = \sum_{v,w \in A_N} {}^{v}_{v}\!\succ_{0}\!\prec^{w}_{w}$ . The normalization was obtained by setting $(id \otimes \mu)\Delta(\mathbb{1}) = \mathbb{1}$ and $(\mu \otimes id)\Delta(\mathbb{1}) = \mathbb{1}$ i.e., $\sum_v t_v < \mu, s_v >= \sum_v < \mu, t_v > s_v = \mathbb{1}$. It is often useful to write this measure in terms of the $e^I$ basis, rather than in terms of the $f^A$ basis: in the example $A_3$, and using the path notation for the dual basis of the basis of horizontal double triangles, one finds

$$\mu = \widehat{v_0 v_0} + \widehat{r_0 r_0} + \widehat{dd} + \widehat{l_1 l_1} + \frac{1}{2}(\widehat{v_1 v_1} + \widehat{\gamma\gamma}) + \widehat{r_1 r_1} + \widehat{gg} + \widehat{l_2 l_2} + \widehat{v_2 v_2}$$

[12]If we had defined $q$ - Racah symbols in 2.3.1 without the sign in front, there would be no sign here either.

## 3.6 Adapted scalar products and associated convolution laws

### 3.6.1 The scalar product $g$ orthonormal for horizontal double triangles

The basis $e_I$ of horizontal double triangles (remember that they are elementary matrices for the product $\circ$ of $\mathcal{B}$) defines a scalar product $g$ for which this basis is orthonormal; it also induces a scalar product, also called $g$, on the dual. So, we have $g(e_I, e_J) = \delta_{IJ}$ and $g(e^I, e^J) = \delta^{IJ}$. However, the basis $f^A$ (vertical double triangles, they are elementary matrices for $\widehat{\circ}$ in $\widehat{\mathcal{B}}$ ) and its dual basis $f_A$ in $\mathcal{B}$, are not orthonormal for this scalar product : $g(f^A, f^B) \neq \delta^{AB}$ and $g(f_A f_B) \neq \delta_{AB}$. The case of $A_3$ is illustrated below; notice that $(g^{AB})$ is orthogonal but not orthonormal : we display the values of $g({}^{a}_{c}\!\succ_{x}\!\prec^{b}_{d}\,, {}^{a}_{c}\!\succ_{x}\!\prec^{b}_{d}\,)$, as three blocks labelled by $x$, lines labelled by ${}^{a}_{c}$ and columns by ${}^{b}_{d}$, exactly as we did with the matrix display of elementary matrices $f^A$. Calculations are very simple if we express the $f^A$ basis vectors in terms of the $e^I$.

$$(g^{AA}) = \begin{array}{ccc} \begin{array}{ccc} 1 & 1 & 1 \\ 1 & 1/2 & 1 \\ 1 & 1 & 1 \end{array} & & \\ & \begin{array}{cccc} 1 & 1 & 1 & 1 \\ 2 & 1 & 1 & 2 \\ 2 & 1 & 1 & 2 \\ 1 & 1 & 1 & 1 \end{array} & \\ & & \begin{array}{ccc} 1 & 1 & 1 \\ 1 & 1/2 & 1 \\ 1 & 1 & 1 \end{array} \end{array} \qquad (g_{AA}) = \begin{array}{ccc} \begin{array}{ccc} 1 & 1 & 1 \\ 1 & 2 & 1 \\ 1 & 1 & 1 \end{array} & & \\ & \begin{array}{cccc} 1 & 1 & 1 & 1 \\ 1/2 & 1 & 1 & 1/2 \\ 1/2 & 1 & 1 & 1/2 \\ 1 & 1 & 1 & 1 \end{array} & \\ & & \begin{array}{ccc} 1 & 1 & 1 \\ 1 & 2 & 1 \\ 1 & 1 & 1 \end{array} \end{array}$$

### 3.6.2 The scalar product $h$ orthonormal for vertical double triangles

In the same way the basis $f^A$ of vertical double triangles defines a scalar product $h$ on $\widehat{B}$ for which this basis is orthonormal and induces a scalar product, also called $h$, on its dual. So, we have $h(e^A, e^B) = \delta_{AB}$ and $h(e_A, e_B) = \delta_{AB}$. However, the basis $e_I$ of vertical double triangles and its dual basis $e^I$ are not orthonormal for this scalar product: $h(e_I, e_J) \neq \delta_{IJ}$ and $h(e^I, e^J) \neq \delta^{IJ}$. The case of $A_3$ is now discussed; $(h_{IJ})$ is orthogonal but not orthonormal and we display the values, as three blocks labelled by $n$, of $h(\,{}^{a\,b}\big|^{n}_{c\,d}\,, {}^{a\,b}\big|^{n}_{c\,d}\,)$.

Lines are labelled by $ab$ and columns by $cd$, exactly as we did with the matrix display of elementary matrices $e_I$. Actually, the display of results, for $h_{II}$ is as above (similar comment for $h^{II}$) because, taking $x = n$ below, we have

$$(h_{II}) = h(\,{}^{a\,b}\big|^{n}_{c\,d}\,, {}^{a\,b}\big|^{n}_{c\,d}\,) = g({}^{a}_{c}\!\succ_{x}\!\prec^{b}_{d}\,, {}^{a}_{c}\!\succ_{x}\!\prec^{b}_{d}\,) = (g^{AA})$$

### 3.6.3 Norms of projectors (example of $A_3$)

Values of projector norms for scalar products $g$ and $h$ can be readily calculated from the above. One finds $g(\pi_n, \pi_n) = d_n$ and $h(\omega^x, \omega^x) = d_x$; for instance, in the case of $A_3$, $g(\pi_0, \pi_0) = h(\omega^0, \omega^0) = 3$, $g(\pi_1, \pi_1) = h(\omega^1, \omega^1) = 4$, $g(\pi_2, \pi_2) = h(\omega^2, \omega^2) = 3$. Of course $g(\mathbb{1}, \mathbb{1}) =$ $dim(\mathcal{H}) = 10$ and $h(\widehat{\mathbb{1}}, \widehat{\mathbb{1}}) = dim(\mathcal{V}) = 10$. However, the projectors are not properly normalized if we use the "other" scalar product; for instance, in $A_3$, $g(\omega^0, \omega^0) = h(\pi_0, \pi_0) = 5/2$, $g(\omega^1, \omega^1) = h(\pi_1, \pi_1) = 4$ and $g(\omega^2, \omega^2) = h(\pi_2, \pi_2) = 5/2$.

### 3.6.4 Musical isomorphisms (Fourier transforms) and convolution laws

The choice of a scalar product induces a particular identification of a vector space with its dual. In the present case we have two adapted scalar products, so two interesting identifications of $\mathcal{B}$ with its dual. Using such an identification allows one to write the product $\widehat{\circ}$, canonically defined on $\widehat{\mathcal{B}}$ as a second product — since we have already the product $\circ$ — on the same vector space $\mathcal{B}$. This new product may be called "convolution product" and denoted by the symbol $\star$. Let us first remind how this is done in general.

As before, we consider the adapted basis $e_I$ in $\mathcal{B}$ together with its dual $e^I$ in $\widehat{\mathcal{B}}$ , and the adapted basis $f^A$ in $\widehat{\mathcal{B}}$. together with its dual $f_A$ in $\mathcal{B}$. Let $k$ be any scalar product in $\mathcal{B}$, it may be one of the two adapted scalar products $g$ or $h$ but at the moment the discussion is general. To the basis vector $e_I$ of $\mathcal{B}$, we associate the form $e_I^\#$ such that $k(e_I, e_J) = k_{IJ} =< e_I^\#, e_J >$ (the angle bracket denotes the canonical pairing with the dual). We have $e_I^\# = k_{IJ} e^J$. So we have a new basis, namely $e_I^\#$ in $\widehat{\mathcal{B}}$. In the same way, to the basis vector $f^A$ of $\widehat{\mathcal{B}}$ we associate the vector $f_\flat^A$ such that $k(f^A, f^B) = k^{AB} =< f^A, f_\flat^A >$. Clearly $f_\flat^A = k^{AB} f_A$ and we have a new basis $f_\flat^A \in \mathcal{B}$. More generally, to $u \in \mathcal{B}$ we associate $u^\# \in \widehat{\mathcal{B}}$ such that $k(u, v) =< u^\#, v >$ for all $v \in \mathcal{B}$. Musical isomorphisms $\# : \mathcal{B} \mapsto \widehat{\mathcal{B}}$ and $\flat : \widehat{\mathcal{B}} \mapsto \mathcal{B}$ rise or lower indices of component vectors: if $u = u^I e_I \in \mathcal{B}$, then $u^\# = u_J^\# e^J \in \widehat{\mathcal{B}}$, with $u_J^\# = u^I k_{IJ}$. Etc. These isomorphisms are inverse of one another and could of course be called Fourier transforms. Using them we define the convolution product on $\mathcal{B}$ as follows (here we "break the symmetry" since we are choosing explicitly one side) : $e_I \star e_J = (e_I^\# \,\widehat{\circ}\, e_J^\#)_\flat$. The direct Fourier transforms read, as usual, $(e_I \star e_J)^\# = (e_I^\# \,\widehat{\circ}\, e_J^\#)$ and the structure constants of the convolution product defined by $e_I \star e_J = \nu_{IJ}^K e_K$ read $\nu_{IJ}^K = k_{II'} k_{JJ'} k^{KK'} \, \nu_{K'}^{I'J'}$ in terms of the structure constants of the $\widehat{\circ}$ product of $\widehat{\mathcal{B}}$ (or in terms of the coproduct $\Delta$ of $\mathcal{B}$). These structure constants were explicitly computed, in terms of cells, in section 3.2.1.

All this is very elementary but the point to remember is that the chosen metric explicitly comes in the definition of the convolution product. Now if $k$ is chosen to be one of the two adapted scalar products, for instance $k = g$, then $k_{IJ} = g_{IJ} = \delta_{IJ}$ in the $e_I$ basis, and the previous equation looks much simpler, but one has to remember that this is now written in a very specific basis which is orthonormal for the chosen scalar product, and also remember that the two scalar products $g$ and $h$, respectively adapted for $\mathcal{B}$ and $\widehat{\mathcal{B}}$ are intrinsically distinct. The conclusion is that, from a unique coproduct $\Delta$ on $\mathcal{B}$ (or a unique multiplication $\widehat{\circ}$ on its dual), we can define as many convolution products as we want by choosing an arbitrary scalar product, and that, even if we restrict to canonical choices, $g$ or $h$, we have in any case at least two possibilities[13].

In those references where authors prefer to study the two algebra structures on the same space, and this is in particular the case for articles from the operator algebra community, there is always an asymmetry between the two products that reflects this fact. Another drawback of working with $\circ$ and $\star$, rather than $\circ$ and $\widehat{\circ}$ or $\circ$ and $\Delta$, is that it is difficult, in such a formalism, to describe the relation between the two products coming from the compatibility existing between product and coproduct. Since one purpose of the present paper is precisely to describe the situation without putting the two products "on the same side", we stop this discussion here.

## 3.7 Algebras of characters for $\circ$ and $\widehat{\circ}$

The quantum groupoid $\mathcal{B}$ associated with a diagram $G$, is neither commutative nor cocommutative, it has therefore two non trivial representation theories and two algebras of characters.

[13]See also the discussion at end of section 3.8.1

Representation theory itself will not be dealt with in this article (see however section 3.8.2) but we shall say a few words about the corresponding algebras of characters and analyze what happens when $G$ is an $A_N$ diagram of the $SU(2)$ system. When $G$ is a member of some Coxeter-Dynkin system (say $SU(n)$) , the algebra of characters $A(G)$, associated with the product $\circ$ in $\mathcal{B}$ is commutative, and it is isomorphic with the graph algebra of the $A$ diagram which has the same Coxeter number $\kappa$ as $G$; it is therefore isomorphic with (or it defines) the fusion algebra of $G$. For the $SU(n)$ system, it is the graph algebra of the Weyl alcove $\mathcal{A}_\ell$, i.e., the Weyl chamber truncated at level $\ell$, and $\kappa = \ell + n$. For the $SU(2)$ system, i.e., usual ADE diagrams, it is $A_N = \mathcal{A}_{\ell=N+1}$. The algebra of characters $Oc(G)$, associated with the product $\widehat{\circ}$ on $\widehat{\mathcal{B}}$ is called the Ocneanu algebra of $G$, or algebra of quantum symmetries. Its structure is very much case dependent and it is not always commutative (see [32] and [12] for members of the $SU(2)$ family and [33], [13], [17], [21] for other results). When $G = \mathcal{A}_\ell$, the situation is simple, since $G = A(G) = Oc(G)$. In general, such algebras of characters can be obtained thank's to a variety of techniques, for instance the modular splitting method ([33], [21]), without having to rely on an explicit knowledge of the bigebra $\mathcal{B}$ — and this is fortunate because this bigebra is not explicitly known in general. Since we have now an explicit description of the algebra $\mathcal{B}(A_N)$, we want to realize explicitly the two algebras of characters, for this family of diagrams.

In the case of a finite group, we may consider that the group algebra acts on itself (on its own columns), and therefore associate irreducible representations with simple components of the group algebra. The usual formula $\rho_x(g) = Tr\, r_x(g)$ defining the character $\rho_x$ of the irreducible representation $r_x$ can be rewritten as $\rho_x(g) = Tr\, \omega_x\, g$ where $\omega_x$ is a minimal central projector of the group algebra. More generally, one can write $< \rho, \sigma >= Tr(\omega_x\, \sigma)$ where $\sigma$ is any element of the group algebra, not necessarily an element of group type obeying $\Delta g = g \otimes g$. Therefore, on general grounds, and at least in the semi-simple situation, we may consider that an irreducible character on a bigebra is a particular element of its dual associated with the choice of one of its simple components via an appropriate generalization of the above relation.

### 3.7.1 Characters for the algebra $\circ$ on $B$ : the fusion algebra

Irreducible characters — we just call them characters, in the sequel — for the semi-simple algebra $\mathcal{B}$ are elements $\widetilde{\pi}_n \in \widehat{\mathcal{B}}$ defined by $< \widetilde{\pi}_n, u >= Tr(\pi_n \circ u)$, where $\pi_n$ is a minimal central projectors (it is represented by the identity matrix on the block $n$ and vanishes on the other simple components) and $u$ is any element in $\mathcal{B}$.

Being elements of $\widehat{\mathcal{B}}$, the $\widetilde{\pi}_n$ can be decomposed on the basis $e^J$ (dual of the basis of matrix units of $\mathcal{B}$) : $\widetilde{\pi}_n = \widetilde{\pi}_{nJ}\, e^J$. Take $u = e_I$ in the defining formula. We have $< \widetilde{\pi}_{nJ}\, e^J, e_I >= Tr(\pi_n \circ e_I)$. The left hand side is $\widetilde{\pi}_{nI}$. Since $e_I$'s are elementary matrices, the right hand side vanishes if $I \neq n$ (if $I$ is not a multi-index relative to the block $n$), and it also vanishes if, although relative to the block $n$, $e_I$ is not a diagonal elementary matrix. In the other cases it is equal to 1. So $\widetilde{\pi}_{nI} = 1$ if $e_I$ is diagonal and of type $n$, it is equal to 0 otherwise. Therefore

$$\widetilde{\pi}_n = \sum_{\substack{I \in n \\ I \text{ is diagonal}}} e^I \qquad \text{to be compared with} \qquad \pi_n = \sum_{\substack{I \in n \\ I \text{ is diagonal}}} e_I$$

This justifies our notation for the character $\widetilde{\pi}_n$ associated with the block characterized by the projector $\pi_n$.

The projectors $\pi_n$ are elements of $\mathcal{B}$ and obey the usual projectors identities : $\pi_m \circ \pi_n = \delta_{mn}$ but the characters $\widetilde{\pi}_n$ are elements of $\widehat{\mathcal{B}}$ and generate, for the product $\widehat{\circ}$ a subalgebra of $\widehat{\mathcal{B}}$ isomorphic with the graph algebra of the diagram $A_N$, i.e., the fusion algebra itself.

$$\widetilde{\pi}_m \,\widehat{\circ}\, \widetilde{\pi}_n = N_{mnp}\, \widetilde{\pi}_p$$

where the structure constants of the algebra are the fusion matrices introduced in section 2.1.2. This algebra has a unit (namely $\widetilde{\pi}_0$) but it does not coincide with the unit $\widehat{\mathbb{1}}$ of $\widehat{\mathcal{B}}$.

**Example of $A_3$ .**

Using tables given in section 3.1.5 one finds

$$\begin{aligned}\widetilde{\pi}_0 &= \widehat{v_0 v_0} + \widehat{v_1 v_1} + \widehat{v_2 v_2}\\ \widetilde{\pi}_1 &= \widehat{r_0 r_0} + \widehat{l_1 l_1} + \widehat{r_1 r_1} + \widehat{l_2 l_2}\\ \widetilde{\pi}_2 &= \widehat{dd} + \widehat{\gamma\gamma} + \widehat{gg}\end{aligned}$$

It is interesting also to compare the above results with those obtained for the projector $\pi_n$ (see section 3.3.1. The fusion algebra of $A_3$ explicitly given in 2.1.2 is recovered :

$$\begin{array}{ll}\widetilde{\pi}_0 \,\widehat{\circ}\, \widetilde{\pi}_n = \widetilde{\pi}_n & \widetilde{\pi}_1 \,\widehat{\circ}\, \widetilde{\pi}_1 = \widetilde{\pi}_0 + \widetilde{\pi}_2\\ \widetilde{\pi}_1 \,\widehat{\circ}\, \widetilde{\pi}_2 = \widetilde{\pi}_1 & \widetilde{\pi}_2 \,\widehat{\circ}\, \widetilde{\pi}_2 = \widetilde{\pi}_0\end{array}$$

Evaluating the counit $\widehat{\epsilon}$ of $\widehat{\mathcal{B}}$ on characters gives in general $\widehat{\epsilon}(\widetilde{\pi}_n) = d_n$, i.e., $(3,4,3)$ for $A_3$.

### 3.7.2 Characters for the algebra $\widehat{\circ}$ on $\widehat{B}$ : the algebra of quantum symmetries

In the same way, characters for the semi-simple algebra $\widehat{\mathcal{B}}$ are elements $\widetilde{\omega}_x \in \mathcal{B}$ defined for all $\sigma \in \widehat{\mathcal{B}}$ by $< \sigma, \widetilde{\omega}_x >= Tr(\sigma \,\widehat{\circ}\, \omega_x)$, where $\omega_x$ are the minimal central projectors for $\widehat{\mathcal{B}}$.

Being elements of $\mathcal{B}$, the $\widetilde{\omega}_x$ can be decomposed on the basis $f_A$ (dual of the basis of matrix units of $\widehat{\mathcal{B}}$) : $\widetilde{\omega}_x = \widetilde{\omega}_{xA}\, f_A$. We can now take $\sigma = f^A$ in the defining formula, and the calculation is the same as before, showing that

$$\widetilde{\omega}_x = \sum_{\substack{A\in x\\ A \text{ is diagonal}}} f_A \qquad \text{to be compared with} \qquad \omega_x = \sum_{\substack{A\in x\\ A \text{ is diagonal}}} f^A$$

The projectors $\omega_x$ are elements of $\widehat{\mathcal{B}}$ and obey the usual projectors identities : $\omega_x \,\widehat{\circ}\, \omega_y = \delta_{xy}$ but the characters $\widetilde{\omega}_x$ are elements of $\mathcal{B}$ and generate, for the product $\circ$ a subalgebra $Oc(G)$ of $\mathcal{B}$ which, in general – i.e., for a diagram $G$ – is called the algebra of quantum symmetries. Its structure constant are denoted $O_{xyz}$ i.e.,

$$\widetilde{\omega}_x \circ \widetilde{\omega}_y = O_{xyz}\, \widetilde{\omega}_z$$

In the case of $A_N$ diagrams, we have the isomorphism $Oc(G) = A(G) = A_N$, so that we again recover the fusion algebra. The algebra of quantum symmetries has a unit (namely $\widetilde{\omega}_0$) but it does not coincide with the unit $\mathbb{1}$ of $\mathcal{B}$.

**Example of $A_3$ .**

Using tables given in section 3.1.5, one finds

$$\begin{aligned}\widetilde{\omega}_0 &= v_0 v_0 + v_1 v_1 + \gamma\gamma + v_2 v_2\\ \widetilde{\omega}_1 &= v_0 v_1 + v_1 v_0 + v_1 v_2 + v_2 v_1\\ \widetilde{\omega}_2 &= v_0 v_2 + v_1 v_1 - \gamma\gamma + v_2 v_0\end{aligned}$$

It is interesting to compare the above results with those obtained for the projector $\omega_x$ (see section 3.3.1. Warning: it would be a mistake to determine the expression of the characters $\widetilde{\omega_x}$ from the expression of the projectors $\omega_x$, just replacing the vectors $e^I$ by vectors $e_I$ (cf the warning (2) of section 3.1.5).

The fusion algebra of $A_3$ explicitly given in 2.1.2 is again recovered :

$$\begin{array}{ll}\widetilde{\omega}_0 \circ \widetilde{\omega}_n = \widetilde{\omega}_n & \widetilde{\omega}_1 \circ \widetilde{\omega}_1 = \widetilde{\omega}_0 + \widetilde{\omega}_2\\ \widetilde{\omega}_1 \circ \widetilde{\omega}_2 = \widetilde{\omega}_1 & \widetilde{\omega}_2 \circ \widetilde{\omega}_2 = \widetilde{\omega}_0\end{array}$$

Evaluating the counit $\epsilon$ of $\mathcal{B}$ on characters gives in general $\epsilon(\widetilde{\omega}_x) = d_x$, i.e., $(3,4,3)$ for $A_3$.

### 3.7.3 Comments

In references [32] and [5], the two products are defined on the same underlying vector space. In [5], the roles of our $e_I$ and $f^A$ are permuted (for several authors, matrix units for an algebra do not belong to it but to its dual), and the normalization is different because our diffusion graphs (the $e_I$ or the $f^A$) denote elementary matrices whereas in the quoted reference (and several others, like [32], [38]), they denote vectors only proportional to elementary matrices. As stated in the introduction, one drawback of defining the two products in the same space, in a situation like this one, where we have a natural bigebra, is that it is very difficult to discuss the homomorphism property of the coproduct in terms of a compatibility between the two products, without introducing the coproduct itself. Starting from a bigebra, one way to obtain the two multiplication laws "on the same side" (like $\circ$ and $\star$ rather than $\circ$ and $\widehat{\circ}$ is to use a scalar product and this is of course what happens when one starts from an a priori given $\mathbb{C}^*$ algebra. Besides this drawback, we have also the fact that one breaks, in this approach, the symmetry between the two structures, even for $A_N$ diagrams, despite of the fact that, for such diagrams, the map $f : \mathcal{B} \mapsto \widehat{\mathcal{B}}$ defined by $e_I \mapsto f^A$ is clearly such that $f(u \circ v) = f(u)\widehat{\circ}f(v)$ and $\widehat{\Delta}(f(u)) = (f \otimes f)\Delta(u)$. In reference [5], the objects corresponding to our minimal central projectors $\pi_n$ and characters $\widetilde{\pi_n}$ are identified (so they can be multiplied with the other multiplication) and are shown to close for this other product. However, the other projectors corresponding to our $\omega_x$ do not obviously correspond to characters that close under the first product. As a result, in such a formalism, the fusion algebra is easy to recognize within the algebra of double triangles, but the algebra of quantum symmetries is not. For this reason, we like to think that the approach presented here is both natural and simple. This may be a question of taste...

## 3.8 Varia

### 3.8.1 Path realization : horizontal (essential) paths and vertical paths

In the first section we started from a graded vector space $\mathcal{H}$ spanned by admissible triangles. These triangles were themselves directly obtained from the table describing the associative multiplication on the diagram $A_N$ (fusion algebra). More generally these triangles $\xi^n_{ab} = (a, n, b)$ could be obtained from the table describing the module structure of a given diagram (typically an $ADE$ diagram) over the fusion algebra of some (appropriate) $A_N$. The algebra $(\mathcal{B}, \circ)$, for its multiplication $\circ$ was obtained at the next step, as the graded endomorphism algebra associated with $\mathcal{H}$. We remember that, with $\mathcal{H} = \bigoplus_n \mathcal{H}_n$, we set $\mathcal{B} = \bigoplus_n \mathcal{B}_n$, with $\mathcal{B}_n = End(H_n)$.

There is another realization of the vector space $\mathcal{H}$, in terms of particular paths (called "essential paths") on the diagram $G$. This is actually the historical approach, and is how the algebra $\mathcal{B}$ was first introduced (see the article [32] by A. Ocneanu). This approach was then repeated and imitated in all subsequent papers. One of the features of the present article is that we do not need to introduce the concept of essential paths (not even the concept of paths) to build the bigebra $\mathcal{B}$ and discuss its properties since $\mathcal{H}$ itself is directly obtained from the multiplication table of $A_N$ or from the module table of a diagram $G$ over $A_N$. The method has also the advantage of being generalizable to higher Coxeter-Dynkin systems, for instance the $SU(3)$ system, in a straightforward way, whereas the concept of essential paths, for higher systems, is not totally obvious. It remains that essential paths are very interesting objects, and their study may ease the determination of cells, for general diagrams. One may want to make contact with a formalism used elsewhere, therefore we give a brief presentation of the subject. See also the footnote, in this section, concerning the relation with preprojective algebras.

**Paths and essential paths on ADE diagrams.** The following definitions are adapted from [32], see also several comments made in [11] and [12].

Call $\beta$ the norm of the graph $G$ i.e., the biggest eigenvalue of its adjacency matrix. The corresponding eigenspace is one dimensional, we call $\mu_a$ the components of the (normalized) eigenvector; for ADE diagrams the normalization is defined by stating that $\mu_0 = 1$ where 0 labels the terminal vertex of the longest branch (select one if there are several). The vector $\mu$ is often called the Perron Frobenius eigenvector, and $\mu_a$ is, by definition, the quantum dimension of the vertex labelled by $a$.

Call $\sigma_a$ the vertices of $G$ and, if $\sigma_b$ is a neighbor of $\sigma_a$, call $\xi_{ab}$ the oriented edge from $\sigma_a$ to $\sigma_b$. If $G$ is unoriented (the case for $ADE$ and affine $ADE$ diagrams), each edge should be considered as carrying both orientations. An *elementary path*, by definition, is a a sequence $(\xi(1)\xi(2)\ldots)$ of consecutive edges. It can be also be written as a finite sequence of consecutive (i.e., neighbors on the graph) vertices, $[\sigma_{a_1}\sigma_{a_2}\sigma_{a_3}\ldots]$ The correspondence between the two notations is of course defined by $\xi(1) = \xi_{a_1a_2} = \sigma_{a_1}\sigma_{a_2}$, $\xi(2) = \xi_{a_2a_3} = \sigma_{a_2}\sigma_{a_3}$, etc. . Vertices are considered as paths of length 0. The length of the possibly backtracking path $(\xi(1)\xi(2)\ldots\xi(p))$ is $p$. We call $r(\xi_{ab}) = \sigma_b$, the range of $\xi_{ab}$ and $s(\xi_{ab}) = \sigma_a$, the source of $\xi_{ab}$. For all edges $\xi(n+1) = \xi_{ab}$ that appear in an elementary path, we set $\xi(n+1)^{-1} \doteq \xi_{ba}$. The vector space generated by elementary paths is called $Paths$. It is graded by the length of the paths.

For every integer $n > 0$, the Ocneanu annihilation operator $C_n$, acting on the vector space generated by elementary paths of length $p$ is defined as follows: if $p \leq n$, $C_n$ vanishes, whereas if $p \geq n+1$ then

$$C_n(\xi(1)\xi(2)\ldots\xi(n)\xi(n+1)\ldots) = \sqrt{\frac{\mu_{r(\xi(n))}}{\mu_{s(\xi(n))}}}\,\delta_{\xi(n),\xi(n+1)^{-1}}(\xi(1)\xi(2)\ldots\hat{\xi}(n)\hat{\xi}(n+1)\ldots)$$

Here, the symbol "hat" ( like in $\hat{\xi}$) denotes omission. The result is therefore either 0 or a linear combination of paths of length $p-2$. Intuitively, $C_n$ chops the round trip that possibly appears at positions $n$ and $n+1$.

A path is called essential if it belongs to the intersection of the kernels of the annihilators $C_n$'s. Example: the following difference of non essential paths of length 2 on $A_3$ is an essential path : $\sigma_1\sigma_2\sigma_1 - \sigma_1\sigma_0\sigma_1$. The vector subspace of all essential paths of length $p$ is called $\mathcal{H}_p = EssPath_p$. We also set $\mathcal{H} = \bigoplus_p \mathcal{H}_p$. Vertices are taken as paths of zero length.

In the context of quantum groupoids discussed in the first section, there is a one-to-one correspondence between essential paths and non-zero entries of the module multiplication table of the diagram $G$. The spaces $\mathcal{H}_p$ are just the horizontal spaces needed in the definition of the algebra $B = \bigoplus_p End(\mathcal{H}_p) = \bigoplus_p \mathcal{H}_p \otimes \widehat{\mathcal{H}}_p$.

The vector space $Paths$ can be given an Hilbert space structure by declaring that the basis of elementary paths is orthonormal. Any vector subspace, in particular the subspace of essential paths, inherits this Hilbert space structure. For any operator acting on its vector space, one can then consider its adjoint (notation †). Creation operators are defined are the adjoint of annihilation operators.

Acting on elementary path of length $p$, the creation operators $C_n^\dagger$ are obtained as follows: if $n > p+1$, $C_n^\dagger$ vanishes, whereas, if $n \leq p+1$, setting $b = r(\xi(n-1))$,

$$C_n^\dagger(\xi(1)\ldots\xi(n-1)\ldots) = \sum_{d(b,c)=1} \sqrt{(\frac{\mu_k}{\mu_j})}(\xi(1)\ldots\xi(n-1)\xi_{bc}\xi_{cb}\ldots)$$

The above sum is taken over the neighbors $\sigma_c$ of $\sigma_b$ on the graph. Intuitively, this operator adds one (or several) small round trip(s) at position $n$. The result is therefore either 0 or a linear combination of paths of length $p+2$. For instance, on paths of length zero (i.e., vertices),

$$C_1^\dagger(\sigma_b) = \sum_{d(b,c)=1} \sqrt{(\frac{\mu_c}{\mu_b})}\xi_{bc}\xi_{cb} = \sum_{d(b,c)=1} \sqrt{(\frac{\mu_c}{\mu_b})}\,[\sigma_b\sigma_c\sigma_b]$$

Essential paths can also be defined as the elements of the vector space which is the intersection of the kernels of all the projectors $e_n$ defined by $e_n \doteq \frac{1}{\beta}C_n^\dagger C_n$

Call $Path_{ab}^p$ the vector subspace of all paths of length $p$ starting from vertex $a$ and ending on vertex $b$. Consider the space of endomorphisms with fixed origin $a$ and fixed length $p$, it is $\oplus_b EndPath_{ab}^p$. One can show that it is generated by the unit and by the corresponding restriction of the projectors $e_1, \ldots e_n$, for $n \leq p-1$. These endomorphisms provide a realization of the algebra of Jones-Temperley-Lieb (for given $a$ and $p$). Using the canonical Hilbert space structure defined previously, one can introduce an orthogonal decomposition of the vector space of paths: $Paths = EssPaths \oplus EssPaths^T$ and determine the corresponding orthogonal projectors.

The above definitions make sense for all bipartite graphs but we have in mind either the $ADE$ diagrams (their norm $\beta$ is smaller than 2, or the corresponding affine diagrams (their norm is equal to 2). In the later case, the diagrams can be interpreted in terms of binary polyhedral groups (McKay correspondence [24]) since they encode tensor multiplication of irreducible representations of these finite groups by the fundamental representation (see [11] for a discussion of essential paths along these lines).

We have given a realization of the graded space $\mathcal{H}$ of horizontal triangles in terms of essential paths on a given $ADE$ diagram $G$. Since we have always an explicit matrix realization of the algebra $\widehat{\mathcal{B}}$, in terms of a direct sum of matrix algebras, we have also a graded space $\mathcal{V}$ of vertical triangles, such that $\widehat{\mathcal{B}}_x$ can be identified with $End(\mathcal{V}_x)$. The vertical triangles can also be realized in terms of other essential paths, but those are paths on $Oc(G)$, the Ocneanu graph of $G$, not on $G$ itself. In this paper we are mostly concerned with the $G = A_N$ cases (where $Oc(G) = G$), so we stop the discussion here.

**Scalar product considerations** Elementary paths generate the vector space Paths. In particular, any essential path is a linear combination of elementary paths. There are two canonical scalar products on $Paths$. The first, is defined by stating that elementary paths are orthonormal. Admissible triangles are identified with essential paths that are normalized for this scalar product. For instance, the admissible triangle of $A_3$ that we called $\gamma = \xi_{11}^2$ in section 3.1.5 can be identified with the normalized essential path $\{|1 \stackrel{2}{\rightharpoonup} 1 >\} = (\sigma_1\sigma_2\sigma_1 - \sigma_1\sigma_0\sigma_1)/\sqrt{2}$. The following basis of $\mathcal{H}$ is therefore orthonormal for this canonical scalar product : $\{v_0, v_1, v_2, r_0, l_1, r_1, l_2, d, g, \gamma\}$.

The other scalar product is obtained by renormalizing elementary paths in a particular way. Let us consider a base made of elementary paths and call $|a \stackrel{n}{\rightharpoonup} b >$ such a basis vector (an elementary path of length $n$ starting from $a$ and ending on $b$). We then define

$$|a \stackrel{n}{\rightarrow} b >= (\frac{\mu_n}{\mu_a \mu_b})^{1/4} |a \stackrel{n}{\rightharpoonup} b >$$

The second scalar product is obtained by declaring that the basis made of these new vectors is orthonormal. The two scalar products also define, by restriction, hermitian structures on the vector subspace of essential paths and induce corresponding structures at the dual level and at the level of the tensor algebra. In particular, the choice of a Hilbert structure for $\mathcal{H}$ induces a Hilbert structure for the bigebra $\mathcal{B}$.

For diagrams of type $A$, there is one and only one essential path with given endpoints and prescribed length. In other words, the space $EssPath^n_{a,b}$ is of dimension 1. If the diagram is not of type $A$, there may be several linearly independent essential paths with the same length and endpoints and some non canonical choice of independent vectors has to be made in each of the vector subspaces $EssPath^n_{a,b}$.

**Frontiers: products on the space of paths** The set of paths has a classical groupoid structure: given two elementary paths $\xi$, $\eta$, such that $r(\xi) = s(\eta)$, one can define their concatenation product in the usual way but multiplication is not always defined. Replacing the set of paths by the vector space they generate allows one to obtain an everywhere defined multiplication by deciding that the product is zero (the null vector) when the paths cannot be concatenated. This product is a graded product: it is compatible with the length (see reference [16] for more considerations about related structures). The associative multiplication law $\circ$ of the bigebra $\mathcal{B}$, being the composition of endomorphisms of the graded vector space $\mathcal{H}$, one could be tempted to deduce the other associative multiplication law $\widehat{\circ}$ from some product defined in the space $\mathcal{H}$, at least in the case of diagrams $A_N$. Such a candidate for the product in $\mathcal{H}$ (call it $\times$) was explicitly considered for $A_N$ cases (see the last section of [15], for $A_2$ and [20] for other examples), but a general construction is not available. The product $\times$ on $\mathcal{H}$, that involves concatenation of paths and $6J$ symbols, is not the graded product but a filtered product[14] (in this respect, notice that concatenation of two essential paths is not essential in general, and one has reproject the result on $EssPath$ to obtain a graded product, which can indeed be shown to be associative [16]). However the product $\times$ of [20] was shown to be associative only in the case of $A_2$. Being however distributive over $+$, it could reproduce, once properly projected, the (associative) multiplication $\widehat{\circ}$ at the level of $\widehat{\mathcal{B}}$. At this point, we should return to the renormalized scalar product defined on $Paths$ (and on $\mathcal{H}$, $\widehat{\mathcal{H}}$, $\mathcal{B}$ and $\widehat{\mathcal{B}}$ ): on horizontal admissible triangles identified with normalized essential paths, it reads $(\xi^m_{ab}, \eta^n_{cd}) = \delta(\xi, \eta)\sqrt{\frac{ab}{n}}$ and for instance, in the case of $A_3$, the following basis is orthonormal for it: $\{v_0, v_1/2^{\frac{1}{4}}, v_2, r_0, l_1, r_1, l_2, d, g, \gamma/2^{\frac{1}{4}}\}$. Its interest lies in the fact that it is compatible with the product found in $\mathcal{H}$, in the sense that $(\xi \times \eta, \zeta) = (\xi, \zeta \times \widetilde{\eta})$ where $\widetilde{\eta}$ is the $Z_2$ image of $\eta$ (just reverse the path); this observation was made in [39]. This interesting bilinear form does not coincide with the two scalar products called $g$ and $h$ in section 3.6.

### 3.8.2 Representations for quantum groupoids

We only mention that, for quantum groupoids, the notion of tensor product of representations $V$ and $W$ is not the usual one: it is **not** $V \otimes W$ but

$$V \widetilde{\otimes} W \doteq \Delta(1) \quad V \otimes W$$

The unit object for this (modified) tensor product is not the usual "trivial" representation, but the subalgebra $\mathcal{B}_t$. As a consequence, the trivial representation is usually not 1-dimensional (see [26].) The action of $\mathcal{B}$ on $\mathcal{B}_t$ is defined as follows: $(u \in \mathcal{B}) \triangleright (s \in \mathcal{B}_t) = \epsilon_t(u \circ s)$.

### 3.8.3 Modular properties

The adjacency matrix of diagrams $A_N$ can be diagonalized, together with all other fusion matrices (they commute) by a matrix constructed from the set of eigenvectors. We first normalize this matrix in such a way that each line is of (naive) norm 1. We then choose some order on the set of vertices (for instance the natural order on vertices of $A_N$) and the corresponding order on the set of eigenvalues. The obtained unitarizing matrix represents the modular generator $S$ of the group $SL(2,\mathbb{Z})$.The so-called Verlinde formula goes backward, since it allows one to recover the fusion matrices in terms of matrix elements of a given representative of the generator $S$. The other generator of the modular group, $T$, is obtained from the exponential of the Casimir of $SU(2)$, up to a shift and a multiplicative constant, that can be fixed by imposing that the $SL(2,\mathbb{Z})$ relation $(ST)^3 = S^2$ hold. With $\kappa = N + 1$, one finds,

$$S_{ij} = \sqrt{\frac{2}{\kappa}} \sin(\pi \frac{(i+1)(j+1)}{\kappa}) \text{ for } 0 \leq i, j \leq \kappa - 2 \quad \text{and} \quad T_{ij} = exp[2i\pi(\frac{(j+1)^2}{4\kappa} - \frac{1}{8})]\, \delta_{ij}$$

These modular properties of $A_N$ graphs are not simply related with properties of their system of cells. We mention them here only for reference.

# 4 The golden quantum groupoid $A_4$

## 4.1 The $A_4$ fusion algebra

The diagram $A_4$ and its quantum dimensions are displayed below. Now $\kappa = 5$ and we set $q = exp(i\frac{\pi}{5})$. Its quantum mass is $m = \sum_n \mu_n^2 = 1 + \varphi^2 + \varphi^2 + 1 = 2(\varphi + 2)$. Here $\varphi =$

[14] The vector space $\mathcal{H}$ of essential paths and the vector space underlying the so-called Gelfand - Ponomarev preprojective algebra associated with the given graph are isomorphic. Indeed, their (graded) dimensions are equal. For essential paths on ADE diagrams these numbers were explicitly computed in [10], [12] and [38]; for preprojecfive algebras, see [25].

$2\,Cos(\pi/5) = \frac{1+\sqrt{5}}{2}$ is the golden number ($\varphi^2 = \varphi + 1$).

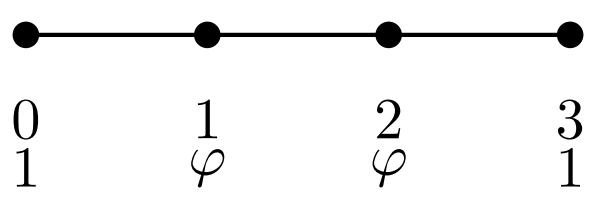

Figure 15: The diagram $A_4$ and its quantum dimensions

### 4.1.1 Fusion matrices and admissible triangles

The four fusion matrices of the diagram $A_4$ and the multiplication table of its fusion algebra:

$$N_0 = \begin{pmatrix} 1 & 0 & 0 & 0 \\ 0 & 1 & 0 & 0 \\ 0 & 0 & 1 & 0 \\ 0 & 0 & 0 & 1 \end{pmatrix} N_1 = \begin{pmatrix} 0 & 1 & 0 & 0 \\ 1 & 0 & 1 & 0 \\ 0 & 1 & 0 & 1 \\ 0 & 0 & 1 & 0 \end{pmatrix} N_2 = \begin{pmatrix} 0 & 0 & 1 & 0 \\ 0 & 1 & 0 & 1 \\ 1 & 0 & 1 & 0 \\ 0 & 1 & 0 & 0 \end{pmatrix} N_3 = \begin{pmatrix} 0 & 0 & 0 & 1 \\ 0 & 0 & 1 & 0 \\ 0 & 1 & 0 & 0 \\ 1 & 0 & 0 & 0 \end{pmatrix}$$

| $\cdot$ | 0 | 1 | 2 | 3 |
|---|---|---|---|---|
| 0 | 0 | 1 | 2 | 3 |
| 1 | 1 | $0+2$ | $1+3$ | 2 |
| 2 | 2 | $1+3$ | $0+2$ | 1 |
| 3 | 3 | 2 | 1 | 0 |

Diffusion graphs for $A_4$ :

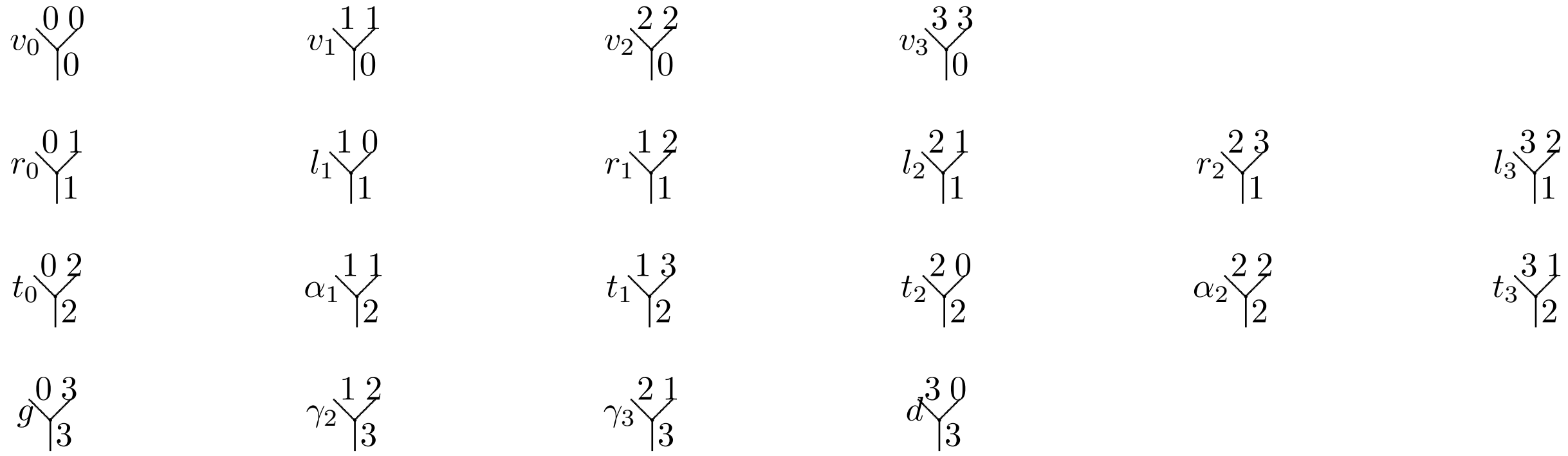

The horizontal graded vector space $\mathcal{H}$, of dimension $d_H = 4+6+6+4 = 20$ is spanned by above graphs. We apply a rotation of $\pi/2$ to them and obtain another copy of the same graphs, that span, by definition, the vertical graded vector space $\mathcal{V}$, of the same dimension.

We do not need to use a realization of admissible triangles in terms of essential paths (see 3.8.1), but, for completeness sake, we give it here:

$$\begin{aligned}
d_0 = 0 \quad &: \quad (0,1,2,3) \\
d_1 = 1 \quad &: \quad (01,10,12,21,23,32) \\
d_2 = 2 \quad &: \quad (012, \frac{1}{\varphi}121 - \frac{1}{\sqrt{\varphi}}101, 123, 210, -\frac{1}{\varphi}212 + \frac{1}{\sqrt{\varphi}}232, 321) \\
d_3 = 3 \quad &: \quad (0123, \frac{1}{\varphi}1012 - \frac{1}{\varphi^{3/2}}1212 + \frac{1}{\varphi}1232, \frac{1}{\varphi}2101 - \frac{1}{\varphi^{3/2}}2121 + \frac{1}{\varphi}2321, 3210)
\end{aligned}$$

We call them, in the same order:

$$( \ v_0, v_1, v_2, v_3 \ ; \ r_0, l_1, r_1, l_2, r_2, l_3 \ ; \ t_0, \alpha_1, t_1, t_2, \alpha_2, t_3 \ ; \ d, \gamma_1, \gamma_2, g \ )$$

We have therefore 20 admissible triangles. Using permutations, we have only 7 distinct classes of triangles. Here are the corresponding values of the triangular function.

| $(a,b,c)$ | $(0,0,0)$ | $(0,1,1)$ | $(0,2,2)$ | $(0,3,3)$ | $(1,1,2)$ | $(1,2,3)$ | $(2,2,2)$ |
|---|---|---|---|---|---|---|---|
| $\theta(a,b,c)$ | $1$ | $-\varphi$ | $\varphi$ | $-1$ | $\varphi$ | $-1$ | $-1/\varphi$ |

## 4.2 Values for quantum 6J and Racah symbols, cells and inverse cells

With $n = 0,1,2,3$ we have $c(0,n) = \{4,6,6,4\}$, $c(1,n) = \{6,14,14,6\}$, $c(2,n) = \{6,14,14,6\}$, $c(3,n) = \{4,6,6,4\}$, i.e., and therefore a total of $20+40+40+20 = 120$ admissible 6J symbols (arrays). Using symmetries, one can see that there are only 19 quantum tetrahedra, given below, with their values:

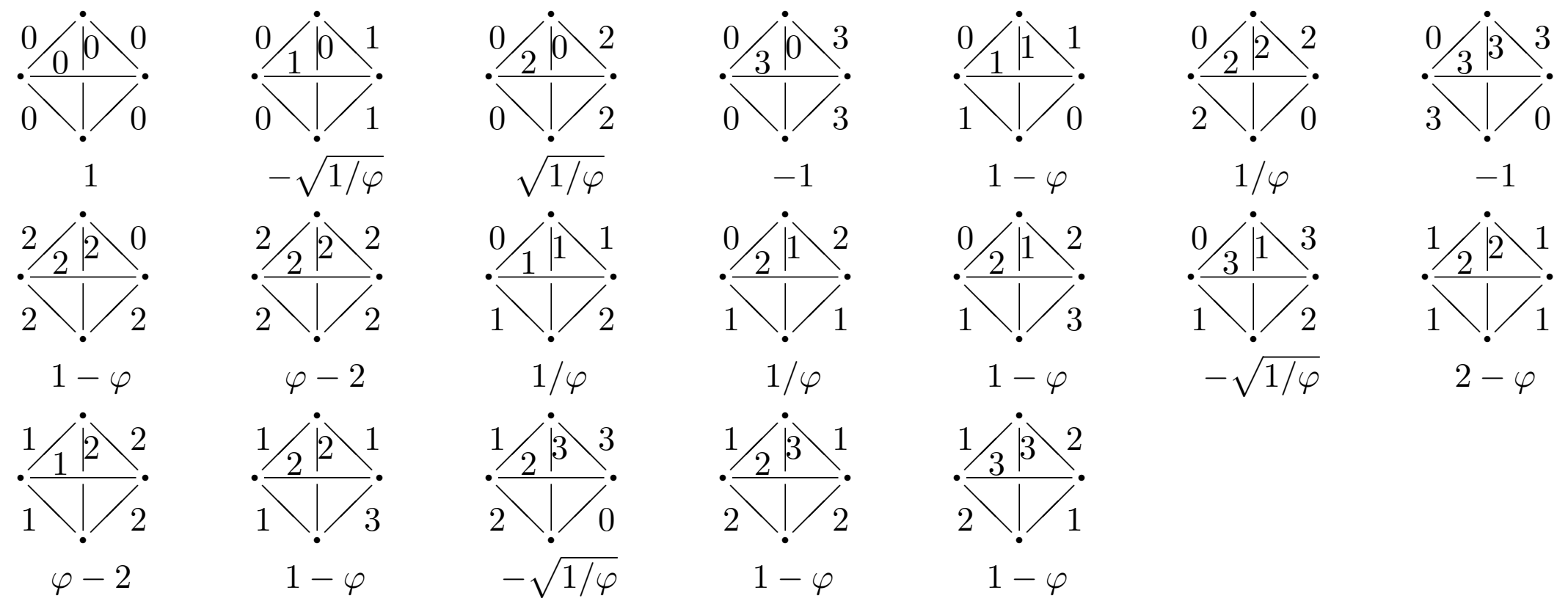

Figure 16: Admissible tetrahedra for the $A_4$ diagram

Using cell notation, we display on figures 17 and 18 the quantum Racah symbols (with their values), and the inverse Racah symbols.

## 4.3 The $A_4$ quantum manifold and its cells

There are 14 basic cells ($n = x = 1$) and 14 inverse basic cells given on figures 19 and 20.

## 4.4 The algebras $B$ and $\widehat{B}$

Elementary matrices $e_I$, for the $\circ$ product of $B(A_4) = EndH_0 \oplus EndH_1 \oplus EndH_2 \oplus EndH_3$, of dimension $d_B = \sum_n d_n^2 = 4^2 + 6^2 + 6^2 + 4^2 = 104$, are given below[15]:

$$\begin{pmatrix} v_0v_0 & v_0v_1 & v_0v_2 & v_0v_3 \\ v_1v_0 & v_1v_1 & v_1v_2 & v_1v_3 \\ v_2v_0 & v_2v_1 & v_2v_2 & v_2v_3 \\ v_3v_0 & v_3v_1 & v_3v_2 & v_3v_3 \end{pmatrix} \oplus \begin{pmatrix} r_0r_0 & r_0l_1 & r_0r_1 & r_0l_2 & r_0r_2 & r_0l_3 \\ l_1r_0 & l_1l_1 & l_1r_1 & l_1l_2 & l_1r_2 & l_1l_3 \\ r_1r_0 & r_1l_1 & r_1r_1 & r_1l_2 & r_1r_2 & r_1l_3 \\ l_2r_0 & l_2l_1 & l_2r_1 & l_2l_2 & l_2r_2 & l_2l_3 \\ r_2r_0 & r_2l_1 & r_2r_1 & r_2l_2 & r_2r_2 & r_2l_3 \\ l_3r_0 & l_3l_1 & l_3r_1 & l_3l_2 & l_3r_2 & l_3l_3 \end{pmatrix} \oplus$$

[15]The double triangle 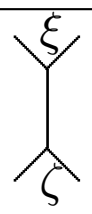 is written $\xi\zeta$

$$
\begin{pmatrix}
t_0t_0 & t_0\alpha_1 & t_0t_1 & t_0t_2 & t_0\alpha_2 & t_0t_3 \\
\alpha_1t_0 & \alpha_1\alpha_1 & \alpha_1t_1 & \alpha_1t_2 & \alpha_1\alpha_2 & \alpha_1t_3 \\
t_1t_0 & t_1\alpha_1 & t_1t_1 & t_1t_2 & t_1\alpha_2 & t_1t_3 \\
t_2t_0 & t_2\alpha_1 & t_2t_1 & t_2t_2 & t_2\alpha_2 & t_2t_3 \\
\alpha_2t_0 & \alpha_2\alpha_1 & \alpha_2t_1 & \alpha_2t_2 & \alpha_2\alpha_2 & \alpha_2t_3 \\
t_3t_0 & t_3\alpha_1 & t_3t_1 & t_3t_2 & t_3\alpha_2 & t_3t_3
\end{pmatrix}
\oplus
\begin{pmatrix}
dd & d\gamma_1 & d\gamma_2 & dg \\
\gamma_1d & \gamma_1\gamma_1 & \gamma_1\gamma_2 & \gamma_1g \\
\gamma_2d & \gamma_2\gamma_1 & \gamma_2\gamma_2 & \gamma_2g \\
gd & g\gamma_1 & g\gamma_2 & gg
\end{pmatrix}
$$

Elementary matrices $f^A$, for the $\widehat{\circ}$ product of $\widehat{\mathcal{B}}$, written in terms of the dual basis $e^I$ of the $e_I$ basis, are immediately obtained from the table of inverse cells (figure 18)

$$
\begin{pmatrix}
\widehat{v_0v_0} & \widehat{r_0r_0} & \widehat{t_0t_0} & \widehat{dd} \\
\widehat{l_1l_1} & \frac{\widehat{v_1v_1}}{\varphi^2} + \frac{\widehat{\alpha_1\alpha_1}}{\varphi} & \frac{\widehat{r_1r_1}}{\varphi} + \frac{\widehat{\gamma_1\gamma_1}}{\varphi^2} & \widehat{t_1t_1} \\
\widehat{t_2t_2} & \frac{\widehat{l_2l_2}}{\varphi} + \frac{\widehat{\gamma_2\gamma_2}}{\varphi^2} & \frac{\widehat{v_2v_2}}{\varphi^2} + \frac{\widehat{\alpha_2\alpha_2}}{\varphi} & \widehat{r_2r_2} \\
\widehat{gg} & \widehat{t_3t_3} & \widehat{l_3l_3} & \widehat{v_3v_3}
\end{pmatrix}
\oplus
\begin{pmatrix}
\widehat{v_0v_1} & \widehat{r_0l_1} & \widehat{r_0r_1} & \widehat{t_0\alpha_1} & \widehat{t_0t_1} & \widehat{d\gamma_1} \\
-\widehat{l_1r_0}\varphi & \widehat{v_1v_0} & -\widehat{\alpha_1t_0}\sqrt{\varphi} & \widehat{r_1r_0}\sqrt{\varphi} & -\widehat{\gamma_1d} & \widehat{t_1t_0}\varphi \\
\widehat{l_1l_2}\sqrt{\varphi} & \widehat{\alpha_1t_2} & \frac{\widehat{v_1v_2}+\widehat{\alpha_1\alpha_2}}{\varphi} & \frac{\widehat{\gamma_1\gamma_2}+\widehat{r_1l_2}}{\varphi} & \widehat{r_1r_2} & \widehat{t_1\alpha_2}\sqrt{\varphi} \\
-\widehat{t_2\alpha_1}\sqrt{\varphi} & \widehat{l_2l_1} & \frac{-\widehat{\gamma_2\gamma_1}-\widehat{l_2r_1}}{\varphi} & \frac{\widehat{v_2v_1}+\widehat{\alpha_2\alpha_1}}{\varphi} & -\widehat{\alpha_2t_1} & \widehat{r_2r_1}\sqrt{\varphi} \\
\widehat{t_2t_3}\varphi & \widehat{\gamma_2g} & \widehat{l_2l_3}\sqrt{\varphi} & \widehat{\alpha_2t_3}\sqrt{\varphi} & \widehat{v_2v_3} & \widehat{r_2l_3}\varphi \\
-\widehat{g\gamma_2} & \widehat{t_3t_2} & -\widehat{t_3\alpha_2} & \widehat{l_3l_2} & -\widehat{l_3r_2} & \widehat{v_3v_2}
\end{pmatrix}
\oplus
$$

$$
\begin{pmatrix}
\widehat{v_0v_2} & \widehat{r_0l_2} & \widehat{r_0r_2} & \widehat{t_0t_2} & \widehat{t_0\alpha_2} & \widehat{d\gamma_2} \\
-\widehat{l_1r_1}\sqrt{\varphi} & \frac{\widehat{v_1v_1}-\widehat{\alpha_1\alpha_1}}{\varphi} & -\widehat{\alpha_1t_1} & \widehat{r_1l_1} & \frac{\widehat{r_1r_1}-\widehat{\gamma_1\gamma_1}}{\varphi} & \widehat{t_1\alpha_1}\sqrt{\varphi} \\
\widehat{l_1l_3}\varphi & \widehat{\alpha_1t_3}\sqrt{\varphi} & \widehat{v_1v_3} & \widehat{\gamma_1g} & \widehat{r_1l_3}\sqrt{\varphi} & \widehat{t_1t_3}\varphi \\
\widehat{t_2t_0}\varphi & -\widehat{l_2r_0}\sqrt{\varphi} & \widehat{\gamma_2d} & \widehat{v_2v_0} & -\widehat{\alpha_2t_0}\sqrt{\varphi} & \widehat{r_2r_0}\varphi \\
-\widehat{t_2\alpha_2}\sqrt{\varphi} & \frac{\widehat{l_2l_2}-\widehat{\gamma_2\gamma_2}}{\varphi} & -\widehat{l_2r_2} & \widehat{\alpha_2t_2} & \frac{\widehat{v_2v_2}-\widehat{\alpha_2\alpha_2}}{\varphi} & \widehat{r_2l_2}\sqrt{\varphi} \\
\widehat{g\gamma_1} & -\widehat{t_3\alpha_1} & \widehat{t_3t_1} & \widehat{l_3l_1} & -\widehat{l_3r_1} & \widehat{v_3v_1}
\end{pmatrix}
\oplus
\begin{pmatrix}
\widehat{v_0v_3} & \widehat{r_0l_3} & \widehat{t_0t_3} & \widehat{dg} \\
-\widehat{l_1r_2} & \frac{\widehat{v_1v_2}}{\varphi^2} - \frac{\widehat{\alpha_1\alpha_2}}{\varphi} & \frac{\widehat{r_1l_2}}{\varphi} - \frac{\widehat{\gamma_1\gamma_2}}{\varphi^2} & \widehat{t_1t_2} \\
\widehat{t_2t_1} & \frac{-\widehat{l_2r_1}}{\varphi} + \frac{\widehat{\gamma_2\gamma_1}}{\varphi^2} & \frac{\widehat{v_2v_1}}{\varphi^2} - \frac{\widehat{\alpha_2\alpha_1}}{\varphi} & \widehat{r_2l_1} \\
-\widehat{gd} & \widehat{t_3t_0} & -\widehat{l_3r_0} & \widehat{v_3v_0}
\end{pmatrix}
$$

The dual basis $f_A$ of the basis of elementary matrices $f^A$, written in terms of the $e_I$ basis, are immediately obtained from the table of direct cells (figure 17).

$$
\begin{pmatrix}
v_0v_0 & r_0r_0 & t_0t_0 & dd \\
l_1l_1 & v_1v_1 + \alpha_1\alpha_1 & r_1r_1 + \gamma_1\gamma_1 & t_1t_1 \\
t_2t_2 & l_2l_2 + \gamma_2\gamma_2 & v_2v_2 + \alpha_2\alpha_2 & r_2r_2 \\
gg & t_3t_3 & l_3l_3 & v_3v_3
\end{pmatrix}
\oplus
\begin{pmatrix}
v_0v_1 & r_0l_1 & r_0r_1 & t_0\alpha_1 & t_0t_1 & d\gamma_1 \\
-\frac{l_1r_0}{\varphi} & v_1v_0 & \frac{-\alpha_1t_0}{\sqrt{\varphi}} & \frac{r_1r_0}{\sqrt{\varphi}} & -\gamma_1d & \frac{t_1t_0}{\varphi} \\
\frac{l_1l_2}{\sqrt{\varphi}} & \alpha_1t_2 & v_1v_2 + \frac{\alpha_1\alpha_2}{\varphi} & \gamma_1\gamma_2 + \frac{r_1l_2}{\varphi} & r_1r_2 & \frac{t_1\alpha_2}{\sqrt{\varphi}} \\
-\frac{t_2\alpha_1}{\sqrt{\varphi}} & l_2l_1 & -\gamma_2\gamma_1 - \frac{l_2r_1}{\varphi} & v_2v_1 + \frac{\alpha_2\alpha_1}{\varphi} & -\alpha_2t_1 & \frac{r_2r_1}{\sqrt{\varphi}} \\
\frac{t_2t_3}{\varphi} & \gamma_2g & \frac{l_2l_3}{\sqrt{\varphi}} & \frac{\alpha_2t_3}{\sqrt{\varphi}} & v_2v_3 & \frac{r_2l_3}{\varphi} \\
-g\gamma_2 & t_3t_2 & -t_3\alpha_2 & l_3l_2 & -l_3r_2 & v_3v_2
\end{pmatrix}
\oplus
$$

$$
\begin{pmatrix}
v_0v_2 & r_0l_2 & r_0r_2 & t_0t_2 & t_0\alpha_2 & d\gamma_2 \\
\frac{-l_1r_1}{\sqrt{\varphi}} & v_1v_1 - \frac{\alpha_1\alpha_1}{\varphi} & -\alpha_1t_1 & r_1l_1 & -\gamma_1\gamma_1 + \frac{r_1r_1}{\varphi} & \frac{t_1\alpha_1}{\sqrt{\varphi}} \\
\frac{l_1l_3}{\varphi} & \frac{\alpha_1t_3}{\sqrt{\varphi}} & v_1v_3 & \gamma_1g & \frac{r_1l_3}{\sqrt{\varphi}} & \frac{t_1t_3}{\varphi} \\
\frac{t_2t_0}{\varphi} & \frac{-l_2r_0}{\sqrt{\varphi}} & \gamma_2d & v_2v_0 & -\frac{\alpha_2t_0}{\sqrt{\varphi}} & \frac{r_2r_0}{\varphi} \\
\frac{-t_2\alpha_2}{\sqrt{\varphi}} & -\gamma_2\gamma_2 + \frac{l_2l_2}{\varphi} & -l_2r_2 & \alpha_2t_2 & v_2v_2 - \frac{\alpha_2\alpha_2}{\varphi} & \frac{r_2l_2}{\sqrt{\varphi}} \\
g\gamma_1 & -t_3\alpha_1 & t_3t_1 & l_3l_1 & -l_3r_1 & v_3v_1
\end{pmatrix}
\oplus
\begin{pmatrix}
v_0v_3 & r_0l_3 & t_0t_3 & dg \\
-l_1r_2 & v_1v_2 - \alpha_1\alpha_2 & r_1l_2 - \gamma_1\gamma_2 & t_1t_2 \\
t_2t_1 & -l_2r_1 + \gamma_2\gamma_1 & v_2v_1 - \alpha_2\alpha_1 & r_2l_1 \\
-gd & t_3t_0 & -l_3r_0 & v_3v_0
\end{pmatrix}
$$

## 4.5 Projectors and characters

From the previous results, we read immediately, the projectors and the characters, as well as the units and counits for algebras $\mathcal{B}$ and $\widehat{\mathcal{B}}$.

The four projectors $\pi_n \in \mathcal{B}$ given on figure 4.5 are obtained by summing the diagonal elements of the four blocks displaying the $e_I$ basis (elementary matrices for $\circ$). They obey $\pi_m \circ \pi_n = \delta_{mn}\pi_n$. The unit is $\mathbb{1} = \sum_n \pi_n$.

The four characters $\widetilde{\pi_n} \in \widehat{\mathcal{B}}$ that generate the fusion algebra of $A_4$ are also also given on figure 4.5, in terms of the vectors $e^I$ of the dual basis. They obey $\widetilde{\pi_m} \widehat{\circ} \widetilde{\pi_n} = N_{mn}^p \, \widetilde{\pi_p}$, where $N_{mn}^p$ are matrix elements of the fusion matrices.

The four projectors $\omega_x \in \widehat{\mathcal{B}}$ given on figure 4.5 are obtained by summing the diagonal elements of the four blocks displaying the $f^A$ basis (elementary matrices for $\widehat{\circ}$). They obey $\omega_x \widehat{\circ} \omega_y = \delta_{xy} \omega_x$. We display these projectors in terms of the $e^I$ basis, which is dual to the $e_I$ basis. The unit is $\widehat{\mathbb{1}} = \sum_x \omega_x$.

The four characters $\widetilde{\omega_x} \in \mathcal{B}$ that generate the algebra of quantum symmetries of $A_4$ are also given on figure 4.5 , in terms of the elementary matrices $e_I$ for $\mathcal{B}$. Like for all $A_N$ diagrams, this algebra is isomorphic with the fusion algebra. The characters $\widetilde{\omega_x}$ obey the relations $\widetilde{\omega_x} \circ \widetilde{\omega_y} = N_{xy}^z \, \widetilde{\omega_z}$, where $N_{xy}^z$ are matrix elements of the fusion matrices.

Counit $\epsilon$ for $\mathcal{B}$ : its value is 1 on all double triangles $e_I$ of type $n = 0$ and it vanishes on all others. Counit $\widehat{\epsilon}$ for $\widehat{\mathcal{B}}$ : its value is 1 on all double triangles $f^A$ of type $x = 0$ and it vanishes on all others. Antipodes : they are obtained like in section 3.4.

# 5 Cells : varia

Here we gather several remarks about cells that did not properly fit in the previous sections.

## 5.1 Cell systems, basic cells, orientation matters, quivers and quantum manifolds

### 5.1.1 Cell systems and basic cells.

The set of all cells with specified values of $n$ and $x$ is called the cell system of type $(n, x)$. We know that the fusion algebra $A(G)$ and the algebra of quantum symmetries $Oc(G)$ both act on the vector space spanned by the vertices of $G$. The associative algebra $A(G)$ has a unit denoted 0 and one generator denoted 1. The associative algebra $Oc(G)$ has usually (but not always) two generators denoted $1_L$ and $1_R$. The two actions of $A(G)$ and $Oc(G)$ on $G$ are described by "annular" and "dual annular" matrices[16] $F_n$ and $S_x$. Since these matrices have positive integer entries, they can be considered as adjacency matrices of particular graphs that we call $\mathcal{F}_n$ or $\mathcal{S}_x$. Matrix $F_1$ coïncides with the adjacency matrix of $G$, so that $\mathcal{F}_1 = G$. The graphs $\mathcal{F}_n$, for $n > 1$ are usually disconnected. In the same way $\mathcal{S}_{1_L}$ is associated with one chiral part of the Ocneanu graph and $\mathcal{S}_{1_R}$ with the other. A cell of type $(n, x)$ has horizontal edges that belong to the graph $\mathcal{F}_n$ and vertical edges that belong to the graph $\mathcal{S}_x$. A general theory of connections on systems of four graphs was described in [29]. A cell system of type $(n, x = 1_L)$ or $(n, x = 1_R)$ is simply encoded by the matrix $F_n$, together with particular lines relating non-zero matrix elements. Each entry of this matrix (hence each vertex of $\mathcal{F}_n$ ) indeed refers to an (admissible) horizontal triangle, i.e., the top or a bottom edge of a cell, sometimes with multiplicity; whereas edges of $\mathcal{F}_n$ encode the cells themselves. Basic cells are of the special kind $n = 1$ and $x = 1_L$ or $x = 1_R$.

### 5.1.2 Orientation matters, quivers and quantum manifolds.

Cells, that we always draw unoriented in the present paper should actually always be thought of as oriented as follows :

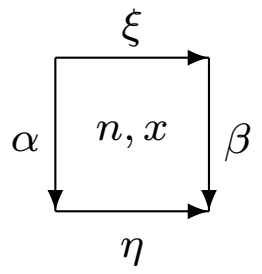

Each cell of type $(n, x)$ has top and bottom edges, as well as left and right edges; using the above orientation, each of the four edges has itself a source and a target (in the simplest case $(A_2)$, such considerations fit in the framework of double categories [1]). When we follow a clockwise loop starting from the source of a cell (its northeast corner), we follow a sequence of four sides but we walk along the first two ones in a way that is compatible with their orientation ( northwest $\mapsto$ northeast and north $\mapsto$ south), but the last two in a way opposite to their orientation. If we compose two cells vertically (composition of cells meaning repeated use of the Racah identity) the third side of the first cell becomes the first side of the next, but two such sides carry opposite orientation. For this reason, each system of cells, in particular the system of basic ones, can be written as the union of two subsets, mirror images of each other, such that vertical composition always uses two cells belonging to each one of the two subsets. For a diagram $G$ and for cells of type $(x, n)$, the two subsystems may be called "quantum manifolds" of type $(x, n)$ associated with the given diagram. The full quantum manifold is, by definition, the union of these two mirror images. Each quantum manifold is associated with a particular way to orient the edges of the given diagram — a quiver. Therefore, a given diagram gives rise to two quivers and the rule is that any oriented edge should be surrounded by edges carrying the opposite orientation.

[16]Again, remember that for $A_N$ diagrams, matrices $N$, $F$ and $S$ coincide.

Although the present article is mostly devoted to $A_N$ diagrams, we shall illustrate the above general comments with several features borrowed from the study of the $E_6$ diagram. In that case we have the following two quivers 5.1.2:

One may consider two diagrammatic descriptions of the cell systems; we call the first a "quantum manifold" description, and the next a "cell diagram" description. Consider the following description of the $E_6$ quantum manifold (figure 24)

From left to right (drawn vertically), we see four copies of the $E_6$ diagram. We can also see it as a Jones' tower, displayed from left to right, associated with a Brateli diagram which, in this case, is the direct quiver of $E_6$, and truncated after four steps (reflect the first $E_6$ diagram with respect to successive vertical lines). If we identify the last vertical line with the first, we "see" the cells: they are sequence of four edges making a closed 4-gon. Clearly, this quantum manifold has 12 vertices ($= 2 \times 6$), 20 edges (10 horizontal and 10 vertical paths of length 1, with $10 = 5 \times 2$ for the two orientations) and 15 cells. The mirror quantum manifold has the same vertices and edges but distinct cells (15 others).

For more general cell systems (not only the basic one), it is easier to display cell diagrams. In the case of $E_6$, the cell diagram, for its basic cell system, is displayed to the right of the quantum manifold, on figure 24; on this diagram we recognize the 10 non-zero matrix elements of the adjacency matrix, but these 10 "points" are now interpreted as sides of cells; we see also the 15 connecting lines (these "lines" represent cells). We therefore recover the 30 cells of the full quantum manifold once we orient the connecting lines.

**The $A_3$ example.** Going back to this simple example we display the two quivers for $A_3$ on figure 25. We also give the corresponding quantum manifold and the cell diagram for its basic cells ($n = 1$ and $x = 1$). The quantum manifold associated with $A_3$ has 6 vertices, 8 edges and 4 cells — or 8 for the full quantum manifold.

## 5.2 Corner and anti-corner matrices

The "corner matrix" $C$ is a direct sum of $N$ blocks (the number of vertices of the given diagram $G$) and each block, labelled by the vertex $v$, is a square matrix whose matrix elements are cells with specified corners (extremities of the main diagonal) both equal to $v$. The "anti-corner matrix" $A$ is obtained from $C$ when we apply an horizontal symmetry to all its cells. It is therefore a direct sum of $N$ blocks, and each block, labelled by $v$, is a matrix whose elements are cells with specified "anti-corners" (extremities of the second diagonal) both equal to $v$. The corner matrix of $A_3$ contains $1^2 + 2^2 + 1^2 = 6$ cells. The corner matrix of $E_6$ contains $1^2 + 2^2 + 3^3 + 2^2 + 1^2 + 1^2 = 20$ cells. Notice that we have one cell for each endpoint of the diagram, $3^2 = 9$ cells for the triple point and $2^2 = 4$ cells for the other interior points. We have the same combinatorics for the anti-corner matrix. For $E_7$ this leads to $3\times1^2+1\times3^2+3\times2^2 = 22$ non trivial cells for $C$ or $A$, with $3\times3\times1+1\times3+3\times2 = 12$ common entries, so $22+22-12 = 32$ basic cells. For $E_8$ we have $3 \times 1^2 + 1 \times 3^2 + 4 \times 2^2 = 26$ non trivial cells for $C$ or $A$, with $3 \times 3 \times 1 + 1 \times 3 + 4 \times 2 = 14$ common entries, so $26 + 26 - 14 = 38$ basic cells Warning: for general graphs (not the $A_N$ case), there are two families of conjugated basic cells: those of of type $n = 1, x = 1_L$ and those of type $n = 1, x = 1_R$, and that the total number of basic cells is twice the previous numbers; all together, for instance, we have $30 + 30 = 60$ basic cells for $E_6$.

Not all basic cells appear in the corner matrix, and not all basic cells appear in the anti-corner matrix either, but all basic cells appear in the union of the two. Actually the set of diagonal elements of (4 for $A_3$, 10 for $E_6$) for $A$ and $C$ is the same, although they do not appear at the same place, and we recover the total number of basic cells by writing $30 = 20 + 20 - 10$ for $E_6$ and $8 = 6 + 6 - 4$. More generally, for $A_N$ diagrams, the number of basic cells will be $(4N - 6) + (4N - 6) - (2N - 2) = 6N - 10$. Of course, this is also equal to the number $c(1, 1)$ calculated from fusion matrices.

Cells belonging to the same column of $C$ (or of $A$) have the same top edges. If we apply the vertical reflection to all the cells of $A$, we recover $C$. If we apply the horizontal reflection to all the cells of $A$, we recover the transpose $C^T$ of $C$. The unitarity properties, for cells, can be expressed in terms of unitarity for matrices $C$ and $C^T$. Matrix $A$ is not unitary.

As an example, we display the corner and anti-corner matrices for $A_3$ and $E_6$. Vertices of $A_3$ are ordered $0, 1, 2$ and those of $E_6$ as $0, 1, 2, 5, 4, 3$.

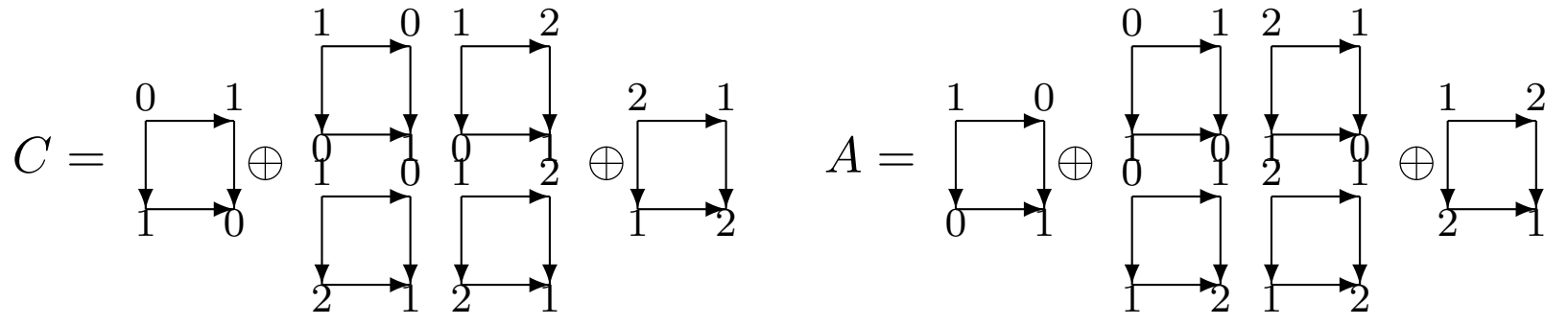

Corner and anti-corner matrices : the $A_3$ case.

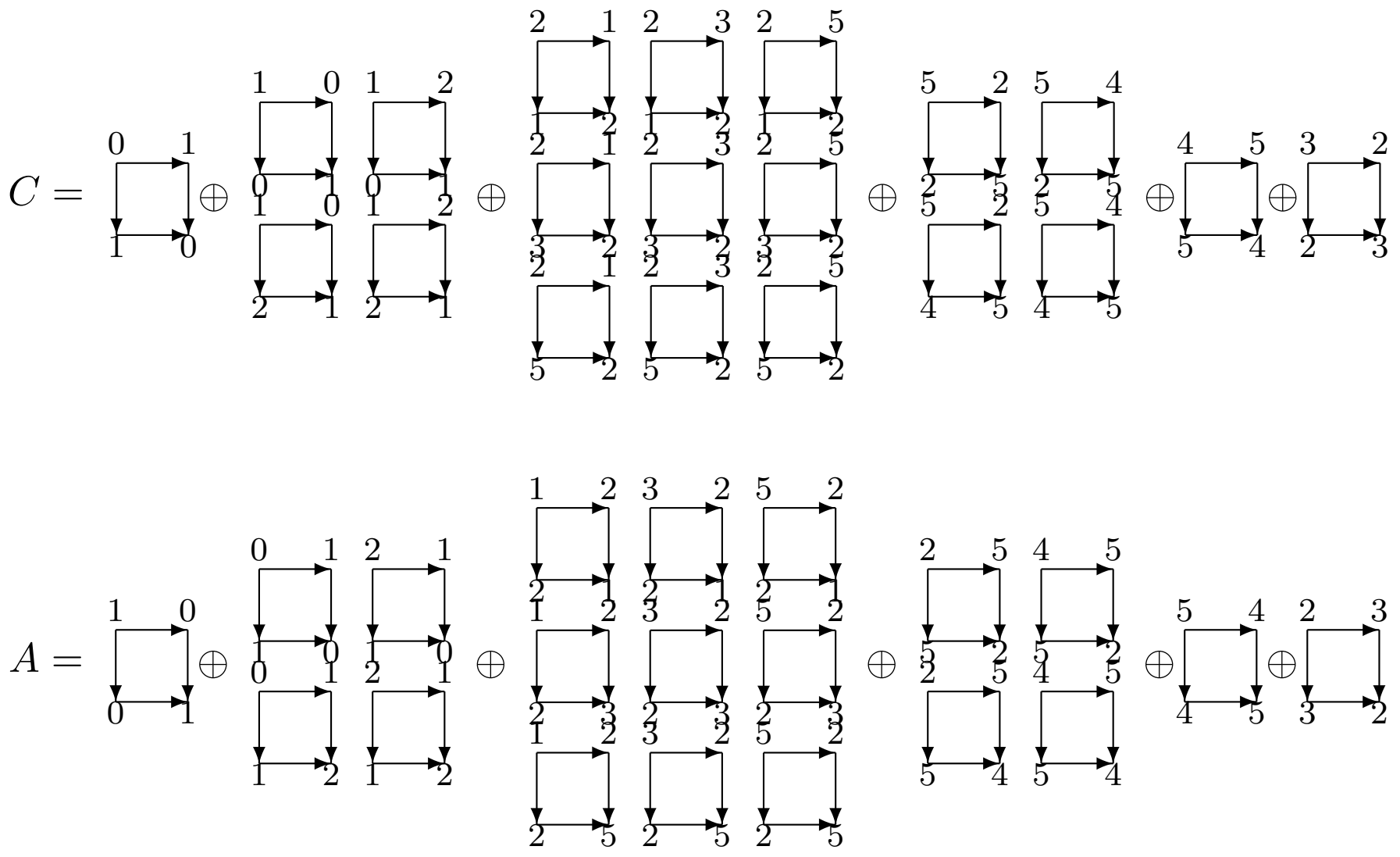

Corner and anti-corner matrices : the $E_6$ case.

### 5.3 The basic top, bottom, left and right matrices $T$, $B$, $L$ and $R$

In the $A_N$ cases, we already know a direct way of computing the values of cells, via the associated quantum Racah symbols, but for more general diagrams, it is useful to consider an analog of the matrices defined below.

**Definition of $T$ and $B$.** Matrix elements of the "top matrix" $T$ are the cells of the quantum manifold (4 cells for $A_3$). Lines of this matrix are labelled by the (oriented) upper horizontal edges of the cells, and they coincide with the arrows of the direct quiver. Its columns are labelled by the (oriented) lower horizontal edges of the cells, and they coincide with the arrows of the reflected quiver. Matrix elements of the "bottom matrix" $B$ are the cells of the other (mirror) quantum manifold of (again 4 for$A_3$). Lines of this matrix are also labelled by the (oriented) upper edges of the cells but they now coincide with the arrows of the reflected quiver. Columns are also labelled by the (oriented) lower edges of the cells, but they now coincide with the arrows of the direct quiver. Matrices $T$ and $B$ are not unitary. The cells of $B$ are obtained from those of $T$ by vertical reflection. Multiplication $T.B$ (or $B.T$) implements vertical composition of cells up to normalization factors.

$$T = \qquad\qquad\qquad\qquad B =$$

$$\begin{Bmatrix} 0 & 1 \\ 1 & 0 \\ 2 & 1 \\ 1 & 0 \end{Bmatrix} \begin{Bmatrix} 0 & 1 \\ 1 & 2 \\ 2 & 1 \\ 1 & 2 \end{Bmatrix} \quad \| \quad \begin{Bmatrix} 1 & 0 \\ 0 & 1 \\ 1 & 2 \\ 0 & 1 \end{Bmatrix} \begin{Bmatrix} 1 & 0 \\ 2 & 1 \\ 1 & 2 \\ 2 & 1 \end{Bmatrix}$$

**Definition of $L$ and $R$.** Matrix elements of the "left matrix" $L$ are the cells of the quantum manifold. Lines of this matrix are labelled by the (oriented) left vertical edges of the cells, and they coincide with the arrows of the direct quiver. Its columns are labelled by the (oriented) right vertical edges of the cells, and they coincide with the arrows of the reflected quiver. Matrix elements of the "right matrix" $R$ are the cells of the other (mirror) quantum manifold. Lines of this matrix are also labelled by the the (oriented) left vertical of the cells but they now coincide with the arrows of the reflected quiver. Columns are also labelled by the (oriented) right vertical edges of the cells, but they now coincide with the arrows of the direct quiver. Matrices $L$ and $R$ are not unitary. Cells of $R$ are obtained from those of $L$ by horizontal reflection. Multiplication $L.R$ (or $R.L$) implements the horizontal composition of cells up to normalization factors.

$$L = \qquad\qquad\qquad\qquad R =$$

$$\begin{Bmatrix} 0 & 1 \\ 1 & 0 \\ 2 & 1 \\ 1 & 0 \end{Bmatrix} \begin{Bmatrix} 0 & 1 \\ 1 & 2 \\ 2 & 1 \\ 1 & 2 \end{Bmatrix} \quad \| \quad \begin{Bmatrix} 1 & 0 \\ 0 & 1 \\ 1 & 0 \\ 2 & 1 \end{Bmatrix} \begin{Bmatrix} 1 & 2 \\ 0 & 1 \\ 1 & 2 \\ 2 & 1 \end{Bmatrix}$$

### 5.4 Glimpse of a general theory of cells and connections

We have two graded vector spaces ($\mathcal{V}$ and $\mathcal{H}$), with involution, coming with two particular basis. One first defines elementary cells as squares with four sides. Sides (horizontal or vertical) are oriented and are taken as vectors belonging to the two given basis. For instance one can take $\mathcal{V}$ and $\mathcal{H}$ as the linear span of admissible triangles. On these particular basis elements, we need also two maps, called "source" $s$ and "target" $t$ that generalize those defined by $s[(a,b,n)] = a$ and $t[(a,b,n)] = b$. Objects such as $a, b$, or $n$, have also quantum dimensions $\mu_a, \mu_b$ or $\mu_n$. One constructs a vector space $Cells$ which is the linear span of all elementary cells. At the moment, a cell has no values, it is just a box. If two cells match horizontally (or vertically), one calls $C_1 \# C_2$ the larger cell obtained by concatenating the upper and lower (or left and right) paths of the two cells and removing the intermediate vertical (or horizontal) path. If the two cells do not match, $C_1 \# C_2$ is defined to be the zero cell. A "connection" $X$ is now defined as a linear form on the space $Cells$ that obeys several constraints : constraints under horizontal or vertical reflections (generalizing those discussed in the first part of this paper for quantum Racah symbols), constraints of bi-unitarity (same comment), and a gluing constraint saying that the value of $X[C]$ of $X$ on an elementary cell $C$ is equal, up to normalization factors, to the sum of products $X[C_1]X[C_2]$ where the sum runs over elementary matching cells such that $C = C_1 \# C_2$. Connections, in the previous sense, can be multiplied (composed) and there is an obvious notion of direct sum. It is then natural to consider irreducible connections [32] . Going back to the cases where the system of cells is determined by the study of the $A_N$ module structure of the vector space spanned by the vertices of a simply laced Dynkin diagram $G$, we may identify such irreducible connections with the different blocks of the algebra $(\widehat{\mathcal{B}}, \widehat{\circ})$; its values on cells with horizontal sides labelled by $n$ are given by pairing the elementary matrices of the block $x$ of $(\widehat{B}, \widehat{\circ})$ with elementary matrices of the block $n$ of $(\mathcal{B}, \circ)$, i.e., by pairing double triangles of type $GGO$ with double triangles of type $GGA$: we recover the Ocneanu cells, and the algebra of irreducible connections is isomorphic with the algebra of quantum symmetries. We already mentioned that for generic $ADE$ diagrams, this algebra has two generators denoted $1_L$ and $1_R$ called "fundamental connections" (or "chiral generators"). Actually, there are several variants of the general theory, reflecting (or generalizing) the freedom that we met in the definition of several possible types of quantum Racah symbols —standard ones or geometrical ones for instance. We have therefore several variants of the notion of connection on a system of cells. For the connections that we have used so far (call them "$U$-connections"), the symmetry reflection constraints that we gave are a bit annoying because of the presence of numerical prefactors. One can impose different types of reflection constraints and accordingly obtains different definitions for the connections. Imposing that reflection constraints are as simple as possible (no numerical factor) defines "$S$-connections", but in that case the unitary constraints are more complicated: unitarity becomes unitarity up to a numerical factor. Still another possibility is to use simple axioms for reflections but to modify the "gluing axiom", i.e., decide to introduce a numerical factor whenever cells are glued together (multiplied) along a given side. On basic cells, $U$ and $S$ connection $X_U$ and $X_S$ are related as follows:

$$\begin{array}{cc} v_1 & v_2 \\ \multicolumn{2}{c}{\boxed{\;X_U\;}} \\ v_4 & v_3 \end{array} \quad = \quad \left(\frac{\mu_2\mu_4}{\mu_1\mu_3}\right)^{\frac{1}{4}} \begin{array}{cc} v_1 & v_2 \\ \multicolumn{2}{c}{\boxed{\;X_S\;}} \\ v_4 & v_3 \end{array}$$

Connections on $A$ diagrams were studied in the first part since this amounts to determine the quantum Racah symbols associated with the chosen $A_N$. For other members of the $ADE$ Dynkin system, not all cells can be made real. One possibility, as discussed in [31] is to start from two complex conjugated "fundamental connections" (or "left and right chiral generators") respectively denoted $X = 1_L$ and $X = \overline{1_R}$ defined as follows (they can be given either as $S$-connections or as $U$-connections):

$$\begin{array}{cc} v_1 & v_2 \\ \multicolumn{2}{c}{\boxed{\;X_S\;}} \\ v_4 & v_3 \end{array} \quad = \quad \delta_{13}\,\epsilon\,F^{\frac{1}{2}} + \delta_{24}\,\overline{\epsilon}\,F^{-\frac{1}{2}} \quad or \quad \begin{array}{cc} v_1 & v_2 \\ \multicolumn{2}{c}{\boxed{\;X_U\;}} \\ v_4 & v_3 \end{array} \quad = \quad \delta_{13}\,\epsilon\,F + \delta_{24}\,\overline{\epsilon}$$

where $\epsilon = -i\exp(\frac{i\pi}{2\times\kappa})$ for a diagram with Coxeter number $\kappa$, and where $F = \sqrt{\frac{\mu_2\mu_4}{\mu_1\mu_3}}$. This is compatible with the bi-unitarity constraints because $\epsilon^2 + \overline{\epsilon}^2 + \beta = 0$ with $\beta = (2)_q = 2\cos(\pi/\kappa)$. The quantity $F$ can be thought as the curvature of a cell since it describes a kind of generalized parallel transport from top to bottom horizontal edges along the vertical edges. The other fundamental connection ($1_R$) is obtained by replacing $\epsilon$ by its complex conjugate $\overline{\epsilon}$ in the above expressions. Values previously given for basic quantum Racah symbols do not look compatible with this (for $A_3$, $\beta = \sqrt{2}$ and $\epsilon = \frac{\sqrt{2-\sqrt{2}}}{2} - i\frac{\sqrt{2+\sqrt{2}}}{2}$), but there is a notion of gauge equivalence in this context (changes of basis, in $\mathcal{B}$ and in $\widehat{\mathcal{B}}$ preserving the bi-unitarity conditions); for $A_N$ diagrams the two generators are equivalent and are identified. A particular gauge choice can make this generator real. It vanishes on cells of type $(n, x \neq 1)$, and its values on cells of type $(n, 1)$ are given by the values of the Racah symbols $\{\begin{smallmatrix} a & n & b \\ d & 1 & c \end{smallmatrix}\}$.

Let us conclude with a few considerations on the $E_6$ diagram. Its quantum dimensions are

$$\{\mu_0 = 1, \mu_1 = \sqrt{2+\sqrt{3}}, \mu_2 = 1+\sqrt{3}, \mu_5 = \sqrt{2+\sqrt{3}}, \mu_4 = 1, \mu_3 = \sqrt{2}\}$$

The values of the fundamental connection $1_L$ (or its conjugate $1_R$) for the 15 directed (basic) cells and the 15 reflected (basic) cells are obtained by the above formula. Using the gluing axiom for cells, one should be able to determine the values of all Ocneanu cells. One should actually determine not only all the cells of this type (associated with tetrahedra with two black and two white vertices), but also the four other types of cells. Remember that $E_6$ enjoys self-fusion, that its fusion algebra is $A_{11}$ (so there are 11 blocks labelled by $n$), and that its algebra of quantum symmetries $Oc(E_6)$ is isomorphic with $E_6 \dot{\otimes} E_6$, where the tensor product is taken above the subalgebra $A_3$ spanned by the three end-point vertices (so we have 12 blocks labelled by $x$). The dimension of the blocks, for $\mathcal{B}, \circ$ and for $\widehat{\mathcal{B}}, \widehat{\circ}$ can be obtained by indirect methods (see for instance [10], [12] or [38]): $d_n = (6, 10, 14, 18, 20, 20, 20, 18, 14, 10, 6)$, $d_x = (6, 8, 6; 10, 14, 10; 10, 14, 10; 20, 28, 20)$ so that $dim\, \mathcal{H} = dim\, \mathcal{V} = 156$ and $dim\, \mathcal{B} = dim\, \widehat{\mathcal{B}} = 2512$.

It would be nice to be able to discuss explicitly the properties of this quantum groupoid. We hope that the general discussion carried out in this last section will trigger the interest of some reader who will undertake such a study, be able to find a simple way to calculate the values of the corresponding cells (16256 of them) and encode the results in a simple way. The structure of the set of non-zero cells, for the $E_6$ diagram, can be understood from the analysis of matrices $F_n.S_x.F_n.S_x$ where $F_n$ are the annular matrices and $S_x$ the dual annular matrices (they are given, for instance, in [10]). In particular, the number of cells of type $(n, x)$, is

$$\text{c(n,x)} = \begin{pmatrix} 6 & 8 & 6 & 10 & 16 & 10 & 10 & 16 & 10 & 30 & 56 & 30 \\ 10 & 16 & 10 & 30 & 56 & 30 & 30 & 56 & 30 & 106 & 208 & 106 \\ 16 & 32 & 16 & 56 & 112 & 56 & 56 & 112 & 56 & 208 & 416 & 208 \\ 26 & 48 & 26 & 86 & 168 & 86 & 86 & 168 & 86 & 314 & 624 & 314 \\ 30 & 56 & 30 & 106 & 208 & 106 & 106 & 208 & 106 & 390 & 776 & 390 \\ 32 & 64 & 32 & 112 & 224 & 112 & 112 & 224 & 112 & 416 & 832 & 416 \\ 30 & 56 & 30 & 106 & 208 & 106 & 106 & 208 & 106 & 390 & 776 & 390 \\ 26 & 48 & 26 & 86 & 168 & 86 & 86 & 168 & 86 & 314 & 624 & 314 \\ 16 & 32 & 16 & 56 & 112 & 56 & 56 & 112 & 56 & 208 & 416 & 208 \\ 10 & 16 & 10 & 30 & 56 & 30 & 30 & 56 & 30 & 106 & 208 & 106 \\ 6 & 8 & 6 & 10 & 16 & 10 & 10 & 16 & 10 & 30 & 56 & 30 \end{pmatrix}$$

Lines are ordered as vertices of $A_{11}$. For columns, we order the Ocneanu graph of $E_6$ (see figure 27) as follows: $0, 3, 4; 1_L, 2_L, 5_L; 1_R, 2_R, 5_R; c_1, c_2, c_3$. As expected from our discussion of the quantum manifold of $E_6$, we recover 30 basic cells for $(n = 1, x = 1_L)$, and also 30 basic cells for $(n = 1, x = 1_R)$.

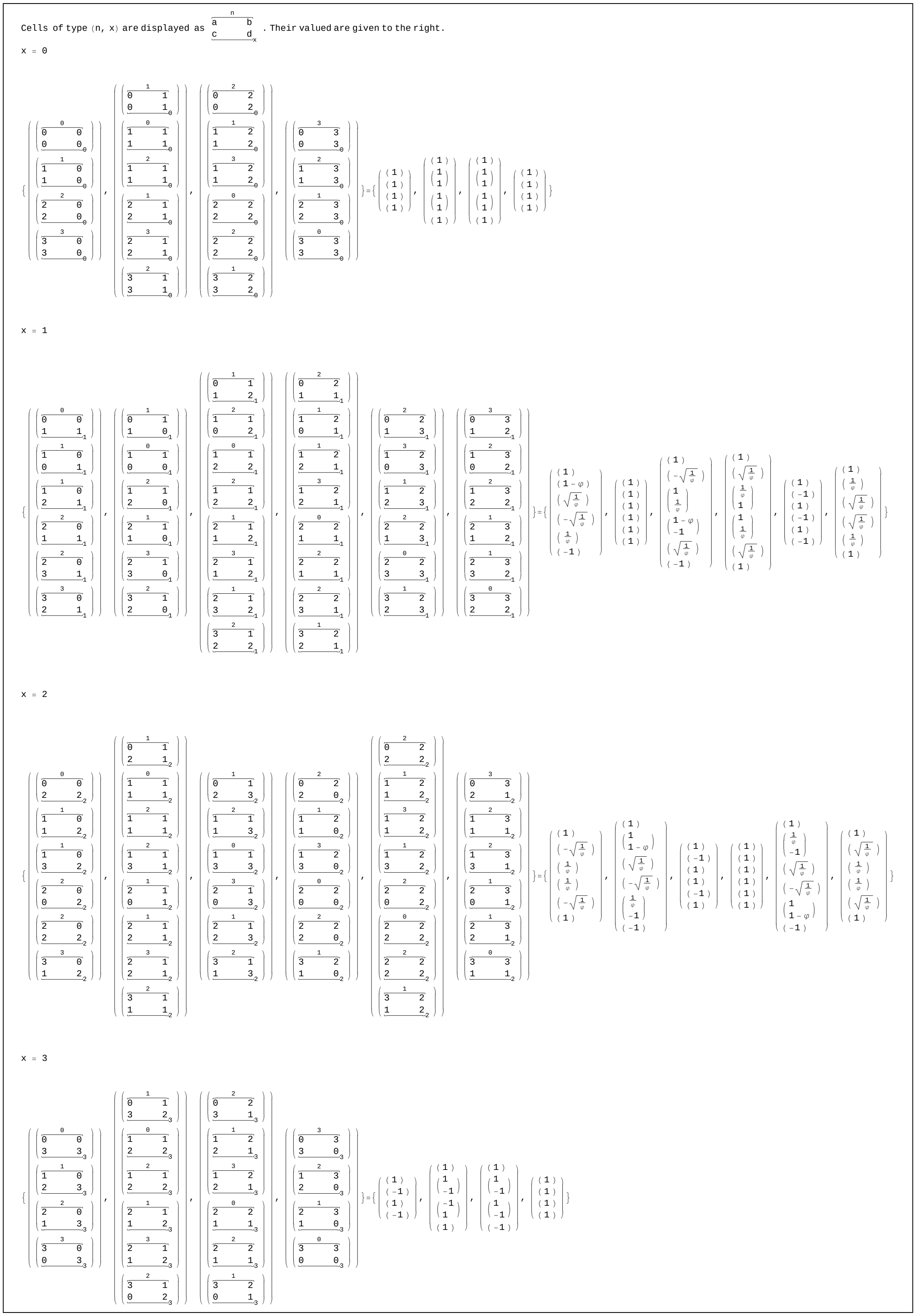

Figure 17: Cells for $A_4$

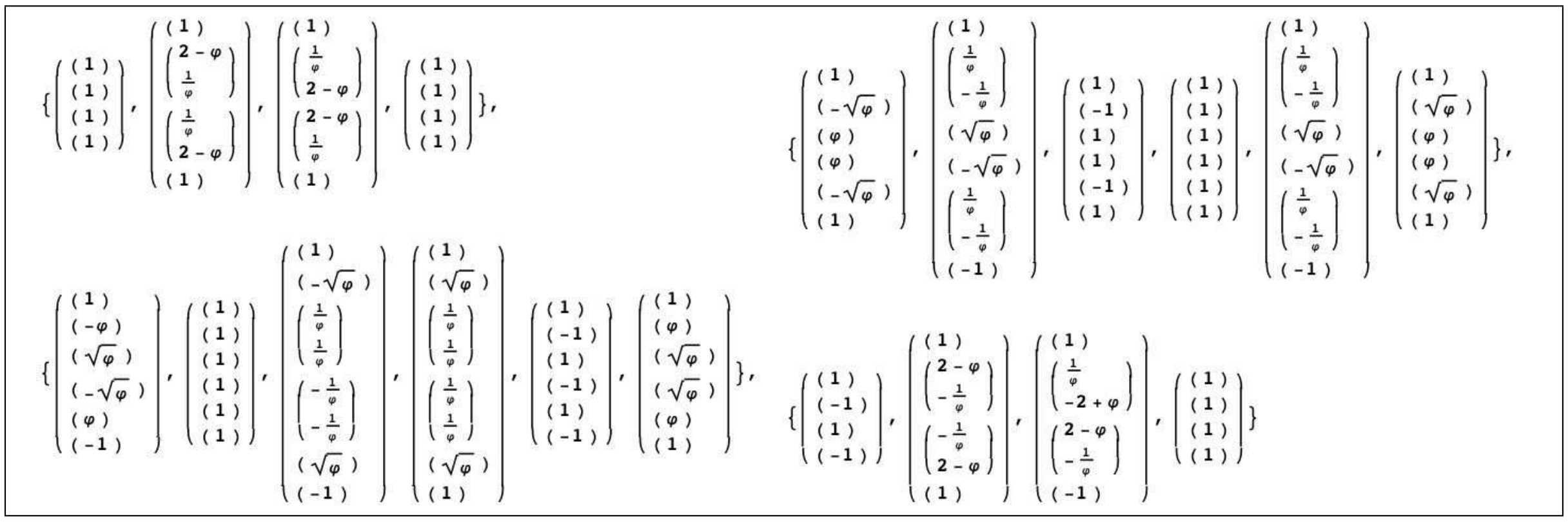

Figure 18: Inverse cells for $A_4$. Arguments are the same as for cells.

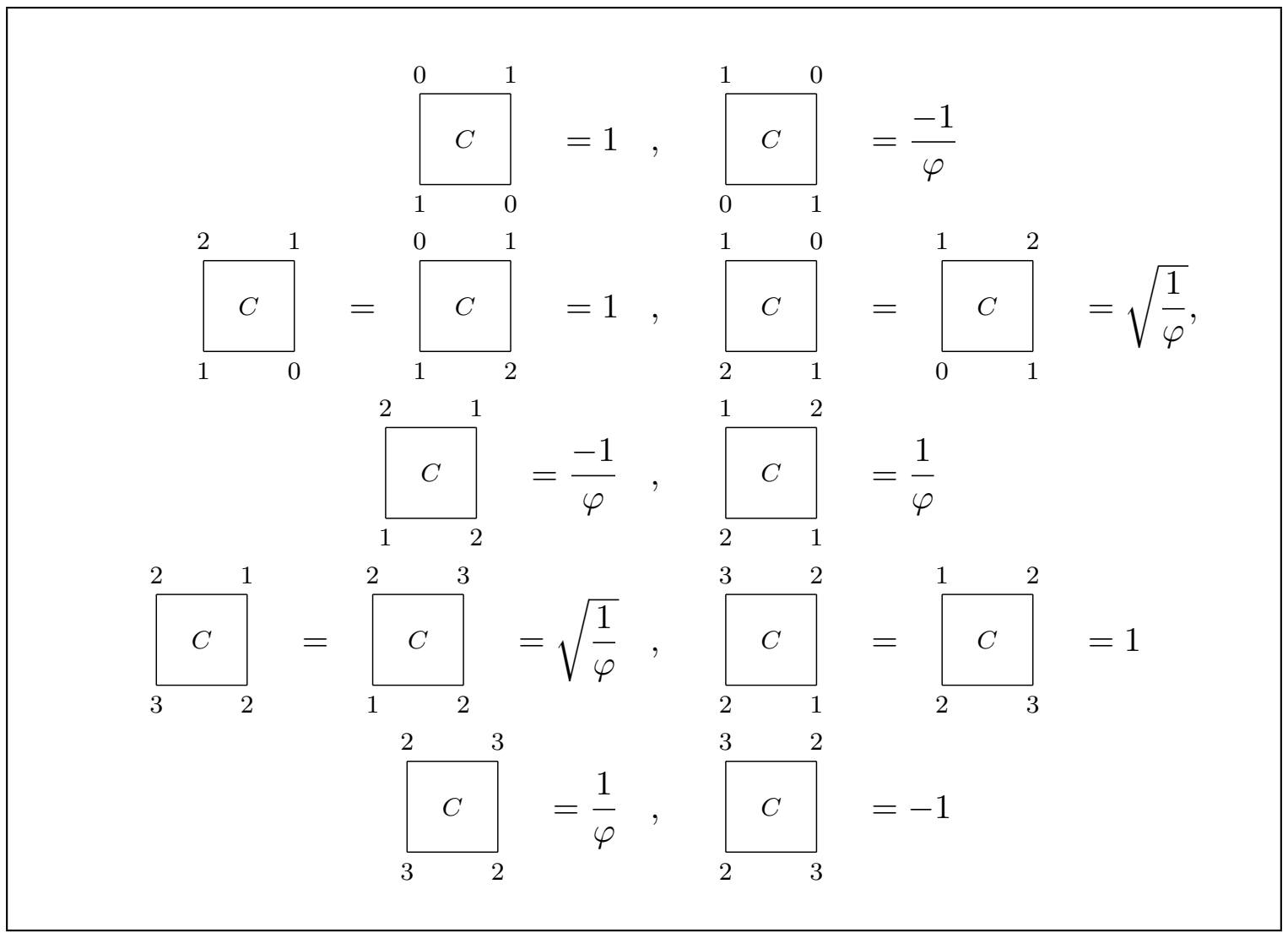

Figure 19: Basic cells for $A_4$

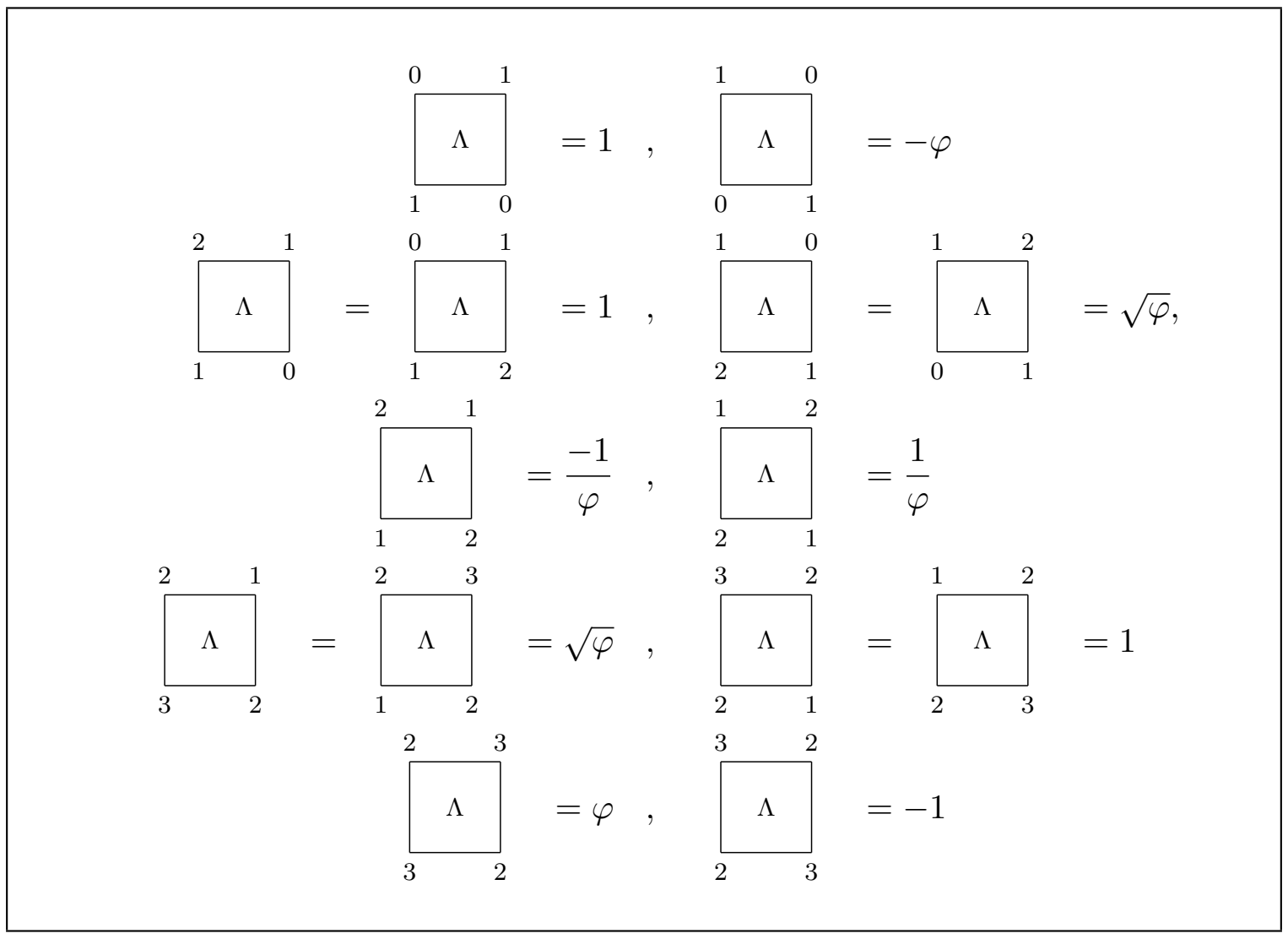

Figure 20: Basic inverse cells for $A_4$

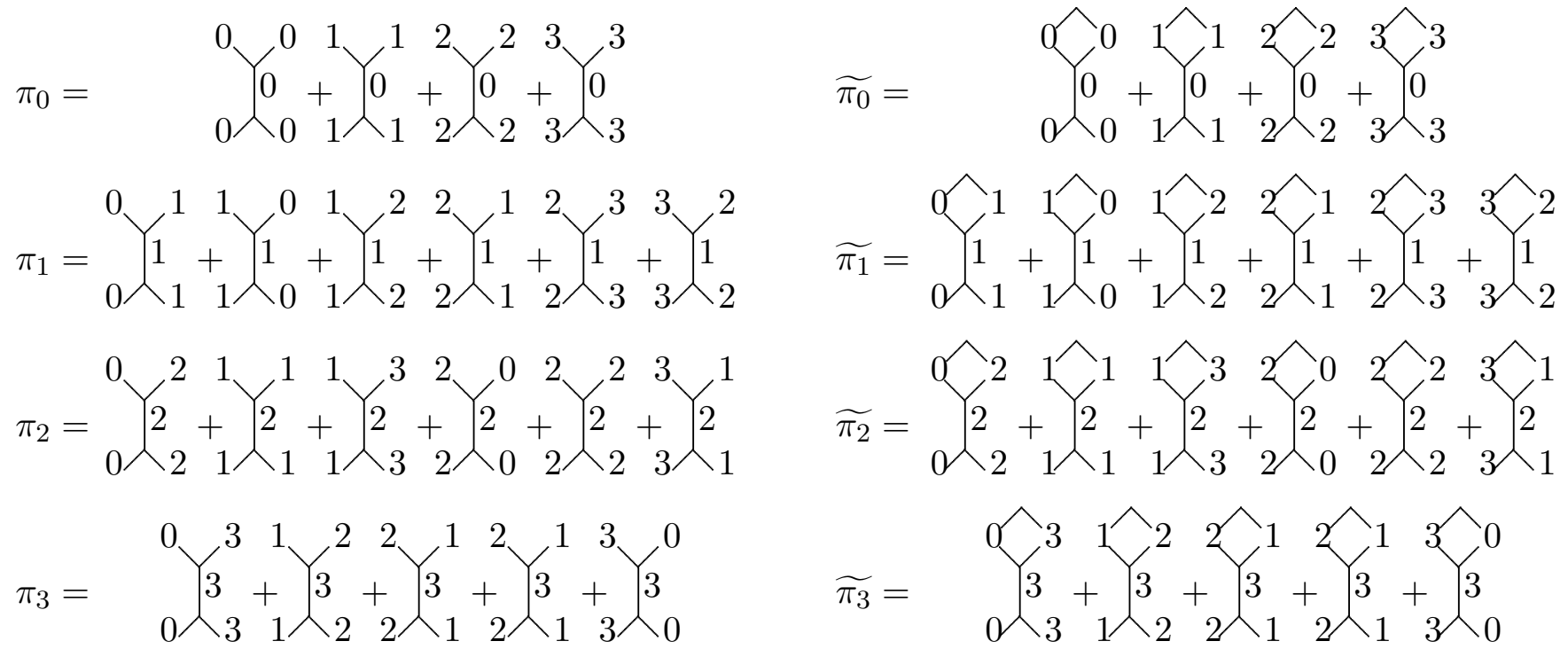

Figure 21: Projectors $\pi_n$ and characters $\widetilde{\pi_n}$ of $\mathcal{B}(A_4)$

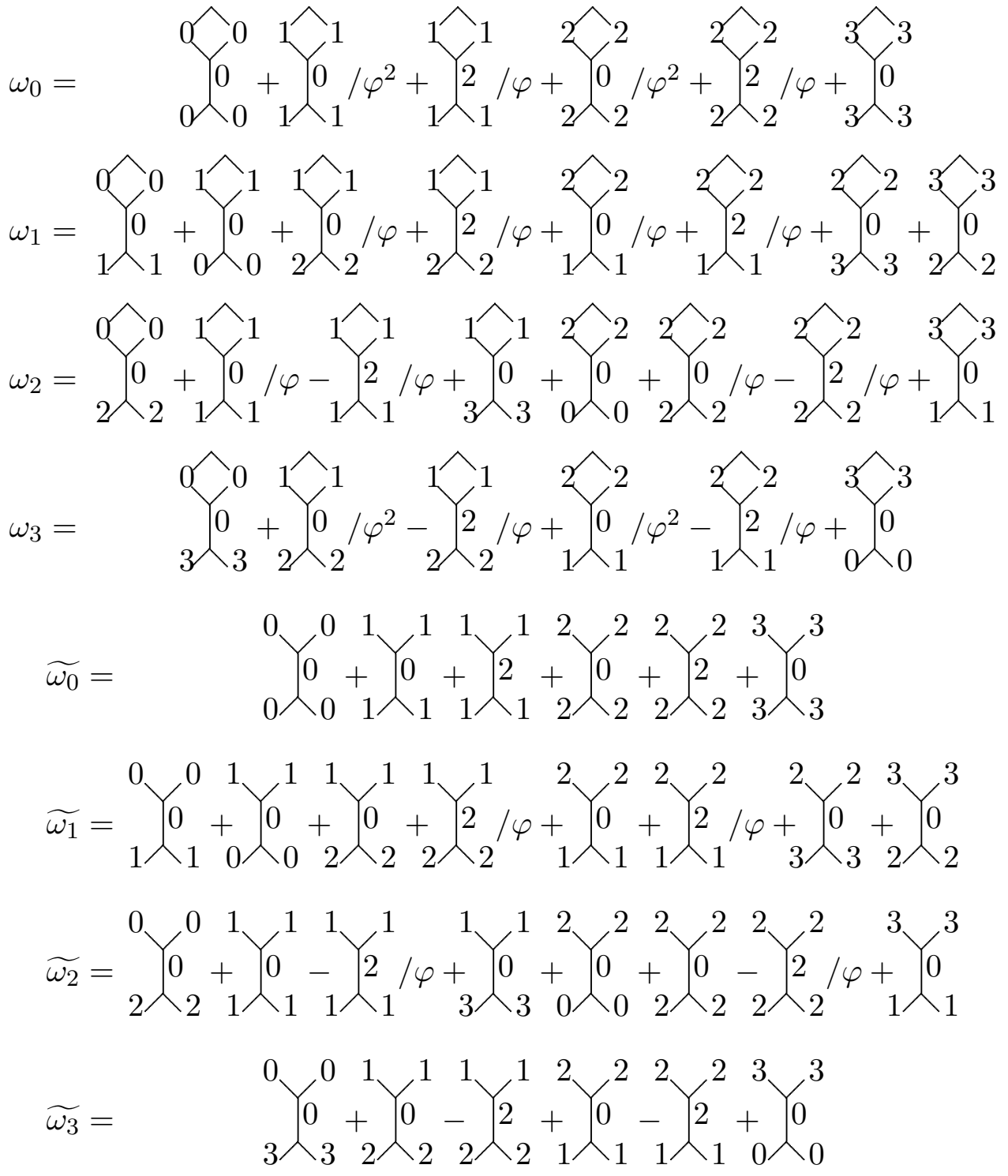

Figure 22: Projectors $\omega_x$ and characters $\widetilde{\omega_x}$ of $\widehat{\mathcal{B}}(A_4)$

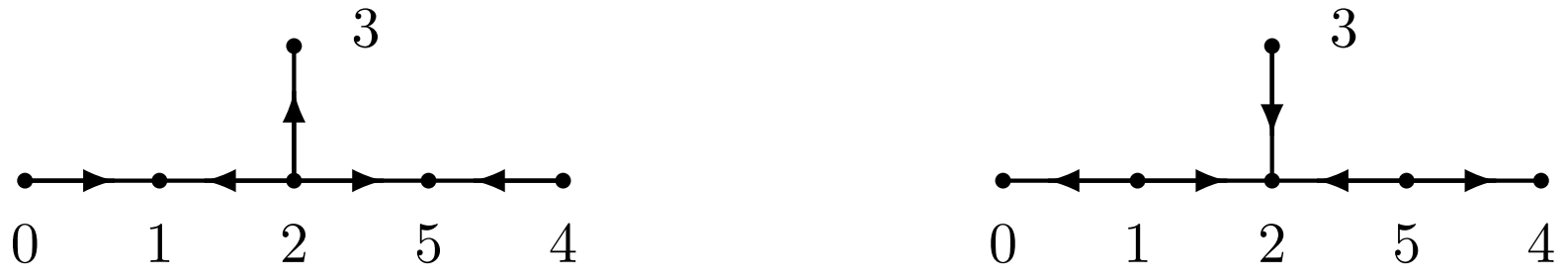

Figure 23: Direct and reflected quivers for $E_6$

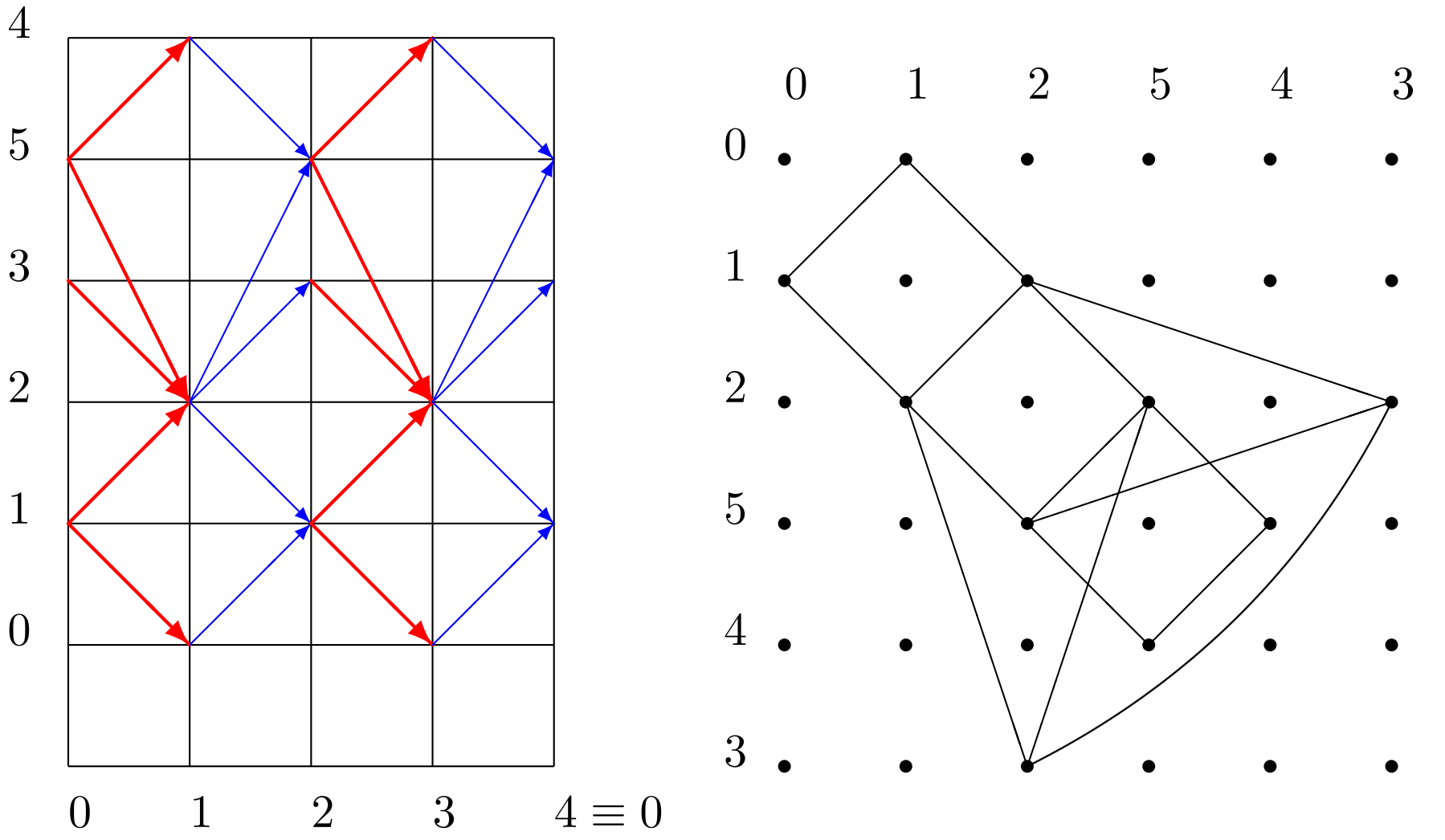

Figure 24: The $E_6$ quantum manifold and its basic cell diagram

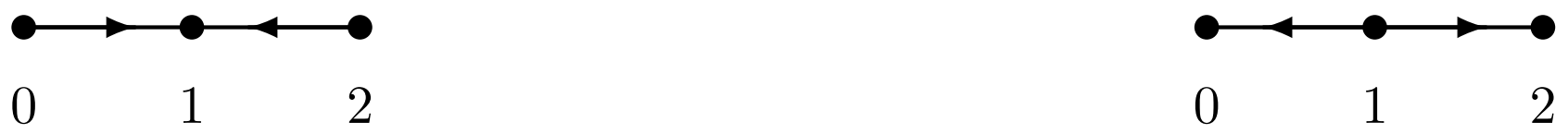

Figure 25: The direct and reflected quiver of $A_3$

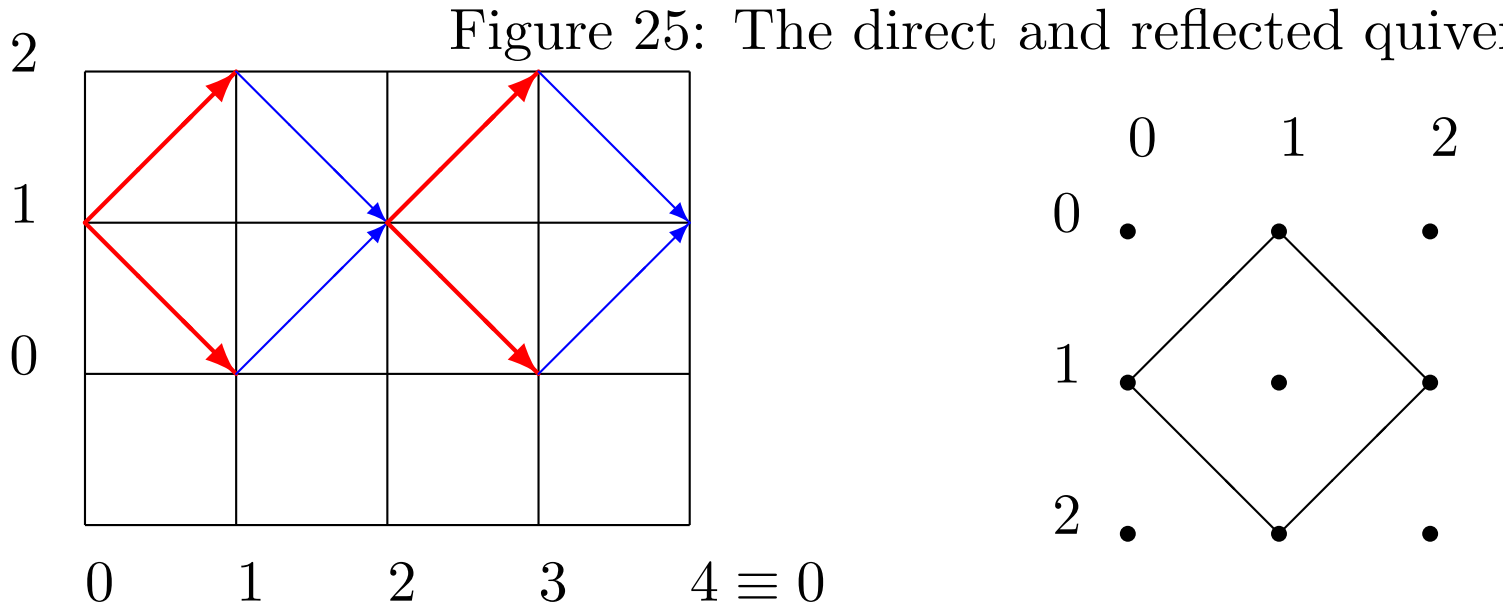

Figure 26: The $A_3$ quantum manifold and its basic cell diagram

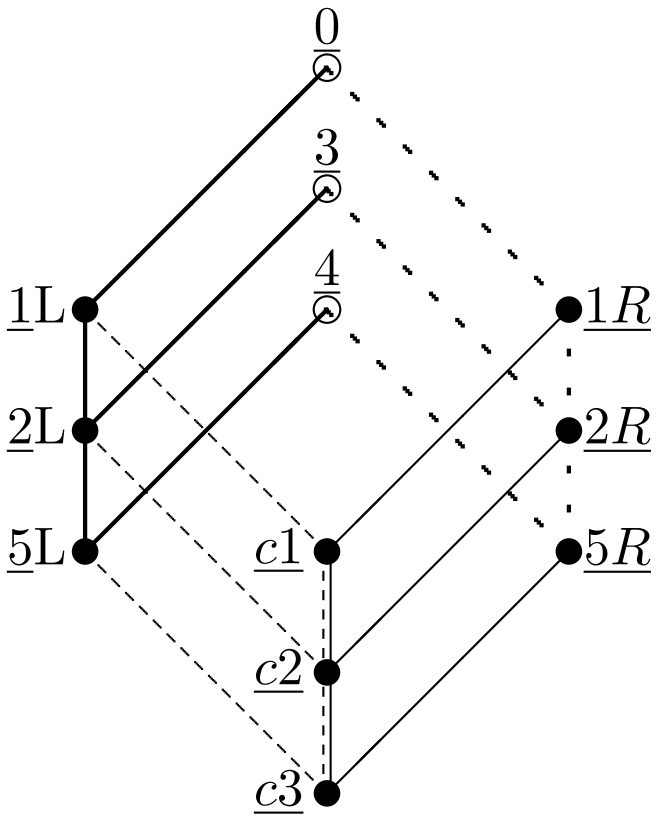

Figure 27: The $E_6$ Ocneanu graph